\documentclass[aps,pra,twocolumn,amsmath,amssymb,superscriptaddress,longbibliography]{revtex4-1}

\usepackage[english]{babel} 

\usepackage[utf8]{inputenc}
\usepackage[T1]{fontenc}

\usepackage{bbm}
\usepackage{amsmath}
\usepackage{verbatim}
\usepackage{graphicx}
\usepackage{times}
\usepackage{amssymb}
\usepackage{epsfig}
\usepackage{graphicx}
\usepackage{bm}
\usepackage{txfonts}
\usepackage{dsfont}
\usepackage{color}

\usepackage{subfigure}

\newcommand{\ket}[1]{|#1\rangle}
\newcommand{\bra}[1]{\langle#1|}
\newcommand{\Deff}{\Delta_{\mathrm{eff}}}
\newcommand{\Defft}{\tilde \Delta_{\mathrm{eff}}}

\newcommand{\fpb}{\partial_\beta}
\newcommand{\fpbs}{{\partial_{\beta^*}}}

\newcommand{\fpbsz}{{\partial^2_{\beta^*}}}

\newcommand{\fpbt}{{\partial_\beta}_t}
\newcommand{\fpbst}{{\partial_{\beta^*_t}}}

\newcommand{\fpbtq}{{(q\partial_{\beta_t}-p\partial_{\beta^*_t})  }}
\newcommand{\fpbstq}{{( q\partial_{\beta_t^*}-p\partial_{\beta_t})}}

\newcommand{\fpbsttq}{{( q\partial_{\beta_{t-\tau}^*}-p\partial_{\beta_{t-\tau}})}}

\newcommand{\ada}{a^\dagger a}

\newcommand{\Alpha}{\Xi}

\newcommand{\slfrac}[2]{\left.#1\middle/#2\right.}

\begin{document}

\selectlanguage{english}

\title{Laser Theory for Optomechanics: Limit Cycles in the Quantum Regime}

\author{Niels L\"orch}
\affiliation{Institut f\"{u}r Gravitationsphysik, Leibniz Universit\"{a}t Hannover and \\ Max-Planck-Institut f\"{u}r Gravitationsphysik (Albert-Einstein-Institut), Callinstra\ss{}e 38, 30167 Hannover, Germany}
\affiliation{Institut f\"ur Theoretische Physik, Leibniz Universit\"at Hannover, Appelstra\ss{}e 2, 30167 Hannover, Germany}

\author{Jiang Qian}
\affiliation{Arnold Sommerfeld Center for Theoretical Physics, Center for NanoScience and Department of Physics, Ludwig-Maximilians-Universit\"at M\"unchen, Theresienstrasse 37, 80333, M\"unchen, Germany}

\author{Aashish Clerk}
\affiliation{Department of Physics, McGill University, Montreal, Quebec, Canada H3A 2T8}

\author{Florian Marquardt}
\affiliation{Friedrich-Alexander-Universit\"at Erlangen-N\"urnberg, Staudtstr. 7, D-91058 Erlangen, Germany}
\affiliation{Max Planck Institute for the Science of Light, G\"unther-Scharowsky-Straße 1/Bau 24, D-91058 Erlangen, Germany}

\author{Klemens~Hammerer}
\affiliation{Institut f\"{u}r Gravitationsphysik, Leibniz Universit\"{a}t Hannover and \\ Max-Planck-Institut f\"{u}r Gravitationsphysik (Albert-Einstein-Institut), Callinstra\ss{}e 38, 30167 Hannover, Germany}
\affiliation{Institut f\"ur Theoretische Physik, Leibniz Universit\"at Hannover, Appelstra\ss{}e 2, 30167 Hannover, Germany}

\date{\today}

\begin{abstract}
Optomechanical systems can exhibit self-sustained limit cycles where the quantum state of the mechanical resonator possesses nonclassical characteristics such as a strongly negative Wigner density, as was shown recently in a numerical study by Qian et al. [Physical Review Letters, {\bf 109}, 253601 (2012)]. Here we derive a Fokker-Planck equation describing mechanical limit cycles in the quantum regime which correctly reproduces the numerically observed nonclassical features. The derivation starts from the standard optomechanical master equation, and is based on techniques borrowed from the laser theory due to Haake and Lewenstein. We compare our analytical model with numerical solutions of the master equation based on Monte-Carlo simulations, and find very good agreement over a wide and so far unexplored regime of system parameters. As one main conclusion, we predict negative Wigner functions to be observable even for surprisingly classical parameters, \emph{i.e.} outside the single-photon strong coupling 
regime, for strong cavity drive, and rather large limit cycle amplitudes. The {\color{black} approach} taken here provides a natural starting point for further studies of quantum effects in optomechanics.
\end{abstract}


\maketitle

\section{Introduction}

Optomechanical systems provide a test bed to study a broad range of paradigmatic quantum optical processes at so far unexplored meso- and macroscopic mass and length scales \cite{Aspelmeyer2013, Meystre2013, Chen2013}. That quantum effects can play an important and even dominating role in the dynamics of these systems has been shown in a number of recent experiments demonstrating cooling to the quantum ground state \cite{Teufel2011,Chan2011a}, ponderomotive squeezing of light \cite{Brooks2012,Safavi-Naeini2013}, back action noise limited position sensing \cite{Murch2008a,Purdy2013}, coherent state transfer \cite{Palomaki2013}, and entanglement \cite{Palomaki2013a}.

In the most elementary optomechanical setup a single cavity mode couples to a single mechanical oscillator through, e.g., radiation pressure or dipole gradient forces. The dynamics of the system depends crucially on the frequency of the external driving field applied to the cavity: For the purpose of position or force sensing as in \cite{Murch2008a,Purdy2013} the driving field is chosen resonant, while for back action cooling or state transfer the field is tuned below (to the red side of) the cavity frequency \cite{Teufel2011,Chan2011a,Palomaki2013}. For blue detuning the system exhibits a rather complex nonlinear behavior. When the driving field is swept from the red to the blue side the nonlinear dynamics sets in as a parametric amplification process where phonons and photons are created correlated in pairs \cite{Hofer2011}. This lies at the heart of the recently reported generation of optomechanical entanglement \cite{Palomaki2013a}. The amplification will finally go over into a regime of self-sustained limit
cycles due to the nonlinearity inherent to the optomechanical coupling. The classical dynamics in this regime has been observed experimentally \cite{Kippenberg2005,Carmon2005,Metzger2008,Anetsberger2009,Grudinin2010,Zaitsev2011} and is well studied theoretically \cite{Marquardt2006,Zaitsev2012,Khurgin2012,Khurgin2012a}. Motivated by the impressive progress towards quantum effects in optomechanical systems also the quantum regime of optomechanical limit cycles received significant attention in theoretical studies \cite{ Ludwig2008,Vahala2008,Rodrigues2010,Armour2012a,Qian2012,Nation2013, Dykman1975}.

In particular, a recent numerical study of the full optomechanical master equation in the limit cycle regime showed that the Wigner function of the mechanical oscillator can become strongly negative \cite{Qian2012}: Negativities of the Wigner function occur for driving fields at the blue sidebands and -- more pronounced -- also for resonant drive. Limit cycle states with negative Wigner density even exist in regions of red detuning where a (simple) classical model would not predict limit cycles at all. The numerical findings were independently confirmed in \cite{Nation2013}. {\color{black} This reference predicts negative Wigner density even on higher sidebands and compares the extend of negativity found for different detunings in more detail.} In view of these findings it is important to strive for a deeper understanding of these effects and the underlying mechanisms on the basis of an appropriate analytical model.

The transition from parametric amplification to optomechanical limit cycles can be understood in analogy to the threshold behaviour of a laser (or maser) cavity \cite{Lax1967,Haake1983,Gardiner2004b} where the roles of the laser cavity and the laser medium are played by, respectively, the mechanical oscillator and the optomechanical cavity \cite{Khurgin2010}.
Along this line a semiclassical rate equation model was derived in \cite{Khurgin2010, Khurgin2012} for optomechanical systems. Rodrigues and Armour \cite{Rodrigues2010,Armour2012a} developed a quantum mechanical treatment employing a truncated Wigner function approach to derive a Fokker-Planck equation (FPE) for the mechanical oscillator. The FPE predicted in particular a sub-Poissonian, or number-squeezed, phonon statistics in the limit cycle when the driving field is blue detuned from the cavity resonance by the mechanical oscillation frequency.

In the present article we apply the laser theory due to Haake and Lewenstein \cite{Haake1983,Gardiner2004b} to describe optomechanical limit cycles in the quantum regime. Our model correctly reproduces the characteristics of limit cycles mentioned above. It identifies general requirements on system parameters (such as coupling strength, driving power, sideband resolution, temperature etc.) for the occurrence of sub-Poissonian phonon statistics and negative Wigner functions, and establishes a tight connection between the two phenomena. We find that negative Wigner functions can be achieved also in rather classical parameter regimes where the coupling per single photon $g_0$ is smaller than the cavity line width, and where the cavity is driven strongly and limit cycle amplitudes are large. The associated small Fano factors are lower bounded by, and can reach, the sideband parameter $\kappa/\omega_{\rm m}$ (ratio of cavity line width to mechanical resonance frequency) for sufficiently strong optomechanical 
cooperativity.

Starting from the standard optomechanical master equation \cite{Aspelmeyer2013, Meystre2013} an effective FPE is derived for the  quasi-probability distribution (such as e.g. the Wigner--, $P$-- or $Q$--function) of the mechanical oscillator under adiabatic elimination of the cavity mode. The nonlinearity of the optomechanical interaction gives rise to nonlinear drift and diffusion coefficients in the FPE which describe, respectively, the (classical) nonlinear physics of limit cycles \cite{Marquardt2006,Zaitsev2012} and the impact of quantum noise of the cavity. The approach taken here permits to work
in a picture which interpolates between the dressed state picture introduced in \cite{Rabl2011,Nunnenkamp2011} through a polaron transformation and the bare state picture of the standard master equation \cite{Wilson-Rae2007,Aspelmeyer2013, Meystre2013,Rodrigues2010, Nation2013}.
{\color{black}Remarkably, in analogy to the polaron picture, this intermediate picture explicitly separates the optical Kerr-nonlinearity inherent to the radiation pressure from the optomechanical interaction. In contrast to the polaron picture, the interaction term is not removed from the master equation and both the mechanical oscillator and the cavity remain separate systems as in the standard master equation picture. The entanglement of cavity and oscillator in the polaron picture would complicate the study of them as separate systems as required in the context of limit cycles.  As we will show, the novel treatment of the optomechanical Kerr nonlinearity presented in this article can become essential to understand the physics of limit cycles.

The effective FPE derived here exactly reproduces the one of Rodriguez and Armour \cite{Rodrigues2010,Armour2012a} when neglecting the different description of the Kerr nonlinearity of the cavity, which is treated in the standard master equation picture there.} In comparison to \cite{Rodrigues2010,Armour2012a} our approach does not require truncation of higher order derivatives, and gives a consistent and natural account of the Kerr nonlinearity.

The article is organized as follows: In Sec.~\ref{sec:Summary} we give an executive summary of the main results, as far as they relate to the appearance of nonclassical mechanical
states. In Sec.~\ref{sec:lasertheory} we introduce the main idea of Haake and Lewenstein's laser theory in the context of optomechanics, and apply it to derive the effective FPE for the mechanical oscillator. Sec.~\ref{OLCR} we discuss the implications of the FPE equation for optomechanical limit cycles in the quantum regime. In principle each of these Sections can be read independently. Readers who are interested only in one particular aspect are encouraged to jump directly to the respective section.

\section{Preview of the main results}\label{sec:Summary}

The aim of this section is to give a preview of our most important results, and to indicate how these results could be derived in a relatively simple (quantum noise) approach. The main idea is to find the width of the mechanical limit cycles in phase space and to deduce from that the spread in phonon numbers. For simplicity, we will assume here that the optomechanical interaction dominates (i.e. formally zero mechanical damping). The full optomechanical laser-theory analysis will go significantly beyond this, but it will reproduce the features discussed here.

In the following, we will find it convenient to characterize the optomechanical coupling in several ways: as the cavity frequency shift per displacement $G$, via the single-photon coupling strength $g_{0}=Gx_{{\rm ZPF}}$, and via the dimensionless ratio $\eta=2g_{0}/\omega_{{\rm m}}$. $x_{{\rm ZPF}}=\sqrt{\hbar/2m\omega_{{\rm m}}}$ is the zero-point amplitude of the mechanical oscillator with mass ${ m}$ and frequency $\omega_{\rm m}$. We start by assuming mechanical oscillations at a fixed amplitude $r$ such that $x(t)=x_{{\rm ZPF}}{\rm Re}[re^{-i\omega_{m}t}]$. At each instant of time, the radiation pressure force ${F}=\hbar G{a}^{\dagger}{a}$ ($a$ is the photon annihilation operator) will feed energy into the mechanical oscillations at a rate (power) ${P}={F}(t)\dot{x}(t)$. Following the classical approach \cite{Marquardt2006}, we can predict the slow drift of the mechanical oscillation amplitude by calculating the average power input, $\overline{\left\langle {P}(t)\right\rangle }$. Here $\left\langle \cdot \right\rangle $ denotes the quantum expectation value, while the overbar averages over a time-window comprising several oscillation periods. {\color{black}We note that the power balance equation is analogous to the loss-gain equations in a laser and that the laser analogy will be heavily used throughout the manuscript. }

In addition to this drift, however, there will be diffusion of the mechanical oscillator's energy, due to the fundamental radiation pressure shot noise fluctuations. The energy diffusion constant is given by
$D_{E}=\frac{1}{2}\int_{-\infty}^{+\infty}d\tau\,\overline{\left\langle \delta{P}(t+\tau)\delta{P}(t)\right\rangle }$,
where $\delta{P}(t)={P}(t)-\overline{\left\langle {P}(t)\right\rangle }$
denotes the fluctuations. In order to discuss the quantum dynamics of optomechanical limit cycles, it turns out to be crucial to study the behaviour of this diffusion constant as a function of cycle amplitude. In particular, we will show that the appearance of nonclassical mechanical states can only be understood by a rather subtle cancellation of some term that would usually dominate, leaving the diffusion constant small and leading to a narrowing of the phonon distribution by the sideband ratio $\kappa/\omega_{{\rm m}}$ (where $\kappa$ denotes the cavity amplitude decay rate).

Our task of calculating this diffusion constant is complicated by the fact that we are dealing with shot noise inside an optical cavity whose resonance frequency oscillates sinusoidally. We thus have to
solve the equation for the light field inside such a mechanically driven cavity, i.e. ${\rm d}{a}/{\rm d}t=[-i(\Delta-Gx(t))-\kappa]{a}+\sqrt{2\kappa}{a}_{{\rm in}}(t)$,
where $\Delta=\omega_{L}-\omega_{{\rm c}}$ is the detuning between the laser at frequency $\omega_L$ and the bare cavity resonance at $\omega_{\rm c}$. The solution for ${a}(t)$ can be expressed via the extra phase $\theta(t)=\eta{\rm Im}[re^{-i\omega_{m}t}]$ accumulated in the cavity field due to the oscillations. It reads

\begin{equation}
{a}(t)=e^{-i\theta(t)}\int_{-\infty}^{t}dt'\,\chi_{c}(t-t')e^{i\theta(t')}{a}_{{\rm in}}(t')\,,\label{eq:amplitudeEquationSolution}
\end{equation}
where $\chi_{c}(t)=\sqrt{2\kappa}\exp[(i\Delta-\kappa)t]$ is the
standard cavity filter function (and $\theta=0$ recovers the usual
case).

The light intensity oscillates at harmonics of the mechanical motion.
$\alpha(t)=e^{i\theta(t)}\left\langle {a}(t)\right\rangle =\sum_{n}\alpha_{n}e^{in\omega_{{\rm m}}t}$
is the average cavity amplitude (modulo the phase), which can be obtained
by evaluating Eq.~(\ref{eq:amplitudeEquationSolution}).
For a constant laser drive, with an amplitude $\sqrt{2\kappa}\left\langle {a}_{{\rm in}}\right\rangle \equiv E$,
we obtain $\alpha_{n}=Ee^{-in\phi}J_{-n}(\eta r)/h_{n}$, where $h_{n}=\kappa+i(n\omega_{{\rm m}}-\Delta)$.
These are the Bessel amplitudes that also determine the appearance
of multiple stable attractors in the classical analysis of the optomechanical
instability \cite{Marquardt2006}. These attractors can be found by noting that the drift of the amplitude $r$ is governed by the power input $\overline{\left\langle {P}(t)\right\rangle }$,
as the energy of the mechanical oscillator is given by $m\omega_{{\rm m}}^{2}r^{2}/2$.
In the regime of interest here, this drift can be approximated as
\begin{align}\label{eq:DriftApproxExpression}
\dot{r} & \equiv \mu(r)=\frac{\overline{\left\langle {P}(t)\right\rangle }}{m\omega_{{\rm m}}^{2}r} \simeq \frac{2\kappa g_{0}E^{2}}{\omega_{{\rm m}}^{2}}\frac{\Delta}{\Delta^{2}+\kappa^{2}}J_{0}(\eta r)J_{1}(\eta r)\,.
\end{align}
The limit cycle amplitude is thus fixed at the zeroes of the Bessel
function, in the absence of additional mechanical damping. This will
be crucial further below.

In addition, there are the electromagnetic vacuum fluctuations $\delta{a}_{{\rm in}}(t)={a}_{{\rm in}}(t)-\langle {a}_{{\rm in}}\rangle$
entering the cavity. In order to evaluate the mechanical energy diffusion
constant that is governed by those fluctuations, we need the force-force
correlator $\left\langle {F}(t){F}(t')\right\rangle $, i.e.
ultimately the shot-noise (irreducible) part of the photon number
correlator. By using the vacuum noise correlator $\langle \delta{a}_{{\rm in}}(t)\delta{a}_{{\rm in}}^{\dagger}(0)\rangle =\delta(t)$,
we find directly
\begin{equation*}
\langle {a}^{\dagger}(t){a}(t){a}^{\dagger}(t'){a}(t')\rangle _{{\rm SN}}=e^{i\Delta(t-t')-\kappa\left|t-t'\right|}\alpha^{*}(t)\alpha(t')\,.
\end{equation*}
For a constant $\alpha$, this reduces to the shot noise correlator
employed in the quantum noise approach to optomechanical cooling \cite{Marquardt2007}. Now, we can proceed to evaluate the energy diffusion constant $D_{E}$ introduced above. The resulting slightly lengthy expression (Eq.~\eqref{DiffW} in Appendix~\ref{adi}) can be simplified in the regime of interest here to
\begin{align}\label{eq:Diffusion}
D_{W} & \equiv  \frac{D_{E}}{(2\hbar\omega_{{\rm m}}{r})^2}
 \simeq \kappa\frac{g_{0}^{2}E^{2}}{\omega_{{\rm m}}^{4}}\left\{ \frac{1}{2}\frac{\omega_{{\rm m}}^{2}}{\kappa^{2}+\Delta^{2}}J_{0}^{2}(\eta r)+J_{1}^{2}(\eta r)\right\}.
\end{align}
Here we have introduced $D_{W}$ as the diffusion constant for the
amplitude $r$ of the limit cycle. This amplitude is connected to
the energy via $E=\hbar\omega_{{\rm m}}r^{2}$, such that one obtains
the relation between $D_{E}$ and $D_{W}$ shown here. It is now crucial to observe that the diffusion constant has a minimum
right at the first limit cycle. This is because the first contribution
in Eq.~(\ref{eq:Diffusion}), which dominates at smaller amplitudes,
is completely suppressed at the limit cycle, where $J_{0}(\eta r)=0$,
see Eq.~(\ref{eq:DriftApproxExpression}). Thus, only the second
term survives, which is suppressed by a factor $\kappa^{2}/\omega_{{\rm m}}^{2}$,
i.e. the sideband ratio squared. We show in the main text that this suppression is caused by squeezing terms that exactly cancel the corresponding incoherent diffusion terms in leading order.

Now we can combine these results to discuss the width $\sigma^{2}$
of the distribution in the amplitude $r$. The compromise between the diffusion at rate $D_{W}$ and the restoring force that drives $r$ back to the limit cycle results in a width $\sigma^{2}=-D_{W}/\mu'$. For a fixed limit cycle amplitude, both diffusion and drift scale as $g_{0}^{2}E^{2}$, such that the laser power and the optomechanical coupling drop out of this expression. This will change in the presence of mechanical damping and thermal fluctuations, but it still correctly describes the behavior once the optomechanical damping rate overwhelms the thermal fluctuations.

In order to estimate when the limit cycle may turn into a nonclassical mechanical quantum state, we will now look at the variance of the phonon number ${\rm Var}(n)$. Since $r$ is already measured in terms of the zero-point amplitude $x_{{\rm ZPF}}$, we have $r^{2}=n$. Thus ${\rm Var}(n)=4\left\langle n\right\rangle \sigma^{2}$. This can be minimized by choosing an optimal detuning ($\Delta=\kappa$), where we find ${\rm Var}(n)=\left\langle n\right\rangle (\kappa/\omega_{{\rm m}})$. In other words, in the resolved sideband regime ($\kappa\ll\omega_{{\rm m}}$),
one can get close to a mechanical Fock state, ${\rm Var}(n)<1$, as
long as the limit cycle is sufficiently small, $\left\langle n\right\rangle <\omega_{{\rm m}}/\kappa$. Note that the optomechanical coupling strength $g_{0}$ enters indirectly here, since (in the absence of mechanical damping) the limit cycle amplitude is determined by $J_{0}(\eta r)=0$, with $\eta=2g_{0}/\omega_{{\rm m}}$. Taking this into account, ${\rm Var}(n)<1$ is equivalent to  $g_{0}^{2}/\omega_{{\rm m}}\kappa>1.4$. However, it turns out that it is easier to produce a nonclassical state, i.e. one where the Wigner density has negative components. For the type of states relevant here, we numerically find that it is sufficient to have ${\rm Var}(n)<0.6\left\langle n\right\rangle ^{0.7}$ for this purpose. Thus, the condition for nonclassicality reads approximately
\begin{equation}\label{g0requ}
\frac{g_{0}}{\kappa}>2\left(\frac{\kappa}{\omega_{\rm m}}\right)^{0.7}\,,
\end{equation}
which is less stringent than the condition for achieving a Fock state, since
one could still admit $g_{0}/\kappa<1$ if the sideband ratio $\omega_{{\rm m}}/\kappa$ is sufficiently large.

In the simplified description given here, we have neglected several
factors which will be discussed
in our full analysis. This
includes the effects of the mechanical damping, which will decrease
the limit cycle amplitude (shifting away from the point of minimum
diffusion constant). In addition, thermal fluctuations will add to
the diffusion. Nevertheless, this effect can be overcome if the scale
of the optically induced damping rate, $\gamma_{{\rm opt}}\simeq\mu'(r)\propto g_{0}^{2}E^{2}/\omega_{{\rm m}}^{3}$, dominates the influx of thermal phonons
\begin{equation}\label{eq:powerrequ}
\left|\gamma_{{\rm opt}}\right|\gg\gamma\bar{n}\,,
\end{equation}
where $\bar{n}$ is the thermal phonon number of the bath, and $\gamma$
is the mechanical damping rate. This is equivalent to the condition
for ground state cooling, but here applied for the instable regime.
It does not involve the coupling per single photon $g_{0}$, but only the linearized coupling $g\propto g_{0}E$, such that Eq.~\eqref{eq:powerrequ} essentially represents a condition on the strength of the driving field.

Another important aspect neglected here is the shift of the cavity
resonance by the Kerr effect.
This leads to an effective detuning $\Delta_{{\rm eff}}$
that will enter all expressions instead of $\Delta$. The impact of
this change is especially large near $\Delta\approx0$, which is precisely
the regime which we find to be optimal for nonclassical states.

The heuristic reasoning applied here and the resulting conditions \eqref{g0requ} and \eqref{eq:powerrequ} for achieving nonclassical mechanical states will receive a rigorous justification in Sec.~\ref{OLCR} on the basis of the Fokker-Planck equation derived in the next Section.

\section{Laser Theory for Optomechanics}\label{sec:lasertheory}

\subsection{Haake-Lewenstein Laser Theory Ansatz in Optomechanics}

\paragraph*{Master Equation ---} The standard master equation of an optomechanical system is \cite{Aspelmeyer2013,Meystre2013}
  \begin{align}\label{eq:MEQRF}
    \frac{\mathrm{d}}{\mathrm{d}t}\,\rho&=\left(L_\mathrm{m}+L_\mathrm{c}+L_\mathrm{int}\right)\rho
  \end{align}
where
    \begin{align}
    L_\mathrm{m}\rho&=-i\left[\omega_\mathrm{m} b^\dagger b,\rho\right]+\gamma(\bar{n}+1)D[b]\rho+\gamma\bar{n}D[b^\dagger]\rho,\label{eq:MEQRF_Lm}\\
    L_\mathrm{c}\rho&=-i\left[-\Delta a^\dagger a-iE\left(a-a^\dagger\right),\rho\right]
    +\kappa D[a]\rho \label {origk} \\
    L_\mathrm{int}\rho&=-i\left[-g_0a^\dagger a\left(b +b^\dagger \right),\rho\right].\label{eq:Lint}
\end{align}
The three Liouvillians $L_\mathrm{m}$, $L_\mathrm{c}$, and $L_\mathrm{int}$ refer to the mechanical oscillator, the cavity, and their interaction respectively. $a$ and $b$ denote the annihilation operators of the cavity and the mechanical oscillator. The frequency of the mechanical oscillator is $\omega_m$, its amplitude damping rate is $\gamma=\omega_\mathrm{m}/Q_\mathrm{m}$, and its mean phonon number in thermal equilibrium $\bar{n}$. We use the notation $D[A]\rho=2A\rho A^\dagger-A^\dagger A\rho-\rho A^\dagger A$ for Lindblad operators.
$\kappa$ is the cavity amplitude decay rate, $\Delta=\omega_\mathrm{L}-\omega_\mathrm{c}$ is the detuning from cavity resonance at $\omega_\mathrm{c}$ of the driving field $E=\sqrt{2\kappa P_L/\hbar\omega_L}$ with power $P_L$ and frequency $\omega_\mathrm{L}$. The master equation is written in a frame rotating at the frequency $\omega_\mathrm{L}$ of the driving field. The optomechanical coupling per single photon is denoted by $g_0$, and essentially determines the dispersive shift of the cavity frequency with
the displacement of the oscillator in units of the
mechanical zero-point amplitude \footnote{The zero point amplitude is $\sqrt{\hbar/m\omega_\mathrm{m}}$ for an oscillator of mass $m$}.

Note that in contrast to e.g. \cite{Qian2012, Marquardt2006, Ludwig2008} the definitions of $\gamma$ and $\kappa$ used here refer to the decay rate of the \emph{amplitude} and will be used for all analytical results, in order to make the equations more readable. The corresponding decay rates for the energy $\kappa_E=2 \kappa$ and $\gamma_E=2 \gamma$ are the standard convention from \cite{Aspelmeyer2013}. For comparison to most experimental and numerical studies, we provide also the energy decay rates in the numerical results.

Our primary aim is to derive an effective equation of motion for the mirror based on the assumption that the dynamics of the cavity adiabatically follows the mechanical oscillator. This will be strictly the case when the cavity decay rate $\kappa$ is larger than the characteristic coupling strength of the oscillator and the cavity mode (\textit{i.e.} $g_0$ or the linear coupling $g=g_0\alpha$ enhanced by the mean cavity field $\alpha$ at the position of the limit cycle). As we will see, the resulting effective equation of motion for the mechanical oscillator gives good results for the stationary state also when this condition is fulfilled barely, and even when it is mildly violated.

\paragraph*{Quasiprobability distribution ---} Most importantly, we will not assume the usual linearization of the optomechanical coupling when we perform the adiabatic elimination. This is achieved by means of an Ansatz inspired by laser theory \cite{Haake1983,Gardiner2004b}, which allows us to use a \emph{different} adiabatic reference state of the cavity field for \emph{each} point in phase space of the mechanical oscillator. The idea is to switch to a phase-space representation for the mechanical degree of freedom. In principle any quasi-probability distribution (\textit{e.g.} $P$-distribution, Wigner function etc.) can be used, but we will in the following mostly focus on the (Husimi) $Q$ function which yields the simplest formulas for the calculation presented below. In this formalism the density operator $\rho$ is replaced
by
\[
\sigma(\beta,\beta^*)=\frac{1}{\pi}\langle\beta|\rho|\beta\rangle
\]
where $|\beta\rangle$ is a coherent state of the mechanical oscillator. In the Appendix we provide an extension and comparison of the present approach based on the $Q$ function to a general ($s$-parameterized) quasi-probability distribution including the $P$-distribution and Wigner function as special cases. $\sigma(\beta,\beta^*)$ is a density operator for the cavity field and a quasi-probability distribution for the oscillator over the complex phase space variables $(\beta,\beta^*)$. The reduced density operator for the cavity is obtained by integrating over phase space,
\[
\rho_\mathrm{c}=\mathrm{tr}_\mathrm{m}\{\rho\}=\int \mathrm{d}^2\beta\,\sigma(\beta,\beta^*),
\]
and the quasi-probability distribution ($Q$ function) for the oscillator follows on taking the trace over the cavity,
\begin{align}\label{eq:Q}
Q(\beta,\beta^*)=\mathrm{tr}_\mathrm{c}\{\sigma(\beta,\beta^*)\}.
\end{align}
$\sigma(\beta,\beta^*)$ itself still contains all information about the state of both systems, and is fully equivalent to the density operator $\rho$. For the $Q$ function the replacement rules \cite{Gardiner2004b}
\begin{align}\label{eq:replrules}
  b^\dagger\rho&\rightarrow \beta^* \sigma(\beta,\beta^*), & b\rho&\rightarrow \left(\beta+\partial_{\beta^*}\right)\sigma(\beta,\beta^*),
\end{align}
and their adjoints can be applied to the master equation~\eqref{eq:MEQRF} in order to arrive at an equivalent description in phase space of the oscillator. We use the notation $\partial_\beta$ to denote the partial derivative with respect to a variable $\beta$. The translated equation of motion is
\begin{align}\label{eq:timedependentcavity}
\partial_t\, \sigma(\beta,\beta^*,t)&=\left(\mathcal L_\mathrm{m} + \mathcal L_\mathrm{c}+\mathcal L_\mathrm{int}\right) \sigma(\beta,\beta^*,t)
\end{align}
with
\begin{align}
\mathcal L_\mathrm{m} \sigma&= \left( \partial_{\beta}(\gamma-i\omega_\mathrm{m}) \beta
+\mathrm{c.c.}  
\right) \sigma+2\gamma (\bar n +1) \partial^2_{\beta\beta^*}  \sigma\label{eq:timedependentcavityLm}\\
\mathcal L_\mathrm{c} \sigma&=L_c \sigma-i \left[-g_0 \left(\beta +\beta^* \right)a^\dagger a, \sigma \right]\nonumber\\
&=-i\left[-\left(\Delta+2g_0\mathrm{Re}(\beta)\right) a^\dagger a-iE\left(a-a^\dagger\right),\sigma\right]
    +\kappa D[a]\sigma\nonumber\\
\mathcal L_\mathrm{int} \sigma&=-ig_0\left( \partial_{\beta} \sigma a^\dagger a-    \partial_{\beta^*}a^\dagger a\sigma\right).\label{eq:timedependentcavityLint}
\end{align}
The Liouvillian $\mathcal L_\mathrm{m}$ affects only the mechanical oscillator, and is just the Fokker-Planck version of Eq.~\eqref{eq:MEQRF_Lm}. A crucial point in this formalism is that the nonlinear optomechanical interaction $L_\mathrm{int}$ from Eq.~\eqref{eq:Lint} makes a contribution to both, the new Liouvillian for the cavity $\mathcal{L}_\mathrm{c}$ and the new interaction $\mathcal{L}_\mathrm{int}$. Parts of the interaction can thus formally be treated as a shift of the detuning by $2g_0 \mathrm{Re}(\beta)$ which depends on the phase space variables $(\beta,\beta^*)$. Note that Eq.~\eqref{eq:timedependentcavity} is still exactly equivalent to \eqref{eq:MEQRF}.

\paragraph*{A Semi-Polaron-Transformation  ---} The parametric dependence of the cavity detuning on the phase space variables can be transformed into one of the driving field $E$ by means of a transformation
      \begin{align}\label{eq:semipolQ}
      \tilde\sigma(\beta,\beta^*,t) &=e^{\eta(\beta-\beta^*) a^\dagger a/2} \sigma(\beta,\beta^*,t) e^{-\eta(\beta-\beta^*) a^\dagger a/2}\\
      &=e^{i\theta(\beta,\beta^*) a^\dagger a} \sigma(\beta,\beta^*,t) e^{-i\theta(\beta,\beta^*) a^\dagger a}\nonumber,
      \end{align}
      with
      \begin{align*}
      \theta(\beta,\beta^*)&=\eta\,\mathrm{Im}(\beta), & \eta&=\frac{2g_0}{\omega_m}.
      \end{align*}
When transforming the equation of motion \eqref{eq:timedependentcavity} care has to be taken on commuting the unitary operators in \eqref{eq:semipolQ} with derivatives with respect to $(\beta,\beta^*)$ in $\mathcal L_\mathrm{m}$ and $\mathcal L_\mathrm{int}$ due to the $\beta$-dependence of $\theta$. Details are given in App.~\ref{App:SemiPolTraf}. The resulting equation of motion for $\tilde\sigma(\beta,\beta^*,t)$ can be written again in the form of Eq.~$\eqref{eq:timedependentcavity}$,
\begin{align}\label{eq:timedependentcavity1}
\partial_t\, {\tilde\sigma}(\beta,\beta^*,t)&=\left(\mathcal L_\mathrm{m} + \tilde{\mathcal L}_\mathrm{c}+\mathcal L_\mathrm{int}\right) {\tilde\sigma}(\beta,\beta^*,t),
\end{align}
where $\mathcal L_\mathrm{m}$ and $\mathcal L_\mathrm{int}$ remain unchanged as in \eqref{eq:timedependentcavityLm} and \eqref{eq:timedependentcavityLint}, and the Liouvillian operator for the cavity becomes
\begin{multline}\label{eq:tildeLc}
\tilde{\mathcal L}_c \tilde\sigma=-i\left[-\Delta a^\dagger a-K(a^\dagger a)^2-iE\left(e^{-i\theta(\beta,\beta^*)}a-\mathrm{h.c.}\right),\tilde\sigma\right]\\
    +\kappa D[a]\tilde\sigma.
\end{multline}
In this picture the phase of the driving field is different for each point in  phase space (via $\theta(\beta,\beta^*)$), and the cavity acquires an effective Kerr nonlinearity of strength
\[
K=\frac{g_0^2}{\omega_\mathrm{m}}.
\]
We point out that the effective Kerr nonlinearity of the optomechanical interaction gives rise to ponderomotive squeezing of light, as was recently observed in \cite{Brooks2012,Safavi-Naeini2013}. 

The equation of motion for $\tilde\sigma(\beta,\beta^*,t)$, Eq.~\eqref{eq:timedependentcavity1}, is an approximation. In principle it contains further terms which are of order $Q_\mathrm{m}^{-1}$ and whose explicit form is given in App.~\ref{semipolcorr}. For high quality oscillators these terms provide only small corrections and, therefore, will be dropped in the following. Apart from this approximation Eq.~\eqref{eq:timedependentcavity1} still contains the full nonlinear dynamics of the system, while the aspect of the optical Kerr nonlinearity is explicitly separated from the nonlinearity in the optomechanical interaction. It is also important to note that the quasiprobability distribution for the reduced state of the oscillator still follows from the transformed state ${\tilde\sigma}(\beta,\beta^*,t)$ in Eq.~\eqref{eq:semipolQ} by taking the partial trace over the cavity
\begin{align}\label{eq:Q_pol}
Q(\beta,\beta^*)=\mathrm{tr}_\mathrm{c}\{\tilde\sigma(\beta,\beta^*)\}.
\end{align}

\paragraph*{Semi-Polaron- versus Polaron-Transformation  ---} The transformation in Eq.~\eqref{eq:semipolQ} has many parallels with the polaron transformation \cite{Mahan2000} which has been applied fruitfully to optomechanical systems in order to describe single-photon strong coupling effects \cite{Rabl2011,Nunnenkamp2011}. The polaron transformation is effected by a unitary transformation of the density operator
\begin{align}
\tilde\rho_\mathrm{pol}&=e^{\eta(b-b^\dagger)a^\dagger a/2}\rho e^{-\eta(b-b^\dagger)a^\dagger a/2}\label{eq:rhotilde}
\end{align}
which should be compared to the transformation in Eq.~\eqref{eq:semipolQ}. Instead of \eqref{eq:MEQRF} the transformed state $\tilde\rho$ fulfills a transformed master equation
\begin{align}
    \dot{\tilde\rho}_\mathrm{pol}&=(L_\mathrm{m}+\tilde{L}_\mathrm{c})\tilde\rho_\mathrm{pol},\label{eq:MEQ_Polaron}\\
    \tilde L_\mathrm{c}\tilde\rho_\mathrm{pol}&=-i\left[-\Delta a^\dagger a-K(a^\dagger a)^2-iE\left(e^{-\eta(b-b^\dagger)/2}a-\mathrm{h.c.}\right),\tilde\rho_\mathrm{pol}\right]\nonumber\\
    &\quad+\kappa D\left[e^{-\eta(b-b^\dagger)/2}a\right]\tilde\rho_\mathrm{pol},\label{eq:Lc_Polaron}
\end{align}
where $L_\mathrm{m}$ is given in Eq.~\eqref{eq:MEQRF_Lm}. This equation is again correct up to terms of order $Q_\mathrm{m}^{-1}$. It is instructive to compare the master equation in the polaron picture \eqref{eq:MEQ_Polaron} to the equation of motion~\eqref{eq:timedependentcavity1} attained in our ``semi-polaron transformation''. In both equations of motion the Liouvillians for the cavity, Eqs.~\eqref{eq:tildeLc} and \eqref{eq:Lc_Polaron} respectively, exhibit a Kerr nonlinearity and contain a driving field whose phase depends on the momentum of the oscillator. Crucially, the polaron transformation  changes the jump operator describing cavity decay from $a$ to $e^{i\eta(b-b^\dagger)/2}a$, and entirely removes the interaction term \eqref{eq:Lint}. Moreover, since the polaron picture corresponds to a transformation into dressed states of the optomechanical system the partial trace of $\tilde\rho_\mathrm{pol}$ over the (dressed) cavity mode does not give the reduced state of the mechanical oscillator, cf. Eq.~\eqref{eq:rhotilde}. In contrast, the semi-polaron transformation introduced here retains a nonlinear interaction $\mathcal L_\mathrm{int}$, Eq.~\eqref{eq:timedependentcavityLint}, leaves the jump operator for cavity decay unchanged, and conserves the important relation \eqref{eq:Q_pol}. These properties are crucial in order to perform second order perturbation theory in $\mathcal L_\mathrm{int}$, and to derive an effective equation of motion for the mechanical oscillator as a separate system. For further comments on the semi-polaron transformation we refer to Appendix~\ref{App:SemiPolTraf}.

\subsection{Fokker-Planck Equation for the Mechanical Oscillator}\label{sec:effectiveequation}

\paragraph*{Interaction picture  ---} Our goal is now to adiabatically eliminate the cavity field from the dynamics, similar to the analysis of sideband-cooling \cite{Wilson-Rae2007}. This requires that the cavity dynamics, governed by $\tilde{\mathcal L}_c$ in \eqref{eq:tildeLc} with dominant characteristic time scale $\kappa$, is fast as compared to all other time scales in $\mathcal{L}_\mathrm{m}$ and $\mathcal{L}_\mathrm{int}$. Since we aim to cover in particular also the resolved sideband regime, $\omega_\mathrm{m}>\kappa$, we move to an interaction picture with respect to the free harmonic motion of the mirror. The equation of motion is still given by Eq.~\eqref{eq:timedependentcavity1}
where $\mathcal L_\mathrm{m}$ describes thermal decay only,
\begin{align*}
\mathcal L_\mathrm{m} \sigma&= \gamma\left( \partial_{\beta}\beta
+\partial_{\beta^*} \beta^*+2(\bar n +1) \partial^2_{\beta\beta^*}\right) \sigma,
\end{align*}
and $\tilde{\mathcal L}_\mathrm{c}$ and $\mathcal L_\mathrm{int}$ become explicitly time-dependent,
\begin{align}
\tilde{\mathcal L}_c \sigma&=-i\left[-\Delta a^\dagger a-K(a^\dagger a)^2-iE\left(e^{-i\theta(\beta,\beta^*,t)}a-\mathrm{h.c.}\right),\sigma\right]\nonumber\\
  &\hspace{5cm}  +\kappa D[a]\sigma,\label{eq:timedependentcavity2}\\
\mathcal L_\mathrm{int} \sigma&=-ig_0\left( e^{i\omega_\mathrm{m}t}\partial_{\beta} \sigma a^\dagger a-    \mathrm{h.c.}\right).\nonumber
\end{align}
The phase of the driving field is $\theta(\beta,\beta^*,t)=\eta\,\mathrm{Im}\left(\beta\, e^{-i\omega_\mathrm{m}t}\right)$.

In the adiabatic elimination it is assumed that the cavity essentially remains in the (quasi) stationary state of its undisturbed (by $\mathcal L_\mathrm{int}$) dynamics,
\begin{align}\label{eq:cavdynamics}
\dot\rho_\mathrm{c}=\tilde{\mathcal L}_\mathrm{c}\rho_\mathrm{c}
\end{align}
with $\tilde{\mathcal L}_\mathrm{c}$ given by \eqref{eq:timedependentcavity2}. This Liouvillian describes the dynamics of a Kerr nonlinear cavity driven by an amplitude- and phase-modulated field,
\begin{align}\label{eq:drive}
Ee^{i\theta(\beta,\beta^*,t)}=E\sum_{n=-\infty}^\infty J_n\left(-\eta|\beta|\right)e^{in\left(\omega_\mathrm{m}t-\phi \right)},
\end{align}
where $J_n$ are Bessel functions. Note that the partial amplitudes depend on the mechanical phase space variable $\beta=|\beta|e^{i\phi}$. We do not attempt to solve Eq.~\eqref{eq:cavdynamics} exactly. While in fact an exact solution for the stationary state of a Kerr nonlinear cavity exists \cite{Drummond1980} for the case of a constant driving field (\textit{i.e.} $\theta\equiv\,$const.), no such state can be expected for the present situation. Due to the periodic modulation of the driving field the cavity will not settle into a strictly stationary state, but rather to quasi-stationary state with a periodic time-dependence. If the Kerr nonlinearity is neglected an exact solution for this quasi-stationary state can be constructed by means of a Floquet series Ansatz \cite{Mari2009}. However, in the present case both aspects, modulated drive and Kerr nonlinearity, are important and shall be taken into account.

In order to arrive at an approximate solution of Eq.~\eqref{eq:cavdynamics} which can serve as a ($\beta$-dependent) reference state for the adiabatic elimination of the cavity we will follow two complementary approaches in the paragraphs below: In the first one we assume the cavity is driven to a state of large mean amplitude, which we determine self-consistently from an essentially classical nonlinear dynamics. The fluctuations around this mean field will be treated in a linearized model as Gaussian noise. The second approach concerns the case of a weak driving fields for which the cavity essentially stays close to its ground (vacuum) state, which corresponds to the regime considered in \cite{Rodrigues2010,Armour2012a}. In this case the master equation Eq.~\eqref{eq:cavdynamics} can be expanded and directly solved on the low lying Fock states.

In both cases we aim to retain a nonlinear dynamics for the mean cavity amplitude, and use a linearized description for fluctuations. Formally this is done by switching to a displaced frame, defining $\tilde{\tilde\sigma}=D(\alpha(t))\tilde\sigma D^\dagger(\alpha(t))$ where $D(\alpha)=\exp(\alpha a^\dagger-\alpha a)$. We choose $\alpha(t) \in \mathds C$ such that the terms of dominant order in $\alpha$ are canceled from the transformed master equation for $\tilde{\tilde\sigma}$.
In the case of $|\alpha(t)| \gg 1$ ($|\alpha(t)| \ll 1$) we cancel the terms of third (up to first) order in $\alpha$ and then neglect the terms up to first (of third) order in $\alpha$. The full equation of this transformation may be found in Eq.~\eqref{hugetrafo} of the Appendix, we proceed here with its most important features:

\paragraph*{Displaced frame for the limit $|\alpha(t)| \gg 1$ ---}  In the limit $|\alpha(t)| \gg 1$ we identify $\alpha(t)$ with the long time solution of
\begin{align}\label{eq:classchi}
\dot\alpha(t)=-\left[\kappa-i\left(\Delta +2K|\alpha(t)|^2  \right)\right]\alpha(t)+E e^{i \theta(\beta,\beta^*,t)} ,
\end{align}
which formally follows from the requirement that terms of third order in $\alpha$ in the resulting master equation for the displaced state $\tilde{\tilde\sigma}$ are canceled.
Due to the Kerr nonlinearity the dynamics described by this equation of motion can be bistable. On assuming a single stable solution we preclude bistable regimes from our description. For a constant phase $\theta$ bistability occurs only for driving fields which are red detuned with respect to the cavity resonance for detunings $\Delta<-\sqrt{3}\kappa$, see \cite{Drummond1980}. For the present case of a modulated drive no such simple condition can be given. However, it is reasonable to expect that bistability will become an issue only when the driving field has sufficient spectral weight for frequencies with a detuning below $-\sqrt{3}\kappa$. In the following we are mainly concerned with the cases of resonant or blue detuned drive, for which it turns out that bistability is not an issue \cite{Genes2008,Ghobadi2011,Aldana2013}. 

From Eq.~\eqref{eq:classchi} and \eqref{eq:drive} we can expect that in the long time limit the cavity amplitude will be of the form
  \begin{align}\label{eq:alphat}
  \alpha(\beta,\beta^*,t)= \sum_{n=-\infty}^\infty \alpha_n(\beta,\beta^*) e^{i n \omega_m t}.
  \end{align}
Inserting this expression into \eqref{eq:classchi} one sees that the effective detuning experienced by the cavity will be dominantly given by the DC component of $|\alpha(t)|^2$, such that it is useful to define an effective detuning
\begin{align}
\Deff(\beta,\beta^*)=\Delta+2 K\sum_n|\alpha_n(\beta,\beta^*)|^2 \label{DeffEq}.
\end{align}
Eq.~\eqref{DeffEq} has to be read as a non-linear equation for $\Deff$. In regimes where more than one solution exists, the system will be bi- or multistable, and we have to expect large photon number fluctuations. The validity of our approach will thus be limited to regions where only a single stable solution for $\Deff$ exists, as discussed above.
We seek an approximate solution to \eqref{eq:classchi} by assuming a fixed effective detuning $\Deff$, such that
  \begin{align}
     \alpha_n&=\frac{E}{h_n}J_n\left(-\eta |\beta|\right)e^{-in \phi},\label{eq:alpha}\\
     h_n&=\kappa+i(n\omega_m-\Deff), \label{eq:hn}
  \end{align}
where we follow the notation of \cite{Rodrigues2010,Armour2012a}.
In total $\alpha(t)$ in \eqref{eq:alphat} depends on the mechanical phase space variable $\beta$ through both $\Deff(\beta,\beta^*)$ and the $\beta$-dependent driving field $E e^{i \theta(\beta,\beta^*,t)}$. We will see that the $\beta$-dependence in $\Deff(\beta,\beta^*)$ is a crucial effect for the case of resonant cavity-drive (for which $\Deff\lesssim\kappa$).

The Liouvillians after the transformation with $D(\alpha(t))$ are
\begin{align}
\mathcal L_\mathrm{m} \sigma&=\gamma\left( \partial_{\beta}\beta
+\partial_{\beta^*} \beta^*+2(\bar n +1) \partial^2_{\beta\beta^*}\right) \sigma\nonumber\\
&\quad -ig_0 \left(\partial_{\beta}e^{i\omega_\mathrm{m}t}|\alpha(t)|^2-\mathrm{h.c.}\right)\sigma,\label{eq:Lmdisplaced}\\
\tilde{\mathcal L}_c \sigma&=-i\left[-\left(\Delta+4K|\alpha(t)|^2\right) a^\dagger a   -K \left( {\alpha(t)^2} a^{\dagger 2}   +h.c.  \right) , \sigma \right]\nonumber\\
  &\quad+\kappa D[a]\sigma,\label{eq:Lcdisplaced}\\
\mathcal L_\mathrm{int} \sigma&=-ig_0 \left(e^{i\omega_\mathrm{m}t}\partial_{\beta} \sigma (\alpha^*(t)a+\alpha(t) a^\dagger)
 +\mathrm{h.c.}\right)\label{eq:Linitfin}
\end{align}
The Liouvillian for the mechanical oscillator, $\mathcal L_\mathrm{m}$, acquires an additional drift term (second line in \eqref{eq:Lmdisplaced}) with a nonlinear nonlinear drift coefficient $\propto e^{i\omega_\mathrm{m}t}|\alpha(\beta,\beta^*,t)|^2$ which contains in particular the nonlinear DC force and dynamic back action effects (\emph{i.e.} optical damping and frequency shifts), as will be discussed below. In the Liouvillian for the cavity, $\tilde{\mathcal L}_\mathrm{c}$, terms of order $\alpha(t)$ and lower have been dropped. The leading terms of order $\alpha^2$ describe squeezing dynamics and an effective detuning.
Finally, in $\mathcal L_\mathrm{int}$ only the term of linear order in $\alpha$ has been kept. Note also that when moving to the displaced frame commutators of the ($\beta$-dependent) displacement operators and derivatives with respect to $\beta$ have been neglected. They would add corrections to the Liouvillians of higher order in $g_0$.
We have now removed the driving field from the
dynamics of the cavity. The remaining Liouvillian \eqref{eq:Lcdisplaced} describe the Gaussian evolution of fluctuations:

The ponderomotive squeezing of the light field is naturally contained in the ${\alpha^2 a^\dagger}^2$-term and its hermitian conjugate. While in this article we will study parameters for which this squeezing is negligible, the effect of ponderomotive squeezing back on the mirror after the adiabatic elimination of the cavity is an interesting perspective for future applications of our new formalism: {\color{black} In the adiabatic elimination of the cavity one would have to use a squeezed reference state, which can introduce possibly additional diffusion terms in the motion of the mirror. Applied to the situation of limit cycles, this may cause the state of the oscillator to become more classical. }

Curiously, the Kerr nonlinearity induces a different effective detuning for the mean field $\alpha$ than for the fluctuations (compare Eqs.~\eqref{eq:classchi} and \eqref{eq:Lcdisplaced}). This is consistent with results for a Kerr nonlinear cavity \cite{Drummond1980}. We therefore define
\begin{align}
&\Defft=\Delta+4 K \sum_n|\alpha_n|^2 \label{DefftEq}.
\end{align}

The fast decay rate to the vacuum is still given by $\kappa$ from the original master Eq.~\eqref{origk}.

This can be used in order to adiabatically eliminate the cavity taking into account second order effects in the optomechanical interaction $\propto g_0$, Eq.~\eqref{eq:Linitfin}, very much in the spirit of laser cooling theory \cite{Wilson-Rae2007}. Details of the calculation can be found in Appendix \ref{adi}. The result is an effective equation of motion for the mechanical oscillator in the form of a Fokker-Planck equation
\begin{align}
\dot Q(\beta,\beta^*)&=g_0^2 \sum_n \left(
\fpbs \fpb \frac{\alpha^*_n \alpha_n}{\tilde h_{n-1}}
 -\fpbs \fpbs  \frac{\alpha^*_{n-2} \alpha_n}{\tilde h_{n-1}}
 \right)Q(\beta,\beta^*) +h.c. \nonumber \\
 &\quad+i g_0 \sum_n \left( \fpbs  \alpha^*_{n-1} \alpha_n \right) Q(\beta,\beta^*) +h.c.\label{FPEQ} \nonumber \\
& \quad +\gamma\left( \partial_{\beta}\beta
+\partial_{\beta^*} \beta^*+2(\bar n +1) \partial^2_{\beta\beta^*}\right)Q(\beta,\beta^*)
\end{align}
for the $Q$-function of the mechanical oscillator. In analogy to $h_n$ we define $\tilde h_{n}=\kappa+i(n \omega_m -\Defft)$ with $\Defft$ given in \eqref{DefftEq}.

The drift and diffusion coefficients in the Fokker-Planck equation \eqref{FPEQ} do not depend on the phase of $\beta$ as a consequence of the rotating wave approximation involved in its derivation. We therefore transform the Fokker-Planck equation to polar coordinates $\beta=r e^{i\phi}$, and focus on the time evolution of the oscillator amplitude $r$ by integrating out the phase variable $\phi$. The time evolution for $r$ is then a one dimensional Fokker-Planck equation (on a half space),
\begin{align}\dot Q(r)= -  \partial_r \mu(r) Q(r) +  \partial_r^2 D(r) Q(r)  \label{FPEpolar}\end{align}
with drift $\mu(r)$ and diffusion coefficient $D(r)$
\begin{align}
&\mu(r)= -\gamma r  +\sum_n g_0E^2   \mathrm{Im}\left[ \frac{J_{n-1}(\eta r) J_n(\eta r)}{h_{n-1} h^*_n}\right]
  \label{driftequa}\\
&D(r)=\frac {\gamma (\bar n+1)} 2+\sum_n \frac{g_0^2E^2}{2} \left(
 \frac{\kappa J_n(\eta r)^2}{|h_{n}|^2|\tilde h_{n-1}|^2}
 -  \mathrm{Re} \left[\frac{J_{n-2}(\eta r) J_n(\eta r)}{\tilde h_{n-1}h^*_{n-2}h_{n}} \right]
 \right). \label{diffschlange}
\end{align}
The details of the transformation may again be found in Appendix \ref{adi}.
Equation \eqref{FPEpolar} admits a potential solution in steady state which is given by (up to normalization)
\begin{align}
&Q(r) \propto \frac {e^{I(r)}}{D(r)}, &I(r):= {\int_0^r \frac {\mu(r')} {D(r')} \mathrm dr'}. \label{finaldistribut}
\end{align}
This solution is valid for any value of $\Delta$, such that it covers both the regime of optomechanical cooling and the regime of self-induced oscillations. In \cite{Wilson-Rae2007} an effective equation of motion for the oscillator was derived in order to study the limits of sideband cooling under linearization of the dynamics and adiabatic elimination of the cavity using a coherent state as a reference state. The present approach generalizes this calculation to the nonlinear regime by using a different reference state for each phase space point of the oscillator. The non-linear quantum dynamics has been described analytically using a method based on the classical theory for limit cycles \cite{Heinrich2011} and by means of Langevin equations \cite{Rodrigues2010}, and has been applied in great detail to limit cycles, but also to the cooling regime \cite{Armour2012a}. The results of our calculation reproduce these results in the regime of a negligible Kerr-Term and provide suitable extensions in those cases
where the Kerr nonlinearity of the cavity becomes a dominant effect. In the next section we will compare the analytical expression for the steady state of the mechanical to numerical solutions of the exact master equation~\eqref{eq:MEQRF} to study the limit cycle regime. {\color{black} We conclude this section by briefly stating the corresponding results for the limit of small intracavity field amplitude, followed by a comparison of limit cycles studied in different laser setups.}

\paragraph*{Displaced frame and adiabatic elimination for the limit $|\alpha(t)| \ll 1$ ---}
In the case of $|\alpha(t)| \ll 1$ all steps can be performed in analogy. The difference is that we need to cancel the terms up to first order in $\alpha$ and then neglect the terms of third order in $\alpha$. The effective detuning now is given by
\begin{equation} \label{delt0}
\Deff=\Delta+K,
\end{equation}
i.e. the bare detuning is just shifted by the constant Kerr term in this extreme regime. No distinction between $\Deff$ and $\Defft$ needs to be made.
 The adequate choice for the displacement amplitude is the long term solution of
\begin{align}\label{eq:classchi0}
\dot\alpha(t)=-\left[\kappa-i \Deff \right]\alpha(t)+E e^{i \theta(\beta,\beta^*,t)} .
\end{align}
The result of the adiabatic elimination is structurally the same, $\alpha_n$ and $h_n$ are given as in equations \eqref{eq:alpha} and \eqref{eq:hn}, but with the effective detuning now given as in \eqref{delt0}. In Eq.~\eqref{diffschlange} the $\tilde h_n$ are simply replaced by $h_n$.

{\color{black}

\paragraph*{Quantum limit cycles in lasers ---}

It seems natural to base a model of optomechanical limit cycles on theory used in the context of lasers \cite{Lax1967,Haake1983,Gardiner2004b}, where quantum limit cycles have been extensively studied most prominently. The standard laser system consists of a reservoir of many atoms which forms a bath for the cavity mode. The pumped atoms will drive the laser mode to a high amplitude limit cycle, where it settles into a coherent state with random phase.

A setup that can be driven to highly sub-Poissonian states is the regularly pumped laser \cite{Sokolov1984, Golubev1995}, where excited atoms fly through a cavity. The mechanism works in the situation, where at each time approximately only one atom interacts with the light mode and the interaction is a swapping of excitations. In the case of more regular than Poissonian statistics of the pump, the fluctuation of the transmitted energy decreases and the light mode will have sub-Poissonian phonon statistics. This setup is sometimes also referred to as one atom laser/maser or micro maser, because the events when more than one atom interacts with the field can be neglected.

The one atom laser is different from the 'one-and-the-same' atom laser \cite{McKeever2003}, where a single atom is trapped inside a cavity and drives the laser mode. Also in this setup a sub-Poissonian steady state can be reached and the explanation again relies on counting the number of interactions exchanged between the atom and the cavity \cite{Kilin2012}.

In our optomechanical system a single laser mode is the bath driving the mechanical oscillator. The bath consisting of only a single mode is in analogy to some extend to the micro maser, as stressed in \cite{Nation2013}, and even more similar to the 'one-and-the-same' atom laser.  Even though we also describe sub-Poissonian boson statistics, the analytical techniques developed e.g. in \cite{Kilin2012} cannot be readily applied to our situation, because they crucially rely one the preservation of total excitations by the interaction, which is not given in the optomechanical setup.  Our analytical model  \ref{sec:lasertheory} is based on \cite{Haake1983}, which was first developed for the standard setup without sub-Poissonian statistics.

For the creation of non-Gaussian states a nonlinearity is required. In the optomechanical setup the nonlinearity stems from the interaction, while in the 'one-and-the-same' atom laser it stems from the two level nature of the bath, which is equivalent to a highly nonlinear cavity.

}

\section{Optomechanical Limit Cycles in the Quantum Regime}
\label{OLCR}

\subsection{Introduction}

\label{introlim}

As an introduction to our study, we sum up some known results on limit cycles that the rest of the article refers to. First, we introduce the theory for the amplitude of classical limit cycles as developed in \cite{Marquardt2006, Ludwig2008}, and then we recapitulate the numerical results on nonclassical states of quantum limit cycles as reported in \cite{Qian2012}. When comparing these findings to our analytical treatment we will be mainly concerned with the special case of close to resonant driving field, $\Deff\simeq\kappa\ll\omega_m$. Therefore, we start out by stating some approximate expressions for this case.

\paragraph*{Close to resonant drive ---} \label{approxexpr}
In the sideband-resolved regime and with a detuning close to the resonance, i.e. $\Deff, \kappa \ll \omega_m$ (but not necessarily $\Deff \ll \kappa$) we keep only the terms with $n=0,1$ in the expression for the drift coefficient, Eq.~\eqref{driftequa}, and approximate
\begin{equation}
\mu(r)\simeq -\gamma r  +   \frac {g_0E^2 } {\omega_m^2} \frac{2 \kappa \Deff(r) }{\Deff^2(r)+\kappa^2}  J_0(\eta r) J_1(\eta r).
\label{muapprox}
\end{equation}
In the sideband-resolved regime, the equation for the effective detuning, Eq.~\eqref{DeffEq}, becomes a third order poynomial in $\Deff$ and in the limit $\Deff \ll \kappa$ it even simplifies to a simple and explicit expression
\begin{equation}\label{Deffapprox}
\Deff (r) \simeq \Delta+2 \frac{K E^2}{\kappa^2} J_0^2(\eta r).
\end{equation}

\begin{figure}[t]
\includegraphics[width=0.45\textwidth]{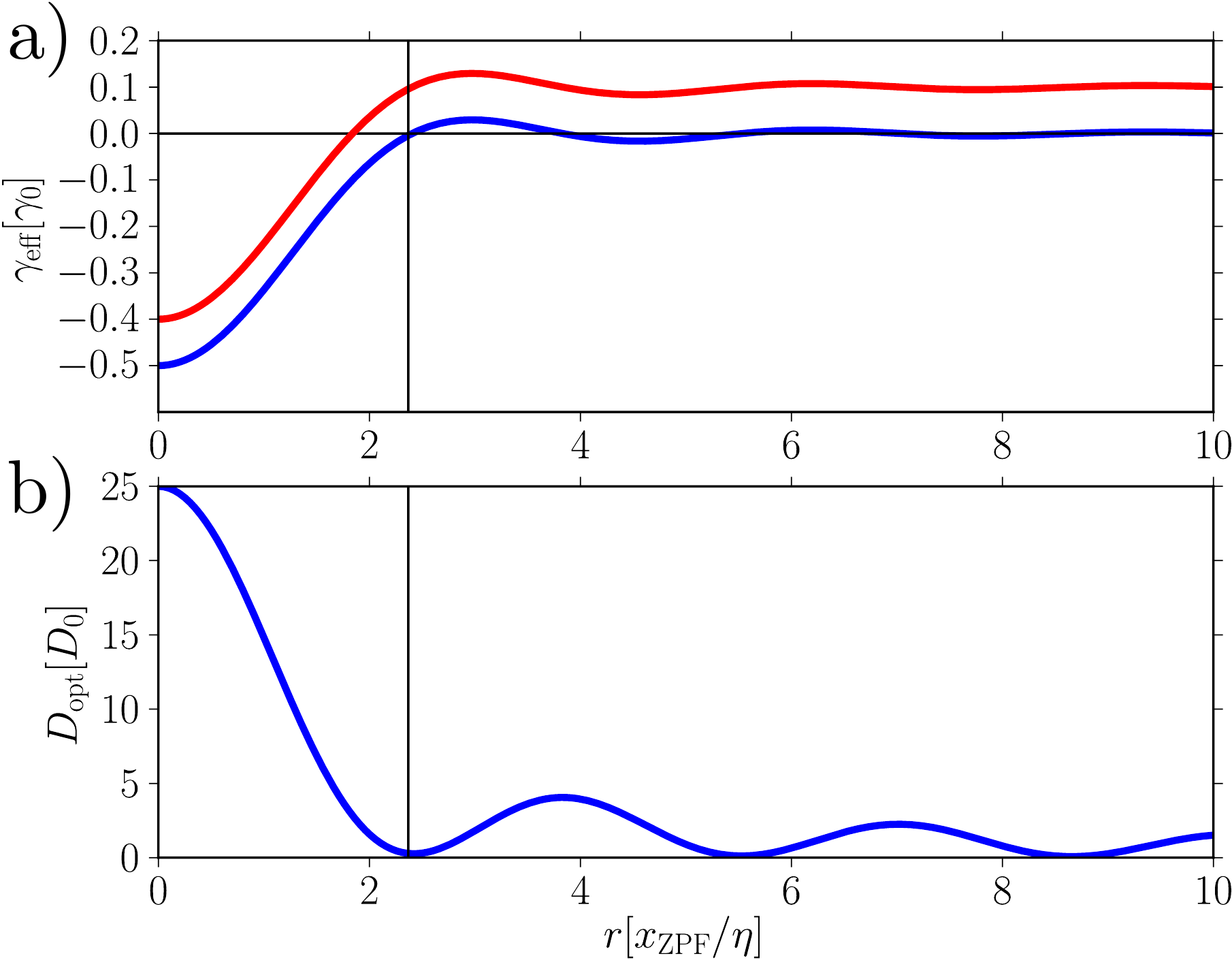}
\caption{\label{intersecfigure} Plot a) shows effective damping $\gamma_{\mathrm{eff}}(r)=\gamma +\gamma_{\mathrm{opt}}(r)$ from Eq.~\eqref{eq:gammaopt} in units of $\gamma_0=\frac {g_0E^2 } {\omega_m^2} \frac{2 \kappa \Deff }{\Deff^2+\kappa^2}$ versus cycle amplitude $r$ in units of zero point fluctuation and $\eta$. The blue and red line are two examples for different intrinsic damping $\gamma$. Limit cycles are stable at roots of the total damping with a positive slope. This happens only once for the red line with $\gamma=0.1 \gamma_0$, corresponding to only one possible amplitude for the oscillation. For the blue line with $\gamma=0$ many such intersections occur and the oscillator amplitude will in general jump between those different meta-stable points.\\
Plot b) shows the optical part of the diffusion from approximation \eqref{DWapprox} for $\kappa/\omega_m=0.1$ and $\Deff=\kappa$ in units of   $D_0={\kappa g_0^2 E^2}/{ \omega_m^4}$. Note that the dominant part of the diffusion from Eq. \eqref{DWapprox} is exactly canceled at the position of the limit cycle for $\gamma=0$, as indictated by the vertical line. This cancellation explains the strongly sub-Poissonian phonon statistics for such parameters.
}
\end{figure}

\paragraph*{Classical limit cycles ---}

The theory for classical optomechanical limit cycles from \cite{Marquardt2006} is reproduced by the drift-part of the Fokker-Planck equation, Eq.~\eqref{driftequa}, when neglecting the diffusion and using a constant effective detuning, $\Deff(r)=\Deff\equiv \mathrm{const}$. Disregarding the diffusion the oscillator amplitude $r(t)$ evolves fully deterministically and obeys
\begin{align*}
\dot r&= \mu(r)= -\gamma_{\mathrm{eff}}(r) r, &
\gamma_{\mathrm{eff}}(r)&=\gamma +\gamma_{\mathrm{opt}}(r).
\end{align*}
Following Eq.~\eqref{muapprox} the combined intrinsic and optically induced damping of the oscillator $\gamma_{\mathrm{eff}}(r)$ close to resonance is then given as the sum of the intrinsic mechanical damping $\gamma$ and the amplitude-dependent optical damping
\begin{equation}\label{eq:gammaopt}
\gamma_{\mathrm{opt}}(r)= -  \frac {g_0E^2 } {\omega_m^2} \frac{2 \kappa \Deff }{\Deff^2+\kappa^2}  \frac{J_0(\eta r) J_1(\eta r)}r.
\end{equation}
Note that the sign of the optically induced damping at $r=0$ coincides with the sign of $\Deff$. For negligible intrinsic damping, $\gamma\ll\gamma_\mathrm{opt}$, one can then expect limit cycles to always start for $\Deff>0$ (whereas the dynamics will be stable for $\Deff<0$). The possible amplitudes $r_0$ for limit cycles are given by the conditions $\gamma_{\mathrm{eff}}(r_0)=0$ and $\gamma'_{\mathrm{eff}}(r_0)>0$. The first condition is equivalent to
\begin{equation}
\label{interseceq}
\frac{J_0(\eta r) J_1(\eta r)}r=\gamma
 \frac {\omega_m^2 } {g_0E^2} \frac{ \Deff^2+\kappa^2}{2 \kappa \Deff}.
\end{equation}
The left hand side of this equation has infinitely many roots as the Bessel functions oscillate at a constant amplitude, cf. Fig.~\ref{intersecfigure} a). The envelope is given by the $r^{-1}$ decay. As illustrated in Fig.~\ref{intersecfigure} a) the exact position of the limit cycle and the number of possible amplitudes is then determined by the right hand side of Eq.~\eqref{interseceq}.

\subsection{Outline}


{\color{black}
In the following subsections we will explain two features of limit cycles on resonance
that can be heavily influenced by the Kerr term:

First, in Sec.~\ref{driftsec} we show that in the strong driving limit $|\alpha|^2 \gg 1$ the phase transition between optomechanical cooling and self-induced oscillations is crucially determined by the dynamical dependence of the effective detuning on the intracavity amplitude and its corresponding nonlinear dependence on the cycle amplitude, cf. Eq.~\eqref{Deffapprox}. This behavior can also be explained in a classical picture.

We then develop an explanation of the interesting numerical result for limit cycles in the quantum regime reported in \cite{Qian2012, Nation2013}: For approximately resonant driving fields, $\Delta \simeq 0$, and at the blue detuned sideband resonance, $\Delta \simeq \omega_m$, the steady state of the mechanical oscillator can have a Wigner function with a negative area.
The requirement on the strength of the optomechanical coupling $g_0$ is more stringent at the sideband than on resonance where non-classical limit cycles appear already for weaker coupling.{\color{black}
 Curiously, on resonance the numerical solution to the master equation predicts (non-classical) limit cycles also for parameters where classically the effective detuning $\Deff<0$, and one would expect a stable cooling dynamics. Fig.~\ref{jiangfigure} shows the steady-state Wigner function of the mechanical oscillator for such parameters.}

\begin{figure}[t]
\includegraphics[width=0.5\textwidth] {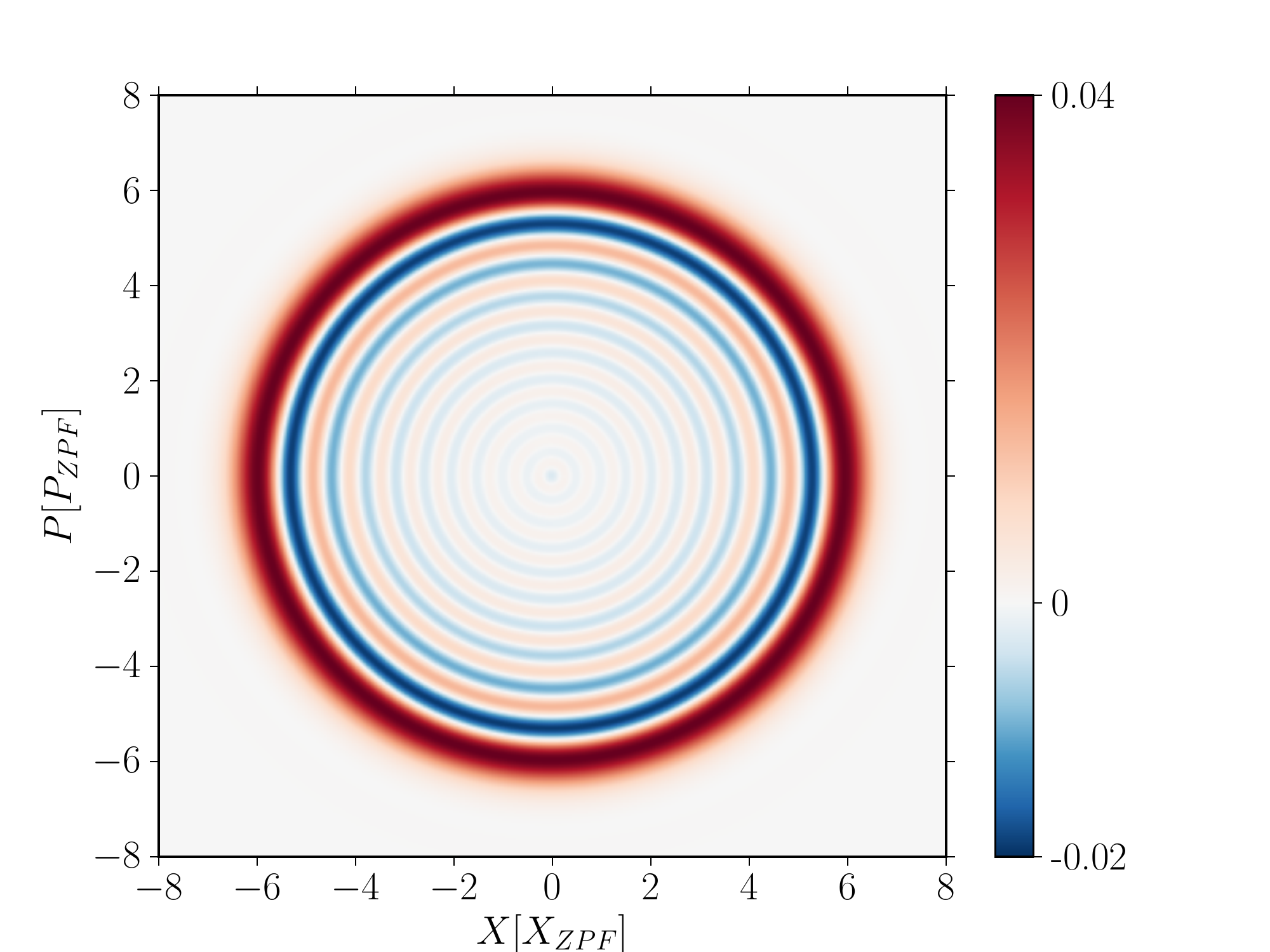}
\caption{\label{jiangfigure} Wigner function $W$ of the lowest metastable limit cycle of the mechanical oscillator for parameters $(g_0,\,\kappa_E=2\kappa,\,\gamma,E,\Delta, K)=(0.275,\,0.1,\,0,\,0.15\, -0.026, 0.076)\times\omega_m$. As there are less than 0.03 photons in the cavity we are in the regime of $|\alpha|^2 \ll 1$, where $\Deff=\Delta+K$, cf. equation \eqref{delt0}. Choosing the bare detuning to minimize the Fano factor ($F$=0.1 at the attractor with lowest amplitude, which is depicted in this plot) implies according to equation \eqref{fullfano}  $\Delta=\kappa-K$, which for the strong optomechanical coupling of this example gives the negative numerical value $\Delta=-0.026$. Note that classically or excluding the Kerr effect a limit cycle would not even start for these parameters. The minimal value of $W$ in this plot is $-0.02$.}
\end{figure}


We will use the analytical description of limit cycles with the Fokker-Planck equation to explain the features displayed in Fig.~\ref{jiangfigure}, and to predict general requirements on system parameters to achieve a non-positive Wigner function. In section \ref{nega} we show that the occurrence of negative Wigner functions in turn is intimately linked to achieving a small variance of the phonon statistics, as characterized by a small Fano factor $F=\langle \Delta n \rangle^2 / \langle n \rangle$, along with a small cycle amplitude $r_0$. We analyze the variance of the phonon number in section \ref{fanosection}
and find that the conditions for small Fano factor are favorable at the $\Delta=0$-resonance.

In section \ref{numericalmethods} we describe the numerical method used to check the analytical predictions. It allows for the first time to numerically study quantum features of optomechanical limit cycles in the regime of large mechanical amplitudes and strong laser drive, populating many states of the cavity. We find that the analytical model can still be applied and even for $g_0 < \kappa$ negativity of the Wigner function can be observed.

}

\subsection{Drift and dynamical detuning}

\label{driftsec}

\label{driftsection}
In this section we study in detail the time evolution of the mean amplitude $\bar r $, which is determined by the drift $\mu(r)$ in \eqref{muapprox}. In particular we show how the dynamical dependence of $\Deff(r)$ on $r$ gives new results which are not observed in any model based on a static detuning (like the one we used above). We focus on the regime where $\bar r$ is larger than its standard deviation $\Delta r$, such that we can derive the time evolution of $\bar r$ via $\dot {\bar r} = \mu(\bar r )$ directly from \eqref{muapprox} as
\begin{align} \label{muEOMapprox}
\dot {\bar r} =  -\gamma \bar r +   \frac {2 \kappa gE^2 } {\omega_m^2}  \frac{ \Deff(\bar  r) }{\Deff(\bar  r)^2+\kappa^2}  J_0(\eta \bar  r ) J_1(\eta \bar  r ).
\end{align}
These assumptions are fulfilled for small $\eta=\frac{2g_0}{\omega_m}$, because $\eta \bar r$ is the argument of the Bessel functions and hence $\bar r  \propto \frac 1 \eta$.

With the oscillator initially in the ground state, it is the sign of $\Deff(0)$ that determines if the limit cycle starts at all: For $\Deff(0)<0$ the optical damping is initially positive and no oscillation starts, but for $\Deff(0)>0$ it is negative and may be larger than the intrinsic damping $\gamma$, so that a self-induced oscillations can start. The oscillator arrives at its steady state, when $\dot{\bar{r}} =0$. Neglecting the small corrections due to $\gamma$, this is equivalent to the condition $\Deff(\bar{r}) J_0(\eta \bar{r})  J_1(\eta \bar{r})=0$. If the effective detuning $\Deff$ is independent of $r$, the smallest root of this product is always the first root of $J_0$. This corresponds to the standard situation (as discussed above) valid for a negligible Kerr parameter or in the weak driving limit, cf. Eq.~\eqref{delt0}.

In the converse case, for large amplitudes $|\alpha|^2 \gg 1$ and non-negligible Kerr parameter, the dynamic nature of the effective detuning can become important: The smallest root of the product $\Deff(\bar{r}) J_0(\eta \bar{r})  J_1(\eta \bar{r})$ is then determined either by $J_0$ or $\Deff$, depending on the sign of $\Delta$. If the bare detuning is on the blue (heating) side, $\Delta \gtrsim 0$, the condition for the limit cycle is still $J_0(\eta  r_0 )=0$ as in the case of a static detuning. However, if the bare detuning is on the red (cooling) side $\Delta<0$ the effective detuning for a small cycle amplitude can still be positive as $\Deff(0)=\Delta+2KE^2/\kappa^2$, cf. Eq.~\eqref{Deffapprox}. This is the case in particular for a driving field $E$ larger than a critical value of $E_{\mathrm{crit}}=\frac \kappa {\sqrt 2 g_0}\sqrt{{|\Delta| \omega_M}}$. The sign of $\Deff(r)$ will then depend on, and ultimately change with, the increasing amplitude $r$ of the oscillation since $\Deff = \Delta<0$ at 
the roots of $J_0^2(\eta r)$. With increasing oscillator amplitude $r$ the DC-component of the cavity occupation and hence (via the Kerr nonlinearity) also the shift of the detuning drops. The steady state amplitude $r_0$ of the limit cycle is reached when $\Deff(r_0)=0$.  Using again approximation \eqref{Deffapprox} the condition $\Deff(r_0)=0$ is equivalent to $J_0(\eta r_0)=\frac \kappa {\sqrt 2 g_0 E} \sqrt{{|\Delta|\omega_M}}$. Thus, the Kerr nonlinearity smoothens the transition from cooling to amplification. This is in contrast to models with a static detuning where a sharp transitions occurs at $\Deff=0$.

\begin{figure}[t]
  \includegraphics[width=0.5\textwidth]{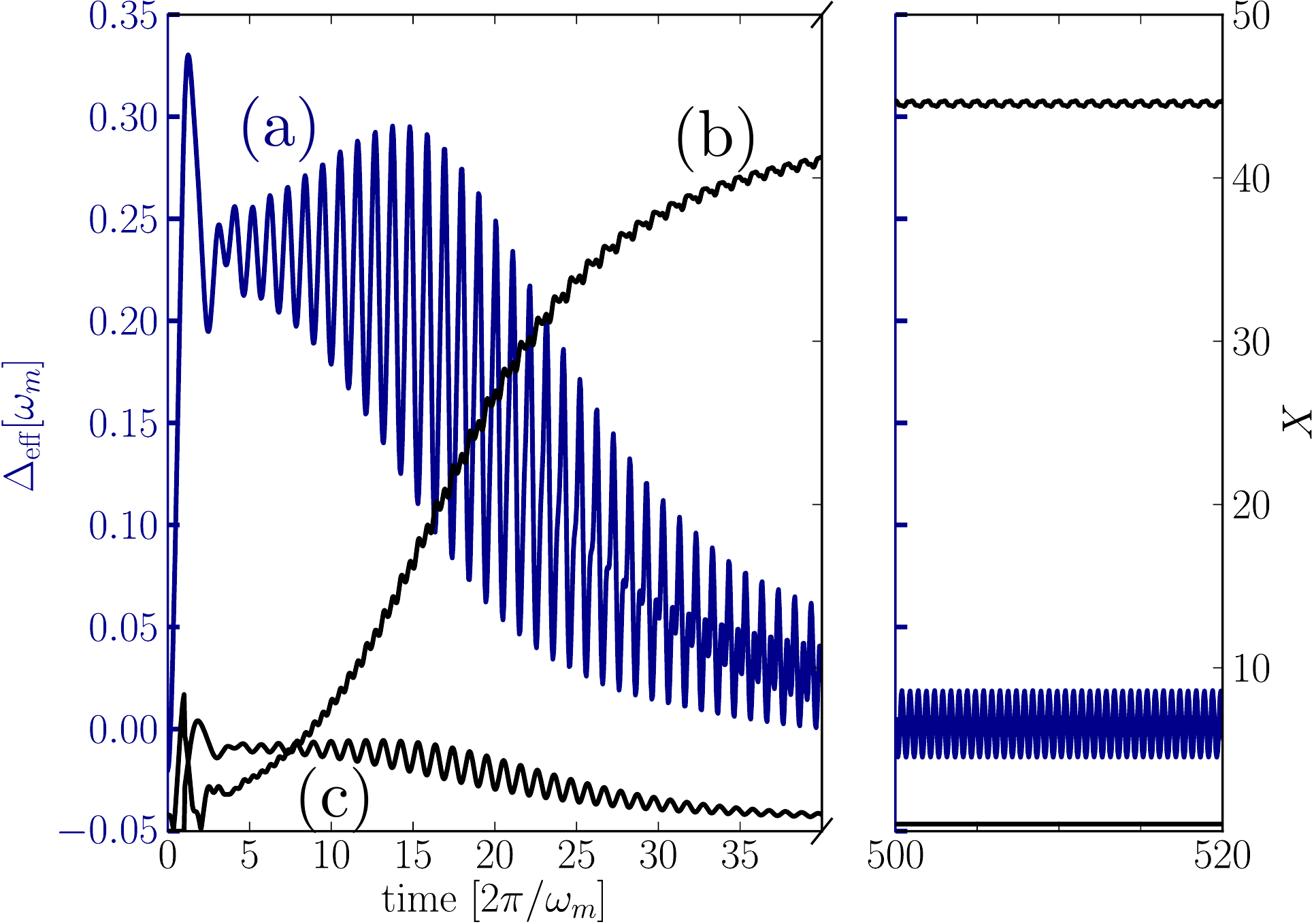}
	\caption{Example of the oscillator time-evolution for the classical equations of motion, see equation \eqref{classical}, with
 initial condition $r=0$ for $\Delta \lesssim 0$ but $\Deff(r=0)>0$. Effective detuning $\Deff(t)$ (a) with scale on the left (blue) axis, oscillator amplitude $r(t)$ (b) and DC-shift in position (c) with scale on the right (black) axis. A positive effective detuning at $r\approx0$ ensures that the limit cycle starts. With increasing oscillator amplitude the intra-cavity photon number $\sum_n|\alpha_n|^2$ from Eq.~\eqref{DeffEq} drops and hence also $\Deff$. As $\mu \propto \Deff$, see Eq.~\eqref{muapprox}, the oscillator settles in steady state as soon as this drop reaches $\Deff=0$. The parameters in this plot are $(E,\,g_0,\,\kappa_E=2\kappa,\,\gamma_E=2\gamma)=(4.0,\,0.05,\,0.3,\,2 \cdot 10^{-5})\times\omega_m$. \label{limittime}}
\end{figure}

\begin{figure}[t]
  \includegraphics[width=0.45\textwidth]{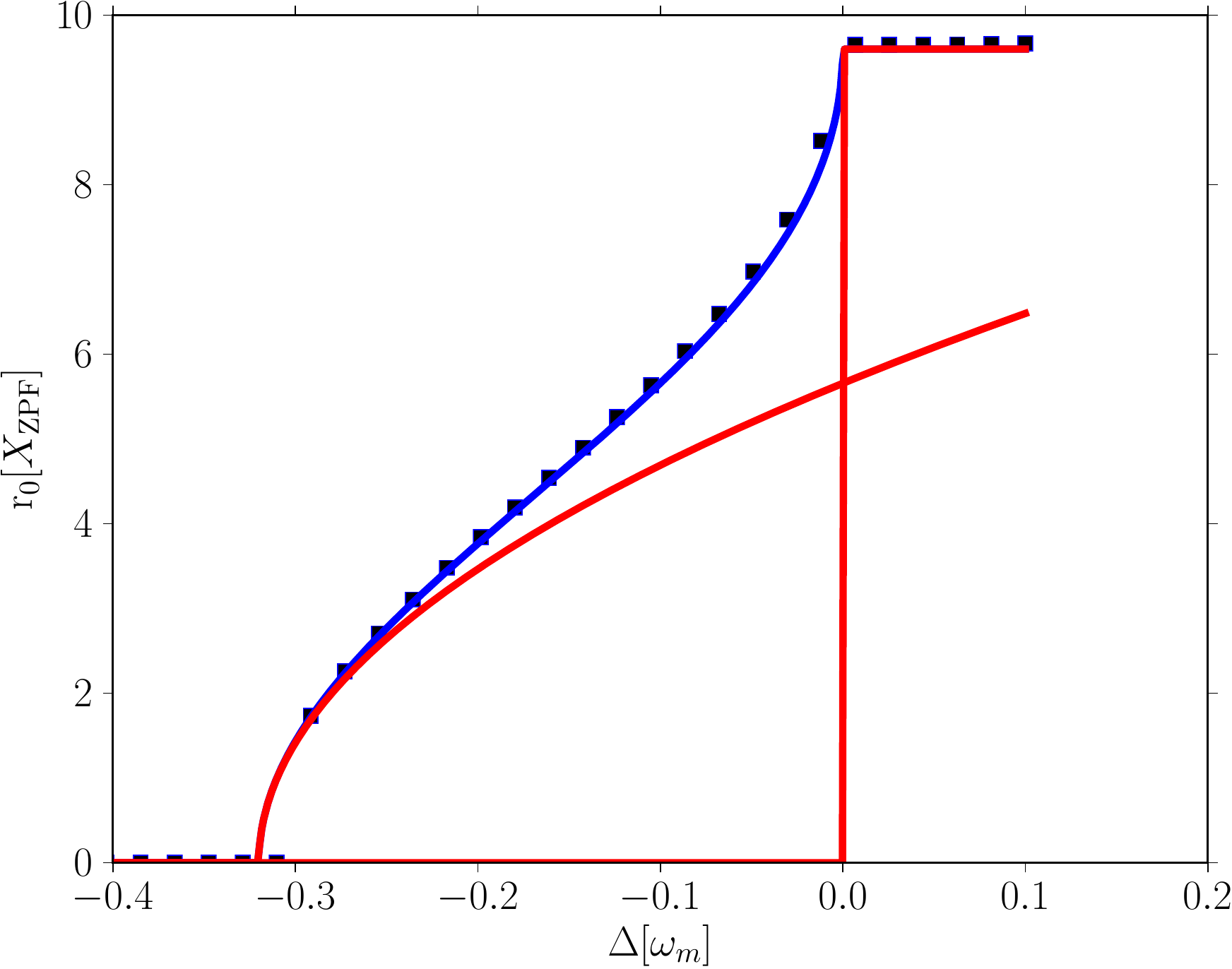}
	\caption{Amplitude $r_0$ for the first stable limit cycle versus bare detuning $\Delta$. In the limit of an amplitude-independent effective detuning (red) the values for large amplitudes are predicted correctly. It is known from \cite{Marquardt2006} that for small amplitudes at the onset of limit cycles, the amplitude follows a square root (red). With inclusion of the dynamical effective detuning $\Deff(r)$ (blue) the limit cycle amplitude $r_0$ follows $J_0(\eta r_0)=\frac \kappa {\sqrt 2 g_0 E} \sqrt{{|\Delta|\omega_M}}$, both limit cases are reproduced, and the whole transition between the regimes of damping and antidamping can be described. In this figure we compare the predictions with the numerical solution (dots) of the classical equation. The parameters of this plot are $(E,\,g_0,\,\kappa_E=2\kappa,\,\gamma)=(0.5,\,0.25,\,0.3,\,0.0)\times\omega_m$.
 \label{phasetrans}}
\end{figure}

We numerically check the dynamical nature of the detuning by integrating the equations of motion
\begin{align}
\label{classical}
&\dot \alpha =i (\Delta+g_0(\beta+\beta^*))\alpha-\kappa \alpha +E,
&\dot \beta=ig_0|\alpha|^2-i\omega_m\beta-\gamma \beta,
\end{align}
which are the classical analogue to the master equation \eqref{eq:MEQRF}. Fig.~\ref{limittime} illustrates the time dependence of the detuning with an example of a time evolution where the bare detuning $\Delta<0$, so that the limit cycle amplitude $r_0$ in steady state is determined by the condition $\Deff(r)=0$. Fig.~\ref{phasetrans} shows that this condition gives a good prediction for $r_0$ as a function of $\Delta$.

An approximation similar to equation \eqref{muEOMapprox} for the case of a laser drive close to the first blue sideband, $\Delta \approx \omega_m$, shows that there the position of the limit cycle does not depend on the exact value of $\Delta$. It is approximately given by the first root of $J_1\left(\eta r\right)$. Thus the limit cycle amplitude is generally smaller on resonance than on the sideband. We will use this observation in section \ref{nega}, where we will see that a small limit cycle amplitude is favorable for the occurrence of a negative area in the Wigner function.

\subsection{Diffusion and Fano factor}
\label{fanosection}

\begin{figure*}[tb]
\includegraphics[width=1.0\textwidth]{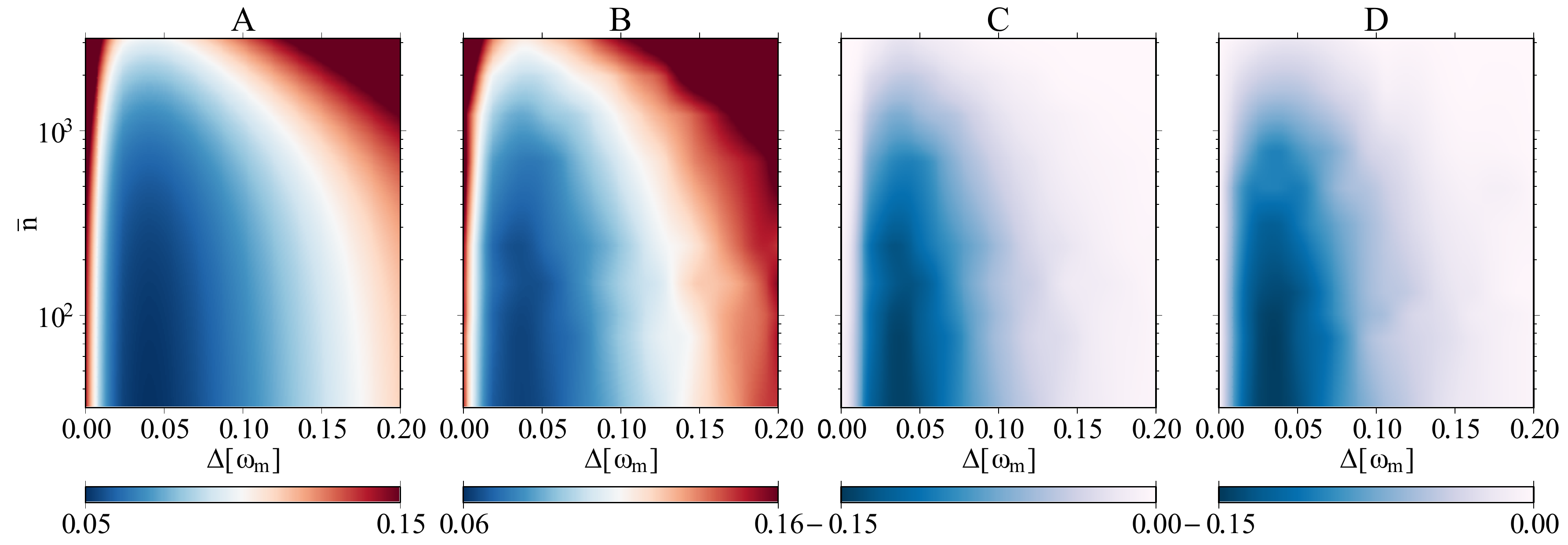}
\caption{Plots A) and B) show the Fano factor $F$ versus (bare) detuning $\Delta$ and bath occupation number $\bar n$. Note that by varying $\bar n$, we automatically vary the crucial quantity $\gamma \bar n$ appearing in equations \eqref{fullfano} and \eqref{eq:fanomini}. Plot A) is a plot of the simple analytical expression \eqref{fullfano}. Plot B) shows the numerical result obtained with Monte-Carlo trajectories for 30000 mechanical oscillations. A) and B) are in good agreement despite the fact that in the parameter regime considered here some of the approximations are barely fulfilled. Note that the colorscale in the numerical prediction for the Fano factor is slightly shifted up by 0.01, hinting possibly at some additional diffusion process not considered in the analytical model. Plot C) shows the prediction for Wigner function negativity (defined as the quotient of the most negative and the most positive value of $W$) obtained by extrapolating the results for $F$ from plot B) using the function from Fig. \ref{negaFANO}. Plot D) shows the Wigner function negativity as directly extracted from the numerical result of the Wigner function. The constant 
parameters in all plots are $(g_0,\,\kappa_E=2\kappa,\,\gamma_E=2\gamma,E)
=(0.05,\,0.1,\,10^{-7},1.56)\times\omega_m$. The approximate average number of photons in the cavity is 1.5 in these plot.
\label{SCANS}
}
\end{figure*}

Having discussed the conditions for a limit cycle to start and having derived the mean amplitude in steady state for the $\Delta \simeq 0$ resonance, we now consider the fluctuations caused by the diffusion $D$ around this mean value to derive a prediction for the Fano factor $F=(\langle n^2 \rangle -\langle n \rangle ^2)/\langle n \rangle=\langle \Delta n \rangle^2 / \langle n \rangle$, which is a measure for number squeezing: For a coherent state the phonon distribution is Poissonian so that $\langle \Delta n \rangle^2 = \langle n \rangle$ and $F=1$. A state with sub-Poissonian phonon variance can hence be characterized by $F<1$.

{\color{black}
We will use the term Fano factor in the context of limit cycles as follows:
For generic parameters an optomechanical system can exhibit several limit cycles, such that the Fano factor of the full density matrix typically is larger than one.
The oscillations at each of these attractors are metastable, such that it is possible to consider the phonon statistics at a particular limit cycle. Especially in the relatively classical regime where $g_0/\omega_m$ is not too large the cycles will be well separated. When we refer to Fano factor, we will implicitly always mean the Fano factor of one particular attractor. 
}

We obtain the mean and variance of the phonon number $n$ via  \cite{Gardiner2004b}
\begin{align}
\langle \left\{ a^r (a^\dagger)^s \right\}_{sym} \rangle=\int \mathrm d^2 \alpha W(\alpha,\alpha^*)\alpha^r (\alpha^*)^s,
\end{align}
where $W(\alpha,\alpha^*)$ is the Wigner function. We use here the Wigner function because it gives better agreement with the numerical analysis for the the statistics of the phonon number than other quasi-probability distributions. Drift and diffusion coefficients for the Wigner function are calculated in App.~\ref{App:PhaseSpace} along the same lines as shown above for the $Q$-function. In particular, close to resonance the radial diffusion coefficient as relevant to the Wigner function is
\begin{equation} \label{DWapprox}
D_W=\frac {\gamma(1+2\bar n)} 4 + \frac{\kappa g_0^2 E^2}{ \omega_m^4} \left( J_1^2(\eta r) +\frac 12 \frac{\omega_m^2}{\kappa^2+\Deff^2}J_0^2 (\eta r) \right),
\end{equation}
where we applied to equation \eqref{DiffW} the same approximations as in Sec.~\ref{introlim} for the drift coefficient.

For most amplitudes the $J_0^2$-term is dominant, as it is enhanced by at least $\left( {\omega_m}/ \kappa \right)^2$ over the $J_1^2$-term. For parameters where the optical anti-damping is much stronger than the intrinsic mechanical decay, a curious cancellation of the diffusion occurs in steady state: The limit cycle will then settle exactly at the first root of $J_0$ as discussed in section \ref{driftsec}. There the term proportional to $J_1^2$, which is suppresed by $\left( \kappa /{\omega_m} \right)^2$,  becomes the only relevant term in the diffusion. 
 This suppression is illustrated in figure \ref{intersecfigure} b) and can be intuitively explained: The last two terms in equation  \eqref{WignerfullDiff} (or equivalently \eqref{DiffW}) are the (coherent) squeezing terms. For $n=1$ they exactly cancel the corresponding (incoherent) diffusion terms $\propto \fpbs \fpbs$ in leading order and only the higher order terms in $\kappa^{2}/\omega_{{\rm m}}^{2}$ remain.
Because of this suppression of diffusion in the sideband-resolved regime one can obtain a very small Fano factor of the mechanical oscillator, as we show below.

The phase space distribution in steady state is given by Eq.~\eqref{finaldistribut}. In the limit of small $ {g_0}/{\omega_m}$, where $\Delta n \ll \langle n \rangle$, and for the case of only a single stable limit cycle centered around a position $r_0$ with $\mu(r_0)=0$, we linearize $\mu(r) \eqsim \mu(r_0) +\mu'(r_0) (r-r_0)$ around this $r_0$ and set $D(r) \simeq D(r_0)$ so that the corresponding solution for $W$ is approximately
\begin{equation} \label{approxW}
W(r) \propto \exp \left(-{ \frac{(r-r_0)^2}{2\sigma^2}  } \right).
\end{equation}
with $\sigma^2=-D(r_0)/\mu'(r_0)$. One can then derive the approximate expression $F \simeq 4 \sigma^2$ for the limit $\omega_m/g_0 >\sigma$. In the sideband-resolved regime and with the limit cycle position at the first root of $J_0$ this gives
\begin{equation}
\label{fullfano}
F \simeq  \slfrac{ \left( \frac {\gamma(1+2\bar n)} 4 + \zeta\frac{\kappa g_0^2E^2}{ \omega_m^4} \right)}{\left(\frac \gamma 4+ \frac{2\kappa \Deff(r_0) }{\Deff(r_0)^2+\kappa^2} \zeta\frac{g_0^2E^2}{\omega_m^3}\right)},
\end{equation}
where $\zeta\simeq 0.27$ is the numerical value of $J_1^2$ at the position of the limit cycle.
The Fano factor is minimal at an effective detuning $\Deff(r_0)=\kappa$ where it takes on the value
\begin{equation}\label{eq:fanomini}
F \simeq  \slfrac{\left( \frac {\gamma(1+2\bar n)} 4 + \zeta\frac{\kappa g_0^2E^2}{ \omega_m^4} \right)}{\left(\frac \gamma 4+ \zeta\frac{g_0^2E^2}{\omega_m^3}\right)}.
\end{equation}

Note first that Eq.~\eqref{eq:fanomini} implies that the Fano factor is lower bounded by the sideband resolution
    \begin{align}\label{Fanobound}
        F>\frac{\kappa}{\omega_m},
    \end{align}
and that this bound is achieved for sufficiently large driving field $E=\sqrt{2\kappa P_L/\hbar\omega_L}$ (laser power $P_L$). Furthermore Eq.~\eqref{eq:fanomini} implies that the condition for sub-Poissonian statistics $1>F$ is \emph{exactly} equivalent to
    $\frac{g_0^2E^2}{\omega_m^3}\left(1-\frac{\kappa}{\omega_m}\right)>\frac{\gamma\bar{n}}{2\zeta}.$
This can be interpreted as a condition for the driving power which for small $\kappa/\omega_m$ becomes
    \begin{align}\label{FanoPowerCond}
        \frac{P_L}{\hbar\omega_L}>\frac{\omega_m^3}{4\zeta\kappa g_0^2}\gamma\bar{n}.
    \end{align}
It is instructive to express this also in terms of the (thermal, linearized) cooperativity parameter
    \begin{align}\label{cooperativity}
      \mathcal{C}=\frac{4g_0^2\alpha^2}{\kappa\gamma(2\bar{n}+1)}
       =\frac{8g_0^2}{\omega_m^2\gamma(2\bar{n}+1)}\frac{P_L}{\hbar\omega_L},
    \end{align}
where we used that the relevant average intracavity amplitude at the optomechanical limit cycles is $\alpha=\alpha_1\simeq E/\omega_{\rm m}$, cf. Eq.~\eqref{eq:alpha}. Condition \eqref{FanoPowerCond} then takes the form (in the limit $\bar{n}\gg1$)
    \begin{align}\label{FanoCoopCond}
      \mathcal{C}>\frac{1}{\zeta}\frac{\omega_m}{\kappa}.
    \end{align}
Note that this is essentially a requirement on the \emph{linearized} optomechanical coupling ($g\propto g_0E$), and not on the coupling per single photon $g_0$. The condition in Eqs.~\eqref{FanoPowerCond} and \eqref{FanoCoopCond}, and the lower bound in Eq.~\eqref{Fanobound} are the main result regarding sub-Poissonian phonon statistics.

The possibility of a sub-Poissonian number distribution was discussed in \cite{Rodrigues2010,Armour2012a} for the resonance at the first (and higher) blue sidebands. The prediction of the analytical model is especially good for the regime with small $g_0$ that results in larger limit cycle amplitudes. In Figure \ref{SCANS}, which compares the Fano factors as derived from our analytical model and from solving the master equation, the good agreement can be seen. For larger $g_0$ {\color{black} (not depicted in Figure \ref{SCANS})} the condition neccesary for adiabatic elimination is less satisfied and also the linear approximation \eqref{approxW} gets worse, because $\Delta n \approx \langle n \rangle$. Thus the quantitative agreement gets worse. Still the resonances for $F$ at $\Delta \approx 0, \omega_m$ are qualitatively reproduced.

{\color{black}
In \cite{Rodrigues2010,Armour2012a} the Fano Factor has been calculated with a derivation using the truncated Wigner function approximation and solving the resulting Langevin equation. If we use the Wigner function as the phase space distribution, our calculation, which does not rely on this truncation, gives the same result in the regime where the Kerr parameter $K$ is negligible.

For limit cycles with the cavity close to its ground state, different approaches to treat the Kerr effect have been taken in the literature: \cite{Nation2013} uses the classical part of the Kerr effect, as derived with the standard master equation approach, to introduce a renormalized detuning with a shift proportional to the cavity occupation. An additional constant (independent of the cavity occupation) shift of the detuning by $K={g_0^2}/{\omega_m}$, was numerically observed in \cite{Armour2012a} and then introduced by hand, to match the numerical data. It is one of the main results of the the semi-polaron approach, that the separate Kerr term for the cavity is naturally derived for limit cycles. It causes exactly the additional quantum shift of $\Delta$ observed in \cite{Armour2012a}, which is most striking in the $|\alpha| \ll 1$ limit, cf. Eq.~\eqref{delt0}.  
}

\subsection{Nonpositive Wigner Function}
\label{nega}

\begin{figure}[t]
  \includegraphics[width=0.5\textwidth]{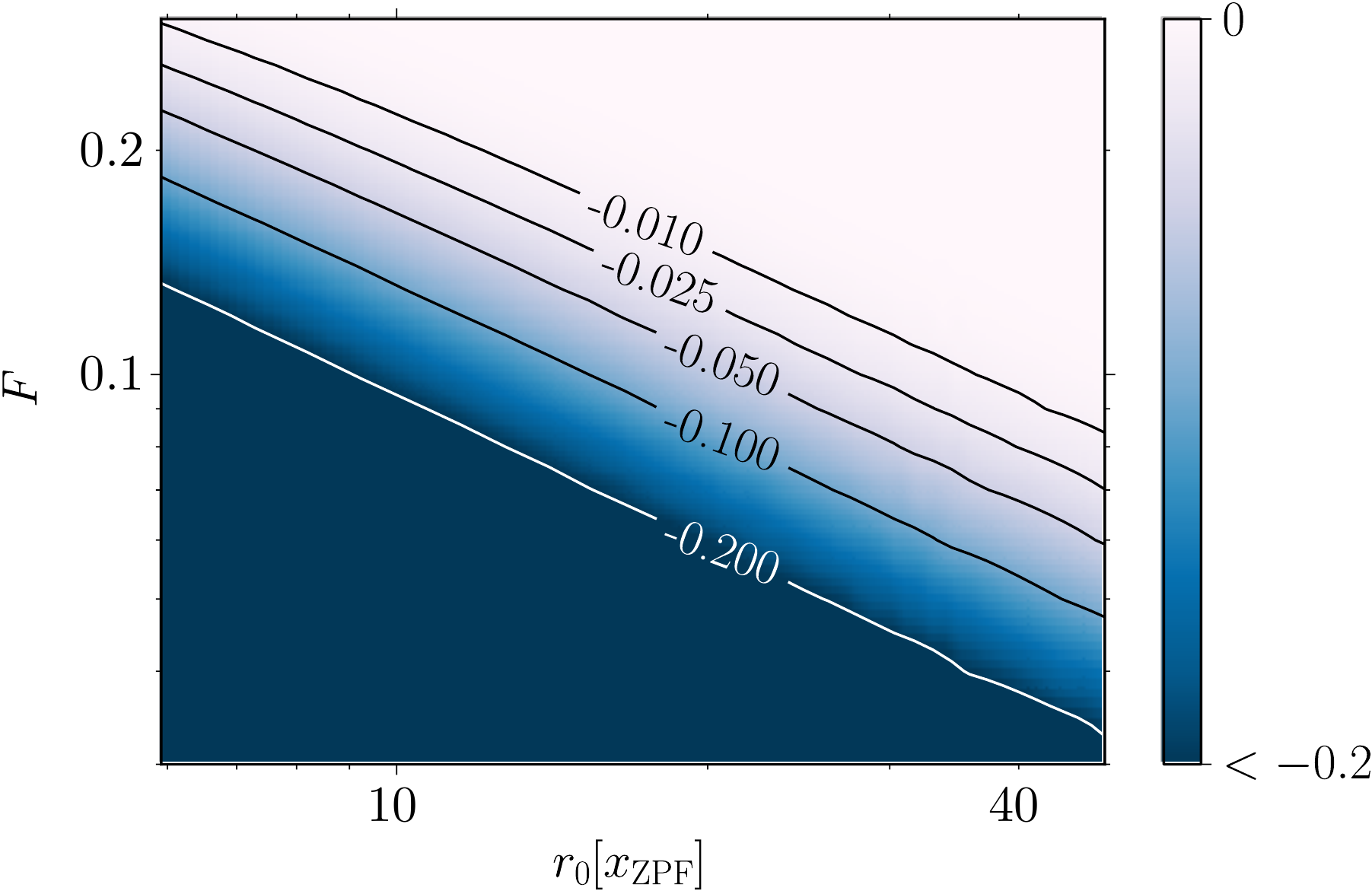}
	\caption{Maximal negativity of the Wigner function (defined as the quotient of the most negative and the most positive value of $W$) as a function of the Fano factor $F$ and the mean amplitude $r_0$ for a phonon distribution as in equation \eqref{ansatz}. From this plot one can read of, how small the Fano factor needs to be for a given $r_0$, to see a negative value in the Wigner function. Implicitely this is also a requirement on ${g_0}$ because $r_0 \propto \frac{\omega_m}{g_0}$, see Section \ref{driftsection}. }
	\label{negaFANO}
\end{figure}

Finally we use the Fano factor to predict the occurence of a negative area in the Wigner function. For a Fock state the Fano factor $F$ is of course zero and, except for the vacuum, all Fock states have a pronounced negativity of the Wigner function. Both $F$ and the Wigner function are continuous functions of the state $\rho$. Hence, for a given mean phonon number $n_0$ there is a critical value $F_c$, such that for a state with $F<F_c$ the Wigner function has a negative area. For simple set of Ansatz states given by a density matrix diagonal in Fock basis with Gaussian probability distribution
\begin{equation}
\label{ansatz}
P(n) \propto \exp \left({-\frac{(n-n_0)^2}{V}}\right),
\end{equation}
we numerically determined the corresponding critical Fano factor $F_c$. The result is illustrated in Fig. \ref{negaFANO}. We use this particular Ansatz, because the typical steady state density matrix of our problem is approximately of this form when $g_0/\omega_m$ is not too large. In \cite{Armour2012a} the steady state as a Gaussian distribution in Fock states is derived in more detail.

Figure \ref{negaFANO} shows that this threshold $F_c$ is smaller for larger amplitude $r_0$. We infer that in order to see negativity of the Wigner function in steady state, small limit cycle amplitudes with small Fano factors are favorable. Applied to the results of \cite{Qian2012, Nation2013} this explains the more favorable condition for negativity at the $\Delta \simeq 0$-resonance as compared to the $\Delta \simeq  \omega_m$-resonance, because the limit cycle there has a smaller amplitude (given by the first root of $J_0(\eta r)$ as compared to $J_1(\eta r)$, as discussed in section \ref{driftsec}). Independent of $\Delta$, the amplitude scales with the inverse of $g_0/\omega_m$, such that for a large ratio $g_0/\omega_m$ a non-positive Wigner function is achieved already for larger Fano factors. More precisely, we can conclude from Fig.~\ref{negaFANO} that
    \begin{align}\label{criticalFano}
      F_c&\simeq\xi r_0^{-s}, & s&\simeq0.6,
    \end{align}
where the constant $\xi$ depends on how negative the Wigner function should be. In order to achieve a ratio of minimal to maximal value of the Wigner function of e.g. $-0.1$ this constant is found to be $\xi\simeq 0.6$. {\color{black}As a comparison, this negativity ratio can reach (approximately) -2.5 for odd Fock states and -0.4 for even Fock states.}

Since the amplitude of the first limit cycle is $r_0\simeq \omega_{\rm m}/g_0$ the condition $F<F_c$, together with Eqs.~\eqref{eq:fanomini} and \eqref{criticalFano}, is equivalent to ($\zeta\simeq 0.27$)
    \begin{align}\label{negativeWigner1}
      \frac{g_0^2E^2}{\omega_m^3}\left[\xi\left(\frac{g_0}{\omega_m}\right)^s-\frac{\kappa}{\omega_m}\right]
      >\frac{\gamma(2\bar{n}+1)}{4\zeta}-\frac{\xi\gamma}{4\zeta}\left(\frac{g_0}{\omega_m}\right)^s.
    \end{align}
Thus, one necessary condition for negative Wigner function is that the square bracket on the left side is positive. This is a condition on the single photon optomechanical coupling $g_0$, that can be written equivalently as both
    \begin{align}\label{g0cond}
     & \frac{g_0}{\omega_m}>\left(\frac{\kappa}{\xi\omega_m}\right)^{1/s}, 
      &\frac{g_0}{\kappa}>\frac{1}{\xi^{1/s}}\left(\frac{\kappa}{\omega_m}\right)^{(1-s)/s}.
    \end{align}
Note that this condition for the occurrence of a quantum state is \emph{weaker} than the condition $g_0/\kappa>1$ which one would have expected naively.

Assuming this condition to be well fulfilled we can drop the second terms on both left and right hand side of \eqref{negativeWigner1} and get the power requirement
    \begin{align}\label{powercondwigner1}
      \frac{P_L}{\hbar\omega_L}>
            \frac{\omega^3_{\rm m}}{4\xi\zeta\kappa g^2_0}\left(\frac{\omega_m}{g_0}\right)^{s}
      \gamma\left(\bar{n}+\textstyle{\frac{1}{2}}\right)
    \end{align}
Note that this is stronger than the requirement \eqref{FanoPowerCond} for sub-Poissonian statistics, as one would expect. In terms of the cooperativity (for any $\bar{n}$) this becomes
    \begin{align}\label{powercondwigner2}
      \mathcal{C}>\frac{1}{\xi\zeta}\left(\frac{\omega_m}{g_0}\right)^{s}.
    \end{align}
Note also that even for zero temperature, $\bar{n}\rightarrow 0$, there is now a threshold for the power (cooperativity) in contrast to the condition for sub-Poissonian statistics. Condition \eqref{g0cond} on the strength of the optomechanical coupling per single photon, and condition \eqref{powercondwigner1} (or \eqref{powercondwigner2}), which reproduce the heuristically derived conditions \eqref{g0requ} and \eqref{eq:powerrequ} from Sec.~\ref{sec:Summary}, are the main results regarding negative Wigner functions.

\subsection{Numerical Analysis}
\label{numericalmethods}

In this section we compare the predictions from the sections above with the numerical result for the master equation of Eq. \eqref{eq:MEQRF}.
To do the calculation for large Hilbert space dimension, we applied the Monte-Carlo wave function method from \cite{Dalibard1992, Dum1992, Molmer1993} as implemented in QuTiP \cite{Johansson2011a, Johansson2013}, the quantum toolbox for python. The advantage is that one needs to simulate only wave functions and not density matrices, so that the Hilbert space dimension required for the simulation scales only with the number of possible pure states $N$ instead of $N^2$. In this method the individual trajectory of an initially pure state is calculated, conditioned on the history of fictive photon and phonon counters measuring the particles leaking out of the system. With this knowledge of the environment an initially pure state stays pure. The density matrix is then retrieved by averaging over a large ensemble of such conditional states.
The ensemble average can be replaced by the time average for calculating a steady state density matrix.

Our implementation was done with an adaptive Hilbert space, where the Fock states are not only limited from above, but also from below and after each mechanical oscillation the Hilbert space is updated so that it is centered around the current state. {\color{black} To make sure that not too much of the Hilbert space is truncated, the number of states to be used is scaled with the standard deviation in energy of the state in the previous step. } This flexibility of the Hilbert space during the calculation allows to run the simulation without much a priori knowledge of the steady state and even fewer basis states are required. 

The solution is obtained in the following steps:
For speed up of the calculation, the initial state is chosen to be a coherent state with an amplitude close to the expected steady state. It is then evolved for some period until at a time $t_0$ the conditional state's amplitude and Fano factor stop to drift and only fluctuate. We then make use of the fact, that in steady state the time average corresponds to the ensemble average, and calculate the steady state of the oscillator as
\begin{align}
\rho_M=\int_{t_0}^{t_0+T} \mathrm{tr}_\mathrm{c} \left(\ket{\psi_t}\bra{\psi_t}\right) \mathrm dt,
\end{align}
where $\ket{\psi_t}$ is the conditional state at time $t$, and $T$ spans many mechanical oscillations.

This procedure is performed many times in parallel on a cluster and the resulting matrices $\rho_M$ are averaged. The deviation of the individual $\rho_M$ provides an error estimate for the method.
As a further benchmark and control, we also calculated the steady state with 
the biconjugate gradient steady-state solver 
from scipy \cite{Oliphant2007}, which is however limited to a comparably small Hilbert space dimension.

\begin{figure}[t]
  \includegraphics[width=0.5\textwidth]{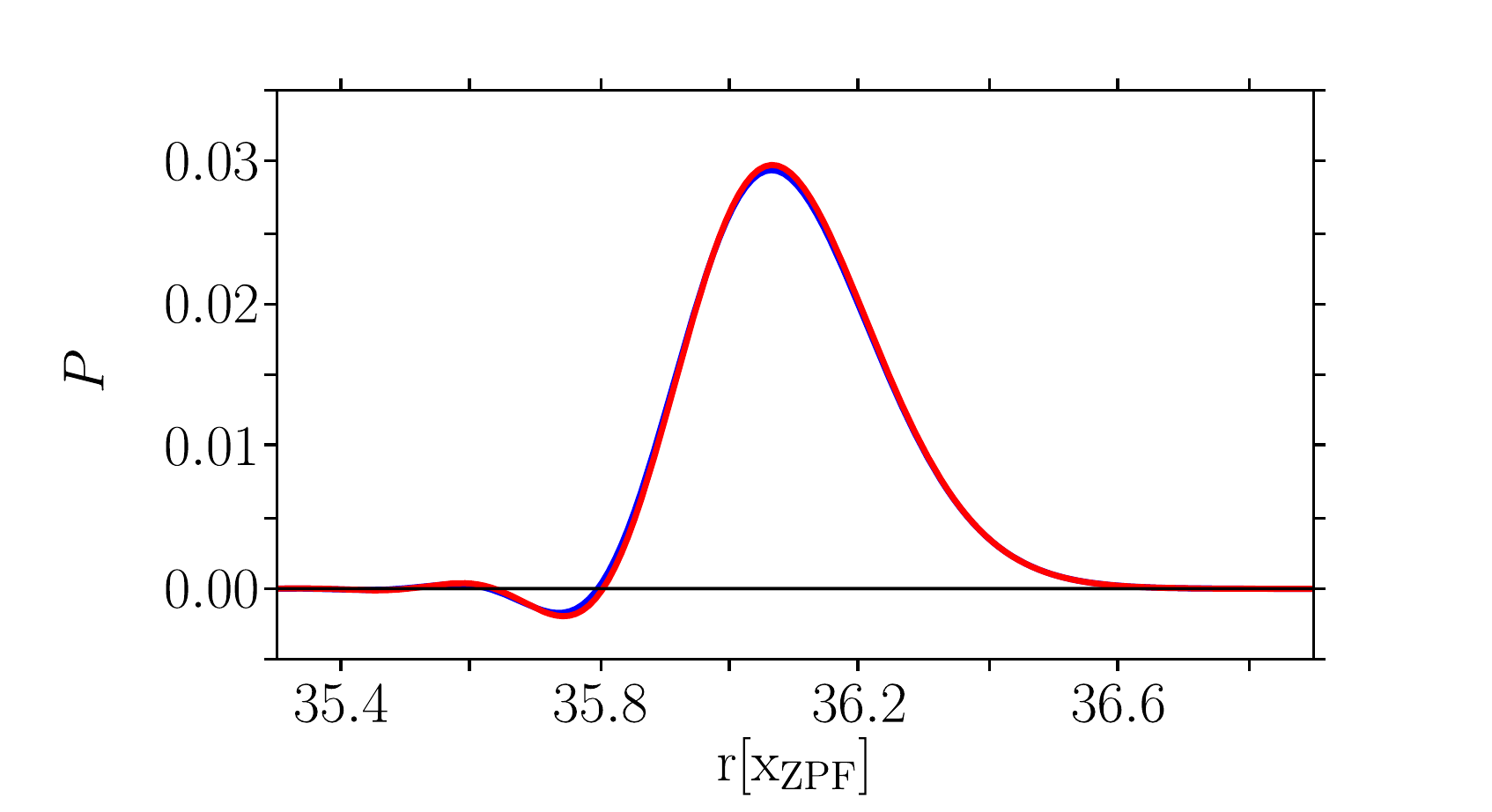}
	\caption{Radial part of a Wigner function for parameters with high amplitude and many photons ($\langle a^\dagger a \rangle \approx 8$) in the cavity featuring a very small Fano factor ($F=0.07$) and some negative density. The parameters are $(g_0,\,\kappa_E=2\kappa,\,\gamma,E,\Delta)=(0.033,0.1,0. ,3.5 , 0.03)\times\omega_m$. The blue and red line are the result of two independent runs (each averaging 5000 mechanical oscillations) of the Monte-Carlo based steady state solver.}
	\label{HighWigner}
\end{figure}

The algorithm described above allows for the first time to numerically study optomechanical limit cycles in the experimentally relevant regime of large amplitudes of the mechanical oscillator (as caused by a relatively small $g_0/\omega_m$) and with more than only a few photons in the cavity. In previous studies the question was posed, whether the analytical theory can be applied to this regime \cite{Armour2012a} and if the nonclassical features survive \cite{Nation2013} for more than one photon in the cavity. We answer this question affirmative: Fig. \ref{HighWigner} shows an example of a Wigner function in this regime with small Fano factor and some negative density.

Strictly speaking the steady state calculated here is only metastable if $\gamma$ is so small that there is more than one attractor for the limit cycle, cf. Fig. \ref{intersecfigure}. The timescale for switching between different attractors is much longer than the time to relax in a given metastable steady state. Thus it is not considered in this article. In order to choose the metastable attractor for the numerical simulation, we choose an initial state in the vicinity of our preferred attractor, in this case the limit cycle with lowest possible amplitude. Also in the analytical expressions for the Fano factor, we always treat possibly metastable states as steady states. For very large $g_0/\omega_m$ different metastable attractors start to merge and the analysis becomes more involved. {\color{black} This merging of attractors and its effect on nonclassical features was studied in detail by \cite{Nation2013}.}

\section{Conclusions}

We studied the quantum regime of optomechanical limit cycles. Based on the Laser theory of Haake and Lewenstein \cite{Haake1983} we derived an effective Fokker-Planck equation for an optomechanical system. The analytical prediction for the oscillator's steady state is in agreement with the work of Rodrigues and Armour \cite{Rodrigues2010, Armour2012a} for driving fields on the first blue sideband.

Our treatment naturally includes also the Kerr effect, which becomes important for large $g_0^2/\omega_m$. One consequence important for the quantum theory of limit cycles is the shift of the detuning of equation \eqref{delt0}, which occurs even without photons in the cavity, and had to be introduced phenomenologically in \cite{Armour2012a}. This shift explains the possibility of limit cycles on the blue sideband in \cite{Qian2012} or for the parameters of Fig. \ref{jiangfigure}.

The effective cavity detuning is usually approximated as a static variable. Within our framework one can describe its dynamical nature, which is a classical phenomenon scaling proportional to the Kerr parameter. Figures \ref{phasetrans} and \ref{limittime} show how this smoothens the phase transition between optomechanical cooling and self-induced oscillations.

We studied the quantum limit cycles on resonance and found the simple analytical expression \eqref{fullfano}, that predicts the possibility of very small values for the Fano factor $F$ of the mechanical oscillator. We found that in the sideband resolved regime a large value of $\frac{g_0^2 E^2}{\omega_m^3 \gamma \bar n}$, i.e. a large \textit{linearized} optomechanical coupling, is required to minimize $F$.

We then established a relation between sub-Poissonian phonon statistics and negativity of the Wigner density for typical parameters of limit cycles: The oscillator's steady state has an approximately Gaussian number distribution at each metastable limit cycle. For these states the requirement on $F$ to see negativity of the Wigner function is given by the function of figure \ref{negaFANO}.

Using a Monte-Carlo method with an adaptive Hilbert space, we numerically checked this scaling even for limit cycles with very large amplitude and many photons in the cavity, where an ordinary steady state solver cannot be applied. The numerical simulation depicted in figure \ref{SCANS} show that indeed the criterion of a small Fano factor can predict the negativity of the Wigner function. For currently more feasible experimental parameters with even smaller ${g_0}/{\omega_m}$, the negativity disappears according to Fig. \ref{negaFANO}, but the very small Fano factors remain.

We believe that the present approach provides a suitable starting point for further studies of optomechanical system in the limit of strong couplings. We point out once more that in the ``semi-polaron picture'' introduced here the Kerr nonlinearity and the optomechanical interaction occur as independent terms. This enables in principle to take into account the squeezed noise of the cavity when deriving effective equations of motion of the mechanical oscillator. {\color{black} While for the parameters considered in this article we could neglect this effect, additional diffusion for the mechanical oscillator is to be expected for very strong laser drive. This would apply to the case of limit cycles, but could also become important in the cooling regime. }

\paragraph*{Acknowledgements}
This work was funded by the Centre for Quantum Engineering and Space-Time Research (QUEST) at the Leibniz University Hannover and by the European Community (FP7-Programm) through iQUOEMS (grant agreement no 323924). We acknowledge the support of the cluster system team at the Leibniz University of Hannover in the production of this work. We thank Denis Vasilyev and Kai Stannigl for fruitful discussions.

\begin{appendix}

\section{Transformations for general phase space distribution}\label{App:PhaseSpace}
\label{generalcalc}
\subsection{Semi-Polaron Transformation}
\label{semipolcorr}
In the main text we introduced the semi-polaron tranformation only for the special case of the Q-function, to make the equations more readable. Here we drop this restriction and assume the more general case of an $s$-parameterized phase space distribution $P_s$ with $s \in [-1, 1]$. For the convenience of the calculation we define $p=\tfrac{s+1}2 \in [0,1]$ and $q=1-p$.
Note that for $q=0$ this corresponds to the Glauber-Sudarshan $P$-representation, for $q=\frac12$ to the Wigner-representation, and for $q=1$ to the Husimi  $Q$-representation.

Starting from the standard optomechanical Hamiltonian and Lindblad operators we first switch to a displaced and rotating frame with frequency $\omega_m$ for the mechanical oscillator so that $b \to \beta_0+b e^{-i \omega_m t}$ and introduce the shorthand notation $b_t=b e^{-i \omega_m t}$. This transformation also leaves the Lindblad operators unchanged and the Hamiltonian transforms to
  \begin{align}
    &H= (\omega_m+i\gamma) \beta_0^* b_t+ (\omega_m-i\gamma) \beta_0 b_t^\dagger\\&
 -g_0(\beta_0+ \beta_0^*)a^\dagger a \nonumber \\&
 -\Delta a^\dagger a
- g_0a^\dagger a\left(b_t +b_t^\dagger \right)
-iE\left(a-a^\dagger\right).\nonumber\nonumber
\end{align}
Using the translation rules
     \begin{align*}
        b \rho &\to (\beta +q \partial_{\beta^*})\sigma &
        b^\dagger \rho &\to (\beta^* -p \partial_{\beta})\sigma
      \end{align*}
we obtain the translated equation of motion
$
\dot \sigma(\beta,\beta^*)=\mathcal L_c \sigma+ \mathcal L_m \sigma+\mathcal L_{int} \sigma
$.
With the shorthands $\beta_t=\beta e^{-i \omega_m t}$ and $\fpbt={\partial_\beta} e^{i \omega_m t}$ this gives
\begin{align}
\label{timedependentcavity}
&\mathcal L_c \sigma=-i \left[-g_0(\beta_0+ \beta_0^*)a^\dagger a       -\Delta a^\dagger a - g_0a^\dagger a\left(\beta_t +\beta_t^* \right), \sigma \right]\nonumber\\
& -i\left[ -iE\left(a-a^\dagger\right), \sigma \right] +L_c \sigma\\
&\mathcal L_{int} \sigma=-ig_0\left( (q\partial_{\beta_t}-p\partial_{\beta^*_t}) \sigma a^\dagger a-    ( q\partial_{\beta_t^*}-p\partial_{\beta_t})a^\dagger a\sigma\right)\\
&\mathcal L_m \sigma=-i   (\omega_m+i\gamma) \beta_0^* \partial_{\beta^*_t}\sigma +  i   (\omega_m-i\gamma) \beta_0   \partial_{\beta_t}\sigma +I_m \sigma.
\end{align}
with
\begin{align}
&I_m=\gamma \left( \partial_{\beta} \beta +  \partial_{\beta^*} \beta^*\right) +2\gamma (\bar n +q^2+pq) \partial_{\beta^*}  \partial_{\beta}\\
&L_c=\kappa D[a].
\end{align}
In analogy to transformation \eqref{eq:semipolQ} we apply the more general
      \begin{align*}
     & \tilde\sigma(t) =\exp\left[-i\theta(t)a^\dagger a\right] \sigma(t) \exp\left[i\theta(t)a^\dagger a\right],
      \end{align*}
 with parameters $\lambda=\lambda_r+i \lambda_i=\tfrac{g_0}{\omega_m+i \gamma}$ and\\ $\theta(t)=i\left(\lambda \beta e^{-i\omega_m t}- \lambda^*\beta^* e^{i\omega_m t}\right)$.
which gives
\begin{align}
&\mathcal L_c \sigma=-i \left[  -\Delta a^\dagger a -g_0 \lambda_r (\ada)^2  -iE\left(e^{i\theta(t)} a-e^{-i\theta(t)}a^\dagger\right), \sigma \right] \nonumber\\
&+L_c \sigma+2\gamma |\lambda|^2(\bar n +q^2+pq+\tfrac{q-p}2)    D[\ada] \sigma
\\
&\mathcal L_{int} \sigma=-ig_0\left( (q\partial_{\beta_t}-p\partial_{\beta^*_t}) \sigma a^\dagger a-    ( q\partial_{\beta_t^*}-p\partial_{\beta_t})a^\dagger a\sigma\right)\nonumber\\
&+2\gamma (\bar n +q^2+pq) \left((\lambda^* \partial_{\beta_t} - \lambda \partial_{\beta^*_t} ) [\ada, \sigma]  \right)
\\
&\mathcal L_m \sigma=I_m \sigma-i   (\omega_m+i\gamma) \beta_0^* \partial_{\beta^*_t}\sigma +  i   (\omega_m-i\gamma) \beta_0   \partial_{\beta_t}\sigma.
\end{align}
and includes terms of order $\frac1Q$. After dropping these terms as an approximation this is, with $K=\tfrac{g_0^2}{\omega_m}$,
\begin{align}
&\mathcal L_c \sigma=-i \left[  -\Delta a^\dagger a -K  (\ada)^2  -iE\left(e^{i\theta(t)} a-e^{-i\theta(t)}a^\dagger\right), \sigma \right]\nonumber \\
&+L_c \sigma
\\
&\mathcal L_{int} \sigma=-ig_0\left( (q\partial_{\beta_t}-p\partial_{\beta^*_t}) \sigma a^\dagger a-    ( q\partial_{\beta_t^*}-p\partial_{\beta_t})a^\dagger a\sigma\right)
\\
&\mathcal L_m \sigma=I_m \sigma-i   (\omega_m+i\gamma) \beta_0^* \partial_{\beta^*_t}\sigma +  i   (\omega_m-i\gamma) \beta_0   \partial_{\beta_t}\sigma.
\end{align}

We now transform to a displaced frame   $\tilde\sigma=D^\dagger(\alpha)\sigma D(\alpha)$ with parameter $\alpha(\beta,t) \in \mathbb C$.
For a master equation of the form
  \[
  \dot\rho=-i\left[-\Delta a^\dagger a-K (\ada)^2 -i\left(E(t)a-E^*(t)a^\dagger\right),\rho\right]+L_c\rho
  \]
  the transformation to a displaced frame $\tilde\rho=D^\dagger(\alpha(t))\rho D(\alpha(t))$ gives
  \begin{align} \label{hugetrafo}
    \dot{\tilde{\rho}}
    &=L\tilde\rho -i [-(\Delta+4K|\alpha|^2) a^\dagger a   -K (\ada)^2  \nonumber\\
&   -K \left( \alpha^2 (a^\dagger)^2 +(\alpha^*a+\alpha a^\dagger) \ada  +h.c.  \right) \nonumber\\
&              -i\left\{\big(\dot\alpha+(\kappa-i\Delta     -i2K|\alpha|^2               )\alpha-E\big)a^\dagger-\mathrm{h.c.}\right\},\tilde\rho ]
  \end{align}

Depending on wether one wants to study the regime $|\alpha| \gg1$ or $|\alpha| \ll  1$, either the terms with low or high order in $\alpha$ can be neglected at this point and a different choice of $\alpha(t)$ is required to cancel all displacement-like terms.

\subsection{Displaced frame for $|\alpha| \gg1$}
  We can cancel the displacement-like terms, which include the terms of order $K |\alpha|^3$, by imposing that $\alpha(t)$ solves
  \[
  \dot\alpha(t)=\left(i(\Delta+2K|\alpha(t)|^2)-\kappa\right)\alpha(t)+E(t)
  \]
  such that in the displaced frame
  \begin{align}
&  \dot{\tilde\rho}=-i[-(\Delta+4K|\alpha|^2) a^\dagger a  \nonumber \\
& -K \left( \alpha^2 (a^\dagger)^2 +(\alpha^*a+\alpha a^\dagger) \ada  +h.c.  \right)   -K (\ada)^2 ,\tilde\rho]+L\tilde\rho.
  \end{align}

Neglecting the terms proportional to $K$ up to first order in $\alpha$, the Liouvillians are
\begin{align}
\mathcal L_c \sigma=&-i\left[-(\Delta+4K|\alpha|^2) a^\dagger a   -K \left( \alpha^2 (a^\dagger)^2   +h.c.  \right) , \sigma \right] \label{Lceq}\nonumber \\
& +L_c \sigma
\\
\mathcal L_{int} \sigma=&-ig_0 (q\partial_{\beta_t}-p\partial_{\beta^*_t}) \sigma (\alpha^*a+\alpha a^\dagger) \nonumber\\
& +i g_0    ( q\partial_{\beta_t^*}-p\partial_{\beta_t})(\alpha^*a+\alpha a^\dagger)\sigma
\\
\mathcal L_m \sigma=&-ig_0 (\partial_{\beta_t}-\partial_{\beta^*_t}) |\alpha|^2 \sigma+I_m \sigma\nonumber \\&
-i   (\omega_m+i\gamma) \beta_0^* \partial_{\beta^*_t}\sigma +  i   (\omega_m-i\gamma) \beta_0   \partial_{\beta_t}\sigma
\end{align}
 Note that now, in analogy to laser theory, $\mathcal L_m$ reproduces the classical drift. We have now $\left|\mathcal L_{int}\right| \propto g_0 |\alpha|  \ll g_0 \langle \ada \rangle  \approx g_0 |\alpha|^2$, where $\langle \ada \rangle$ refers to the average before the transformation.

\subsection{Displaced frame for $|\alpha| \ll 1$}
If we restrict the analysis to only the lowest two Fock states, the operators consisting of three creation/annihlation-operators resulting from transformation \eqref{hugetrafo} can be approximated with just one operator, e.g. $a a^\dagger a \approx a$. This time we neglect the terms proportional to $K$ of third order in $\alpha$.
By imposing that $\alpha(t)$ solves this time
\begin{equation} \label{alpsol0}
  \dot\alpha(t)=\left(i(\Delta+K)-\kappa\right)\alpha(t)+E(t),
\end{equation}
we can cancel the remaining displacement-like terms.
The Liouvillians are now
\begin{align}
\mathcal L_c \sigma=&-i\left[-\Delta a^\dagger a   -K \left( \ada   \right)^2   -K \left( \alpha^2 (a^\dagger)^2   +h.c.  \right)   , \sigma \right] \label{Lceq2}
 +L_c \sigma
\\
\mathcal L_{int} \sigma=&-ig_0 (q\partial_{\beta_t}-p\partial_{\beta^*_t}) \sigma (\alpha^*a+\alpha a^\dagger+\ada) \nonumber\\
& +i g_0    ( q\partial_{\beta_t^*}-p\partial_{\beta_t})(\alpha^*a+\alpha a^\dagger+\ada)\sigma
\\
\mathcal L_m \sigma=&-ig_0 (\partial_{\beta_t}-\partial_{\beta^*_t}) |\alpha|^2 \sigma+I_m \sigma \nonumber\\&
-i   (\omega_m+i\gamma) \beta_0^* \partial_{\beta^*_t}\sigma +  i   (\omega_m-i\gamma) \beta_0   \partial_{\beta_t}\sigma
\end{align}

\section{Derivation of Fokker-Planck equation}
\label{adi}
In order to obtain the approximate Fokker-Planck equation for the mechanical oscillator, we now eliminate the cavity in second order pertubation theory. We show this in detail for $|\alpha| \gg1$ and then briefly write down the results for $|\alpha| \ll1$.
\subsection{Adiabatic elimination of the cavity in the $|\alpha| \gg1$ regime}

Let us for now ignore $\mathcal{L}_m$ and reinclude it later. Defining $\sigma_{ij}:=\bra{i} \sigma \ket{j}$ and cutting off after index $(i,j)=(1,1)$ we get the equation of motion
\begin{align}
&\dot \sigma_{00}=2 \kappa \sigma_{11}\nonumber\\&
+i g_0 \left(\fpbstq \alpha^*_t \sigma_{10} - \fpbtq \alpha_t \sigma_{01}  \right)\nonumber\\&
+i g_0 \left(\fpbst \alpha^*_t \alpha_t \sigma_{00} - \fpbt \alpha^*_t \alpha_t \sigma_{00}  \right)\\
&\dot \sigma_{11}=-2 \kappa \sigma_{11}\nonumber\\&
+i g_0 \left(\fpbstq \alpha_t \sigma_{01} - \fpbtq \alpha^*_t \sigma_{10}  \right)\nonumber\\&
+i g_0 \left(\fpbst  \sigma_{11} - \fpbt  \sigma_{11}  \right)\nonumber
+i g_0 \left(\fpbst \alpha^*_t \alpha_t \sigma_{11} - \fpbt \alpha^*_t \alpha_t \sigma_{11}  \right)\\
&\dot \sigma_{10}=-\kappa \sigma_{10}\nonumber
+i \Defft \sigma_{10}
+i g_0 \fpbst \sigma_{10}\nonumber \\&
+i g_0 \left(\fpbstq \alpha_t \sigma_{00} - \fpbtq \alpha_t \sigma_{11}  \right)\nonumber\\&
+i g_0 \left(\fpbst \alpha^*_t \alpha_t \sigma_{10} - \fpbt \alpha^*_t \alpha_t \sigma_{10}  \right)
\end{align}
We now adiabatically eliminate $\sigma_{10}$ to first order in $g_0$ (note that $\sigma_{11}$ is already of order $g_0^2$):
\begin{align}
\sigma_{10}(t)&= \int_0^\infty \mathrm d \tau e^{-\kappa \tau+i  \Defft \tau} i g_0  \fpbsttq \alpha_{t-\tau} \sigma_{00}(t)
\\&=i g_0 \sum_n (q\fpbs e^{i(n-1)\omega_M t}  \frac{\alpha_n}{\tilde h_{n-1}}-p\fpb e^{i(n+1)\omega_M t}  \frac{\alpha_n}{\tilde h_{n+1}}) \sigma_{00}(t)
\end{align}
where $h_n=\kappa +i(n \omega_M -\Defft)$ and
$  \alpha= \sum_{n=-\infty}^\infty \alpha_n e^{i n \omega_m t}$ with  $\alpha_n=\Alpha_n e^{-in \phi}$.
Now the derivative of the phase space distribution is approximately given by $\dot P_s(\beta,\beta^*) \approx \mathrm{Tr}( \dot \sigma_{00}(\beta,\beta^*)+\dot \sigma_{11}(\beta,\beta^*))$, which gives to second order in $g_0$
\begin{align}
\dot P_s&=  \sum_n
q^2 g_0^2  \left(
\fpbs \fpb \frac{\alpha^*_n \alpha_n}{\tilde h_{n-1}}
 -\fpbs \fpbs  \frac{\alpha^*_{n-2} \alpha_n}{\tilde h_{n-1}}
 \right)P_s  \\
&+p^2 g_0^2   \left(
\fpbs \fpb \frac{\alpha^*_n \alpha_n}{\tilde h_{n+1}}
 -\fpb \fpb  \frac{\alpha^*_{n+2} \alpha_n}{\tilde h_{n+1}}
 \right)P_s  \\
&+pq g_0^2   \left(
\fpbs \fpb  \left(\frac{\alpha^*_n \alpha_n}{\tilde h_{n+1}}+\frac{\alpha^*_n \alpha_n}{\tilde h_{n-1}}\right)
 -\fpb^2   \frac{\alpha^*_{n+2} \alpha_n}{\tilde h_{n+1}}
-\fpbsz   \frac{\alpha^*_{n-2} \alpha_n}{\tilde h_{n-1}}
 \right)P_s  \\
&+i g_0 \left( \fpbs  \alpha^*_{n-1} \alpha_n \right) P_s \\
&+h.c.,
\end{align}
where we neglected terms $\propto \frac1r$ as they are negligible at the position of the limit cycle.
Note that the drift term does not depend on the choice of phase-space distribution.
For the $Q$-function the equation simplies to
\begin{align}
\dot Q&=g_0^2 \sum_n \left(
\fpbs \fpb  \frac{2\kappa \alpha^*_n \alpha_n}{|\tilde h_{n-1}|^2}
 -\fpbs \fpbs  \frac{\alpha^*_{n-2} \alpha_n}{\tilde h_{n-1}}
 -\fpb \fpb  \frac{\alpha_{n-2} \alpha^*_n}{\tilde h^*_{n-1}}
 \right)Q\nonumber \\&
+i g_0 \sum_n \left( \fpbs  \alpha^*_{n-1} \alpha_n
 -\fpb  \alpha_{n-1} \alpha^*_n
 \right) Q,
\end{align}
and for the Wigner function to
\begin{align}
\dot W&= \sum_n
  \frac{g_0^2 \kappa}{|\tilde h_{n+1}|^2} \left(
 \fpbs \fpb \left({|\alpha_n|^2}+{|\alpha_{n+2}|^2}\right)
 -\fpb^2   {\alpha^*_{n+2} \alpha_n}
-\fpbsz   {\alpha^*_{n} \alpha_{n+2}}
 \right)W \nonumber\\&
+i g_0 \sum_n \left( \fpbs  \alpha^*_{n-1} \alpha_n
 -\fpb  \alpha_{n-1} \alpha^*_n
 \right) W
\label{WignerfullDiff}
\end{align}

\subsection{Transformation to polar coordinates}
We are finally interested in the EOM for polar coordinates $(r, \phi)$. When previously $\int \mathrm d \alpha \mathrm d \alpha^* P_s(\alpha,\alpha^*)=1$ the new normalization is $\int r \mathrm d r \mathrm d \phi  P_s(r,\phi)=1$.
With
\begin{align}
&\partial_x= \cos (\phi) \partial_r - \frac{\sin(\phi)}{r} \partial_\phi
\\&\partial_y=\sin (\phi) \partial_r + \frac{\cos(\phi)}{r} \partial_\phi
\end{align}
we get
\begin{align}
&\partial_\beta=\tfrac 12 e^{-i \phi} \left( \partial_r -\frac ir \partial_\phi \right)
\\&\partial_{\beta^*}=\tfrac 12 e^{i \phi} \left( \partial_r +\frac ir \partial_\phi \right)
\end{align}
and
\begin{align}
&(2\partial_\beta)^2=e^{-2i\phi} \left( \partial_r^2-2 \frac ir \partial_{r \phi}+2\frac i{r^2} \partial_\phi -\frac 1r \partial_r -\frac 1{r^2} \partial_\phi^2
\right)
\\&(2\partial_\beta)(2\partial_\beta^*)=  \partial_r^2 +\frac 1r \partial_r +\frac 1{r^2} \partial_\phi^2
\end{align}
Integrating out $\phi$ and again neglecting terms $\propto \frac 1r$ we get e.g. for the $Q$-distribution
\begin{align}
\dot Q&= \sum_n \frac{g_0^2}{2}  \partial_r^2\left(
 \frac{\kappa \Alpha^*_n \Alpha_n}{|\tilde h_{n-1}|^2}
 -  \mathrm{Re} \left[\frac{\Alpha^*_{n-2} \Alpha_n}{\tilde h_{n-1}} \right]
 \right)Q \\
&+ g_0 \partial_r \left(  \mathrm{Im}\left[ \Alpha_{n-1} \Alpha^*_n\right]
 \right) Q
- \frac 1r \partial_r \frac{g_0^2}{2} \left(
 \frac{\kappa \Alpha^*_n \Alpha_n}{|\tilde h_{n-1}|^2}
 -  \mathrm{Re} \left[\frac{\Alpha^*_{n-2} \Alpha_n}{\tilde h_{n-1}} \right]
 \right) Q
\end{align}
or, in compact form with $J_n:=J_n\left(-\eta r\right)$ and including $\mathcal{L}_m$, this gives the parameters
\begin{align}
&D_Q=\frac {\gamma(1+\bar n)} 2+\sum_n \frac{g_0^2E^2}{2} \left(
 \frac{\kappa J_n J_n}{|h_{n}|^2|\tilde h_{n-1}|^2}
 -  \mathrm{Re} \left[\frac{J_{n-2} J_n}{\tilde h_{n-1}h^*_{n-2}h_{n}} \right]
 \right) \\
&\mu_Q=-\gamma r  -\sum_n g_0E^2 \left(  \mathrm{Im}\left[ \frac{J_{n-1} J_n}{h_{n-1} h^*_n}\right]
 \right)
\end{align}
for the FPE:
\begin{equation}
\dot P_s= -  \partial_r \mu_s P_s +  \partial_r^2 D_s P_s
\end{equation}
that can be solved (up to normalization) as
\begin{align}
\label{fullsolution}
&P_s(r) \propto \frac {e^{I_s(r)}}{D_s(r)}, &I_s(r):= {\int_0^r \frac {\mu_s(r')} {D_s(r')} \mathrm dr'}.
\end{align}
The corresponding equation for the Wigner function has the same drift coefficient and a diffusion of
\begin{align}
\label{DiffW}
D_W(r)&=\frac {\gamma (2\bar n+1)} 4 \nonumber \\
&\quad+\sum_n \frac{\kappa g_0^2E^2}{4|\tilde h_{n+1}|^2}\left(\left|\frac{J_{n+2}}{h_{n+2}}\right|^2+\left|\frac{J_{n}}{h_{n}}\right|^2-\frac{J_{n}J_{n+2}}{h_{n}h^*_{n+2}} -\frac{J_{n}J_{n+2}}{h^*_{n}h_{n+2}} \right).\nonumber
\end{align}
In both cases we assume in steady state that $P_s(r,\phi)=P_s(r)$, i.e. the distributions are independent of $\phi$.

\subsection{Fokker-Planck equation for $|\alpha|^2 \ll 1$}
The procedure of the adiabatic elimination is in complete analogy to $|\alpha|^2 \gg 1$. One only has to replace $\Deff$ and $\Defft$ with $\Delta_K=\Delta+K$ and adjust the solution of $\alpha$ as in Eq.~\eqref{alpsol0}. With $h_n=\kappa+i(n\omega_m-\Delta_K)$ The final coefficients for the Fokker-Planck equation then have the same structure but without the distinction between $h_n$ and $\tilde h_n$. E.g. for the $Q$-function one obtains
\begin{align}
&D_Q=\frac {\gamma(1+\bar n)} 2+\sum_n \frac{g_0^2E^2}{2} \left(
 \frac{\kappa J_n J_n}{|h_{n}|^2| h_{n-1}|^2}
 -  \mathrm{Re} \left[\frac{J_{n-2} J_n}{ h_{n-1}h^*_{n-2}h_{n}} \right]
 \right) \\
&\mu_Q=\ -\gamma r  -\sum_n gE^2 \left(  \mathrm{Im}\left[ \frac{J_{n-1} J_n}{h_{n-1} h^*_n}\right]
 \right)
\end{align}

\section{Semi-Polaron Transformation}\label{App:SemiPolTraf}
The semi-polaron transformation, Eq.~\eqref{eq:semipolQ} in Sec.~\ref{sec:lasertheory}, is introduced in terms of the formalism of quasiprobability distributions. In view of the similarities of this transformation with the polaron transformation in Eq.~\eqref{eq:rhotilde} the question arises how the semi-polaron transformation in  Eq.~\eqref{eq:semipolQ} can be expressed in terms of an ordinary operator representation. The transformed state $\tilde\sigma$ in \eqref{eq:semipolQ} fulfills
\begin{align*}
\partial_\eta\tilde\sigma&=\left[\frac{1}{2}\left(\beta-\beta^*\right)a^\dagger a,\tilde\sigma\right]=\frac{1}{2}\left[a^\dagger a,\tilde\sigma\beta-\beta^*\tilde\sigma\right].
\end{align*}
When written in the second form we can apply the replacement rules~\eqref{eq:replrules} to write the last equation in operator representation
\begin{align*}
\partial_\eta\tilde\rho&=\frac{1}{2}\left[a^\dagger a,\tilde\rho b-b^\dagger\tilde\rho\right]\\
&=\frac{1}{4}\left[(b-b^\dagger)a^\dagger a,\tilde\rho\right]+\frac{1}{4}\left({D}[b^\dagger+a^\dagger a]-{D}[b^\dagger]-{D}[a^\dagger a]\right)\tilde\rho\\
&\equiv L_\mathrm{s-pol}\tilde{\rho}
\end{align*}
In the second line we expressed the generator for the semi polaron transformation in terms of a commutator with a Hamiltonian and three Lindblad terms. The semi polaron transformation in operator representation is thus
$$
\tilde\rho=\exp\left(\eta L_\mathrm{s-pol}\right)\rho.
$$
It becomes equivalent to the polaron transformation if the Lindblad terms in the generator $L_\mathrm{s-pol}$ are dropped. Thus, the semi polaron transformation is non-unitary. In the context of adiabatic elimination of a cavity mode in the bad cavity limit a similar transformation to a ``dissipation picture'' was employed in \cite{Barnett1986,Cirac1992}.

\end{appendix}

\bibliography{/home/niloer/Documents/mendeleybibtex/LimitCycles}

\begin{thebibliography}{57}%
\makeatletter
\providecommand \@ifxundefined [1]{%
 \@ifx{#1\undefined}
}%
\providecommand \@ifnum [1]{%
 \ifnum #1\expandafter \@firstoftwo
 \else \expandafter \@secondoftwo
 \fi
}%
\providecommand \@ifx [1]{%
 \ifx #1\expandafter \@firstoftwo
 \else \expandafter \@secondoftwo
 \fi
}%
\providecommand \natexlab [1]{#1}%
\providecommand \enquote  [1]{``#1''}%
\providecommand \bibnamefont  [1]{#1}%
\providecommand \bibfnamefont [1]{#1}%
\providecommand \citenamefont [1]{#1}%
\providecommand \href@noop [0]{\@secondoftwo}%
\providecommand \href [0]{\begingroup \@sanitize@url \@href}%
\providecommand \@href[1]{\@@startlink{#1}\@@href}%
\providecommand \@@href[1]{\endgroup#1\@@endlink}%
\providecommand \@sanitize@url [0]{\catcode `\\12\catcode `\$12\catcode
  `\&12\catcode `\#12\catcode `\^12\catcode `\_12\catcode `\%12\relax}%
\providecommand \@@startlink[1]{}%
\providecommand \@@endlink[0]{}%
\providecommand \url  [0]{\begingroup\@sanitize@url \@url }%
\providecommand \@url [1]{\endgroup\@href {#1}{\urlprefix }}%
\providecommand \urlprefix  [0]{URL }%
\providecommand \Eprint [0]{\href }%
\providecommand \doibase [0]{http://dx.doi.org/}%
\providecommand \selectlanguage [0]{\@gobble}%
\providecommand \bibinfo  [0]{\@secondoftwo}%
\providecommand \bibfield  [0]{\@secondoftwo}%
\providecommand \translation [1]{[#1]}%
\providecommand \BibitemOpen [0]{}%
\providecommand \bibitemStop [0]{}%
\providecommand \bibitemNoStop [0]{.\EOS\space}%
\providecommand \EOS [0]{\spacefactor3000\relax}%
\providecommand \BibitemShut  [1]{\csname bibitem#1\endcsname}%
\let\auto@bib@innerbib\@empty
\bibitem [{\citenamefont {Aspelmeyer}\ \emph {et~al.}(2013)\citenamefont
  {Aspelmeyer}, \citenamefont {Kippenberg},\ and\ \citenamefont
  {Marquardt}}]{Aspelmeyer2013}%
  \BibitemOpen
  \bibfield  {author} {\bibinfo {author} {\bibfnamefont {Markus}\ \bibnamefont
  {Aspelmeyer}}, \bibinfo {author} {\bibfnamefont {Tobias~J.}\ \bibnamefont
  {Kippenberg}}, \ and\ \bibinfo {author} {\bibfnamefont {Florian}\
  \bibnamefont {Marquardt}},\ }\bibfield  {title} {\enquote {\bibinfo {title}
  {{Cavity Optomechanics}},}\ }\href {http://arxiv.org/abs/1303.0733}
  {\bibfield  {journal} {\bibinfo  {journal} {arXiv:1303.0733}\ } (\bibinfo
  {year} {2013})}\BibitemShut {NoStop}%
\bibitem [{\citenamefont {Meystre}(2013)}]{Meystre2013}%
  \BibitemOpen
  \bibfield  {author} {\bibinfo {author} {\bibfnamefont {Pierre}\ \bibnamefont
  {Meystre}},\ }\bibfield  {title} {\enquote {\bibinfo {title} {{A short walk
  through quantum optomechanics}},}\ }\href
  {http://doi.wiley.com/10.1002/andp.201200226} {\bibfield  {journal} {\bibinfo
   {journal} {Annalen der Physik}\ }\textbf {\bibinfo {volume} {525}},\
  \bibinfo {pages} {215--233} (\bibinfo {year} {2013})}\BibitemShut {NoStop}%
\bibitem [{\citenamefont {Chen}(2013)}]{Chen2013}%
  \BibitemOpen
  \bibfield  {author} {\bibinfo {author} {\bibfnamefont {Yanbei}\ \bibnamefont
  {Chen}},\ }\bibfield  {title} {\enquote {\bibinfo {title} {{Macroscopic
  quantum mechanics: theory and experimental concepts of optomechanics}},}\
  }\href {http://stacks.iop.org/0953-4075/46/i=10/a=104001} {\bibfield
  {journal} {\bibinfo  {journal} {Journal of Physics B: Atomic, Molecular and
  Optical Physics}\ }\textbf {\bibinfo {volume} {46}},\ \bibinfo {pages}
  {104001} (\bibinfo {year} {2013})}\BibitemShut {NoStop}%
\bibitem [{\citenamefont {Teufel}\ \emph {et~al.}(2011)\citenamefont {Teufel},
  \citenamefont {Donner}, \citenamefont {Li}, \citenamefont {Harlow},
  \citenamefont {Allman}, \citenamefont {Cicak}, \citenamefont {Sirois},
  \citenamefont {Whittaker}, \citenamefont {Lehnert},\ and\ \citenamefont
  {Simmonds}}]{Teufel2011}%
  \BibitemOpen
  \bibfield  {author} {\bibinfo {author} {\bibfnamefont {J~D}\ \bibnamefont
  {Teufel}}, \bibinfo {author} {\bibfnamefont {T}~\bibnamefont {Donner}},
  \bibinfo {author} {\bibfnamefont {Dale}\ \bibnamefont {Li}}, \bibinfo
  {author} {\bibfnamefont {J~W}\ \bibnamefont {Harlow}}, \bibinfo {author}
  {\bibfnamefont {M~S}\ \bibnamefont {Allman}}, \bibinfo {author}
  {\bibfnamefont {K}~\bibnamefont {Cicak}}, \bibinfo {author} {\bibfnamefont
  {A~J}\ \bibnamefont {Sirois}}, \bibinfo {author} {\bibfnamefont {J~D}\
  \bibnamefont {Whittaker}}, \bibinfo {author} {\bibfnamefont {K~W}\
  \bibnamefont {Lehnert}}, \ and\ \bibinfo {author} {\bibfnamefont {R~W}\
  \bibnamefont {Simmonds}},\ }\bibfield  {title} {\enquote {\bibinfo {title}
  {{Sideband cooling of micromechanical motion to the quantum ground state.}}}\
  }\href {http://dx.doi.org/10.1038/nature10261} {\bibfield  {journal}
  {\bibinfo  {journal} {Nature}\ }\textbf {\bibinfo {volume} {475}},\ \bibinfo
  {pages} {359--63} (\bibinfo {year} {2011})}\BibitemShut {NoStop}%
\bibitem [{\citenamefont {Chan}\ \emph {et~al.}(2011)\citenamefont {Chan},
  \citenamefont {Alegre}, \citenamefont {Safavi-Naeini}, \citenamefont {Hill},
  \citenamefont {Krause}, \citenamefont {Gr\"{o}blacher}, \citenamefont
  {Aspelmeyer},\ and\ \citenamefont {Painter}}]{Chan2011a}%
  \BibitemOpen
  \bibfield  {author} {\bibinfo {author} {\bibfnamefont {Jasper}\ \bibnamefont
  {Chan}}, \bibinfo {author} {\bibfnamefont {T~P~Mayer}\ \bibnamefont
  {Alegre}}, \bibinfo {author} {\bibfnamefont {Amir~H}\ \bibnamefont
  {Safavi-Naeini}}, \bibinfo {author} {\bibfnamefont {Jeff~T}\ \bibnamefont
  {Hill}}, \bibinfo {author} {\bibfnamefont {Alex}\ \bibnamefont {Krause}},
  \bibinfo {author} {\bibfnamefont {Simon}\ \bibnamefont {Gr\"{o}blacher}},
  \bibinfo {author} {\bibfnamefont {Markus}\ \bibnamefont {Aspelmeyer}}, \ and\
  \bibinfo {author} {\bibfnamefont {Oskar}\ \bibnamefont {Painter}},\
  }\bibfield  {title} {\enquote {\bibinfo {title} {{Laser cooling of a
  nanomechanical oscillator into its quantum ground state.}}}\ }\href
  {http://dx.doi.org/10.1038/nature10461} {\bibfield  {journal} {\bibinfo
  {journal} {Nature}\ }\textbf {\bibinfo {volume} {478}},\ \bibinfo {pages}
  {89--92} (\bibinfo {year} {2011})}\BibitemShut {NoStop}%
\bibitem [{\citenamefont {Brooks}\ \emph {et~al.}(2012)\citenamefont {Brooks},
  \citenamefont {Botter}, \citenamefont {Schreppler}, \citenamefont {Purdy},
  \citenamefont {Brahms},\ and\ \citenamefont {Stamper-Kurn}}]{Brooks2012}%
  \BibitemOpen
  \bibfield  {author} {\bibinfo {author} {\bibfnamefont {Daniel W~C}\
  \bibnamefont {Brooks}}, \bibinfo {author} {\bibfnamefont {Thierry}\
  \bibnamefont {Botter}}, \bibinfo {author} {\bibfnamefont {Sydney}\
  \bibnamefont {Schreppler}}, \bibinfo {author} {\bibfnamefont {Thomas~P}\
  \bibnamefont {Purdy}}, \bibinfo {author} {\bibfnamefont {Nathan}\
  \bibnamefont {Brahms}}, \ and\ \bibinfo {author} {\bibfnamefont {Dan~M}\
  \bibnamefont {Stamper-Kurn}},\ }\bibfield  {title} {\enquote {\bibinfo
  {title} {{Non-classical light generated by quantum-noise-driven cavity
  optomechanics.}}}\ }\href {http://dx.doi.org/10.1038/nature11325} {\bibfield
  {journal} {\bibinfo  {journal} {Nature}\ }\textbf {\bibinfo {volume} {488}},\
  \bibinfo {pages} {476--80} (\bibinfo {year} {2012})}\BibitemShut {NoStop}%
\bibitem [{\citenamefont {Safavi-Naeini}\ \emph {et~al.}(2013)\citenamefont
  {Safavi-Naeini}, \citenamefont {Groeblacher}, \citenamefont {Hill},
  \citenamefont {Chan}, \citenamefont {Aspelmeyer},\ and\ \citenamefont
  {Painter}}]{Safavi-Naeini2013}%
  \BibitemOpen
  \bibfield  {author} {\bibinfo {author} {\bibfnamefont {Amir~H.}\ \bibnamefont
  {Safavi-Naeini}}, \bibinfo {author} {\bibfnamefont {Simon}\ \bibnamefont
  {Groeblacher}}, \bibinfo {author} {\bibfnamefont {Jeff~T.}\ \bibnamefont
  {Hill}}, \bibinfo {author} {\bibfnamefont {Jasper}\ \bibnamefont {Chan}},
  \bibinfo {author} {\bibfnamefont {Markus}\ \bibnamefont {Aspelmeyer}}, \ and\
  \bibinfo {author} {\bibfnamefont {Oskar}\ \bibnamefont {Painter}},\
  }\bibfield  {title} {\enquote {\bibinfo {title} {{Squeezing of light via
  reflection from a silicon micromechanical resonator}},}\ }\href
  {http://arxiv.org/abs/1302.6179} {\bibfield  {journal} {\bibinfo  {journal}
  {arXiv:1302.6179}\ } (\bibinfo {year} {2013})}\BibitemShut {NoStop}%
\bibitem [{\citenamefont {Murch}\ \emph {et~al.}(2008)\citenamefont {Murch},
  \citenamefont {Moore}, \citenamefont {Gupta},\ and\ \citenamefont
  {Stamper-Kurn}}]{Murch2008a}%
  \BibitemOpen
  \bibfield  {author} {\bibinfo {author} {\bibfnamefont {Kater~W.}\
  \bibnamefont {Murch}}, \bibinfo {author} {\bibfnamefont {Kevin~L.}\
  \bibnamefont {Moore}}, \bibinfo {author} {\bibfnamefont {Subhadeep}\
  \bibnamefont {Gupta}}, \ and\ \bibinfo {author} {\bibfnamefont {Dan~M.}\
  \bibnamefont {Stamper-Kurn}},\ }\bibfield  {title} {\enquote {\bibinfo
  {title} {{Observation of quantum-measurement backaction with an ultracold
  atomic gas}},}\ }\href {http://dx.doi.org/10.1038/nphys965} {\bibfield
  {journal} {\bibinfo  {journal} {Nature Physics}\ }\textbf {\bibinfo {volume}
  {4}},\ \bibinfo {pages} {561--564} (\bibinfo {year} {2008})}\BibitemShut
  {NoStop}%
\bibitem [{\citenamefont {Purdy}\ \emph {et~al.}(2013)\citenamefont {Purdy},
  \citenamefont {Peterson},\ and\ \citenamefont {Regal}}]{Purdy2013}%
  \BibitemOpen
  \bibfield  {author} {\bibinfo {author} {\bibfnamefont {T~P}\ \bibnamefont
  {Purdy}}, \bibinfo {author} {\bibfnamefont {R~W}\ \bibnamefont {Peterson}}, \
  and\ \bibinfo {author} {\bibfnamefont {C~A}\ \bibnamefont {Regal}},\
  }\bibfield  {title} {\enquote {\bibinfo {title} {{Observation of radiation
  pressure shot noise on a macroscopic object.}}}\ }\href {\doibase
  10.1126/science.1231282} {\bibfield  {journal} {\bibinfo  {journal} {Science
  (New York, N.Y.)}\ }\textbf {\bibinfo {volume} {339}},\ \bibinfo {pages}
  {801--4} (\bibinfo {year} {2013})}\BibitemShut {NoStop}%
\bibitem [{\citenamefont {Palomaki}\ \emph
  {et~al.}(2013{\natexlab{a}})\citenamefont {Palomaki}, \citenamefont {Harlow},
  \citenamefont {Teufel}, \citenamefont {Simmonds},\ and\ \citenamefont
  {Lehnert}}]{Palomaki2013}%
  \BibitemOpen
  \bibfield  {author} {\bibinfo {author} {\bibfnamefont {T~A}\ \bibnamefont
  {Palomaki}}, \bibinfo {author} {\bibfnamefont {J~W}\ \bibnamefont {Harlow}},
  \bibinfo {author} {\bibfnamefont {J~D}\ \bibnamefont {Teufel}}, \bibinfo
  {author} {\bibfnamefont {R~W}\ \bibnamefont {Simmonds}}, \ and\ \bibinfo
  {author} {\bibfnamefont {K~W}\ \bibnamefont {Lehnert}},\ }\bibfield  {title}
  {\enquote {\bibinfo {title} {{Coherent state transfer between itinerant
  microwave fields and a mechanical oscillator.}}}\ }\href
  {http://dx.doi.org/10.1038/nature11915} {\bibfield  {journal} {\bibinfo
  {journal} {Nature}\ }\textbf {\bibinfo {volume} {495}},\ \bibinfo {pages}
  {210--4} (\bibinfo {year} {2013}{\natexlab{a}})}\BibitemShut {NoStop}%
\bibitem [{\citenamefont {Palomaki}\ \emph
  {et~al.}(2013{\natexlab{b}})\citenamefont {Palomaki}, \citenamefont {Teufel},
  \citenamefont {Simmonds},\ and\ \citenamefont {Lehnert}}]{Palomaki2013a}%
  \BibitemOpen
  \bibfield  {author} {\bibinfo {author} {\bibfnamefont {T~A}\ \bibnamefont
  {Palomaki}}, \bibinfo {author} {\bibfnamefont {J~D}\ \bibnamefont {Teufel}},
  \bibinfo {author} {\bibfnamefont {R~W}\ \bibnamefont {Simmonds}}, \ and\
  \bibinfo {author} {\bibfnamefont {K~W}\ \bibnamefont {Lehnert}},\ }\bibfield
  {title} {\enquote {\bibinfo {title} {{Entangling Mechanical Motion with
  Microwave Fields}},}\ }\href
  {http://www.sciencemag.org/content/early/2013/10/02/science.1244563}
  {\bibfield  {journal} {\bibinfo  {journal} {Science}\ }\textbf {\bibinfo
  {volume} {342}},\ \bibinfo {pages} {710--713} (\bibinfo {year}
  {2013}{\natexlab{b}})}\BibitemShut {NoStop}%
\bibitem [{\citenamefont {Hofer}\ \emph {et~al.}(2011)\citenamefont {Hofer},
  \citenamefont {Wieczorek}, \citenamefont {Aspelmeyer},\ and\ \citenamefont
  {Hammerer}}]{Hofer2011}%
  \BibitemOpen
  \bibfield  {author} {\bibinfo {author} {\bibfnamefont {Sebastian~G.}\
  \bibnamefont {Hofer}}, \bibinfo {author} {\bibfnamefont {Witlef}\
  \bibnamefont {Wieczorek}}, \bibinfo {author} {\bibfnamefont {Markus}\
  \bibnamefont {Aspelmeyer}}, \ and\ \bibinfo {author} {\bibfnamefont
  {Klemens}\ \bibnamefont {Hammerer}},\ }\bibfield  {title} {\enquote {\bibinfo
  {title} {{Quantum entanglement and teleportation in pulsed cavity
  optomechanics}},}\ }\href
  {http://link.aps.org/doi/10.1103/PhysRevA.84.052327} {\bibfield  {journal}
  {\bibinfo  {journal} {Physical Review A}\ }\textbf {\bibinfo {volume} {84}},\
  \bibinfo {pages} {052327} (\bibinfo {year} {2011})}\BibitemShut {NoStop}%
\bibitem [{\citenamefont {Kippenberg}\ \emph {et~al.}(2005)\citenamefont
  {Kippenberg}, \citenamefont {Rokhsari}, \citenamefont {Carmon}, \citenamefont
  {Scherer},\ and\ \citenamefont {Vahala}}]{Kippenberg2005}%
  \BibitemOpen
  \bibfield  {author} {\bibinfo {author} {\bibfnamefont {T.}~\bibnamefont
  {Kippenberg}}, \bibinfo {author} {\bibfnamefont {H.}~\bibnamefont
  {Rokhsari}}, \bibinfo {author} {\bibfnamefont {T.}~\bibnamefont {Carmon}},
  \bibinfo {author} {\bibfnamefont {A.}~\bibnamefont {Scherer}}, \ and\
  \bibinfo {author} {\bibfnamefont {K.J.}\ \bibnamefont {Vahala}},\ }\bibfield
  {title} {\enquote {\bibinfo {title} {{Analysis of Radiation-Pressure Induced
  Mechanical Oscillation of an Optical Microcavity}},}\ }\href
  {http://link.aps.org/doi/10.1103/PhysRevLett.95.033901} {\bibfield  {journal}
  {\bibinfo  {journal} {Physical Review Letters}\ }\textbf {\bibinfo {volume}
  {95}},\ \bibinfo {pages} {033901} (\bibinfo {year} {2005})}\BibitemShut
  {NoStop}%
\bibitem [{\citenamefont {Carmon}\ \emph {et~al.}(2005)\citenamefont {Carmon},
  \citenamefont {Rokhsari}, \citenamefont {Yang}, \citenamefont {Kippenberg},\
  and\ \citenamefont {Vahala}}]{Carmon2005}%
  \BibitemOpen
  \bibfield  {author} {\bibinfo {author} {\bibfnamefont {Tal}\ \bibnamefont
  {Carmon}}, \bibinfo {author} {\bibfnamefont {Hossein}\ \bibnamefont
  {Rokhsari}}, \bibinfo {author} {\bibfnamefont {Lan}\ \bibnamefont {Yang}},
  \bibinfo {author} {\bibfnamefont {Tobias}\ \bibnamefont {Kippenberg}}, \ and\
  \bibinfo {author} {\bibfnamefont {Kerry~J.}\ \bibnamefont {Vahala}},\
  }\bibfield  {title} {\enquote {\bibinfo {title} {{Temporal Behavior of
  Radiation-Pressure-Induced Vibrations of an Optical Microcavity Phonon
  Mode}},}\ }\href {http://link.aps.org/doi/10.1103/PhysRevLett.94.223902}
  {\bibfield  {journal} {\bibinfo  {journal} {Physical Review Letters}\
  }\textbf {\bibinfo {volume} {94}},\ \bibinfo {pages} {223902} (\bibinfo
  {year} {2005})}\BibitemShut {NoStop}%
\bibitem [{\citenamefont {Metzger}\ \emph {et~al.}(2008)\citenamefont
  {Metzger}, \citenamefont {Ludwig}, \citenamefont {Neuenhahn}, \citenamefont
  {Ortlieb}, \citenamefont {Favero}, \citenamefont {Karrai},\ and\
  \citenamefont {Marquardt}}]{Metzger2008}%
  \BibitemOpen
  \bibfield  {author} {\bibinfo {author} {\bibfnamefont {Constanze}\
  \bibnamefont {Metzger}}, \bibinfo {author} {\bibfnamefont {Max}\ \bibnamefont
  {Ludwig}}, \bibinfo {author} {\bibfnamefont {Clemens}\ \bibnamefont
  {Neuenhahn}}, \bibinfo {author} {\bibfnamefont {Alexander}\ \bibnamefont
  {Ortlieb}}, \bibinfo {author} {\bibfnamefont {Ivan}\ \bibnamefont {Favero}},
  \bibinfo {author} {\bibfnamefont {Khaled}\ \bibnamefont {Karrai}}, \ and\
  \bibinfo {author} {\bibfnamefont {Florian}\ \bibnamefont {Marquardt}},\
  }\bibfield  {title} {\enquote {\bibinfo {title} {{Self-Induced Oscillations
  in an Optomechanical System Driven by Bolometric Backaction}},}\ }\href
  {http://link.aps.org/doi/10.1103/PhysRevLett.101.133903} {\bibfield
  {journal} {\bibinfo  {journal} {Physical Review Letters}\ }\textbf {\bibinfo
  {volume} {101}},\ \bibinfo {pages} {133903} (\bibinfo {year}
  {2008})}\BibitemShut {NoStop}%
\bibitem [{\citenamefont {Anetsberger}\ \emph {et~al.}(2009)\citenamefont
  {Anetsberger}, \citenamefont {Arcizet}, \citenamefont {Unterreithmeier},
  \citenamefont {Rivi\`{e}re}, \citenamefont {Schliesser}, \citenamefont
  {Weig}, \citenamefont {Kotthaus},\ and\ \citenamefont
  {Kippenberg}}]{Anetsberger2009}%
  \BibitemOpen
  \bibfield  {author} {\bibinfo {author} {\bibfnamefont {G.}~\bibnamefont
  {Anetsberger}}, \bibinfo {author} {\bibfnamefont {O.}~\bibnamefont
  {Arcizet}}, \bibinfo {author} {\bibfnamefont {Q.~P.}\ \bibnamefont
  {Unterreithmeier}}, \bibinfo {author} {\bibfnamefont {R.}~\bibnamefont
  {Rivi\`{e}re}}, \bibinfo {author} {\bibfnamefont {A.}~\bibnamefont
  {Schliesser}}, \bibinfo {author} {\bibfnamefont {E.~M.}\ \bibnamefont
  {Weig}}, \bibinfo {author} {\bibfnamefont {J.~P.}\ \bibnamefont {Kotthaus}},
  \ and\ \bibinfo {author} {\bibfnamefont {T.~J.}\ \bibnamefont {Kippenberg}},\
  }\bibfield  {title} {\enquote {\bibinfo {title} {{Near-field cavity
  optomechanics with nanomechanical oscillators}},}\ }\href
  {http://dx.doi.org/10.1038/nphys1425} {\bibfield  {journal} {\bibinfo
  {journal} {Nature Physics}\ }\textbf {\bibinfo {volume} {5}},\ \bibinfo
  {pages} {909--914} (\bibinfo {year} {2009})}\BibitemShut {NoStop}%
\bibitem [{\citenamefont {Grudinin}\ \emph {et~al.}(2010)\citenamefont
  {Grudinin}, \citenamefont {Lee}, \citenamefont {Painter},\ and\ \citenamefont
  {Vahala}}]{Grudinin2010}%
  \BibitemOpen
  \bibfield  {author} {\bibinfo {author} {\bibfnamefont {Ivan~S.}\ \bibnamefont
  {Grudinin}}, \bibinfo {author} {\bibfnamefont {Hansuek}\ \bibnamefont {Lee}},
  \bibinfo {author} {\bibfnamefont {O.}~\bibnamefont {Painter}}, \ and\
  \bibinfo {author} {\bibfnamefont {Kerry~J.}\ \bibnamefont {Vahala}},\
  }\bibfield  {title} {\enquote {\bibinfo {title} {{Phonon Laser Action in a
  Tunable Two-Level System}},}\ }\href
  {http://link.aps.org/doi/10.1103/PhysRevLett.104.083901} {\bibfield
  {journal} {\bibinfo  {journal} {Physical Review Letters}\ }\textbf {\bibinfo
  {volume} {104}},\ \bibinfo {pages} {083901} (\bibinfo {year}
  {2010})}\BibitemShut {NoStop}%
\bibitem [{\citenamefont {Zaitsev}\ \emph {et~al.}(2011)\citenamefont
  {Zaitsev}, \citenamefont {Pandey}, \citenamefont {Shtempluck},\ and\
  \citenamefont {Buks}}]{Zaitsev2011}%
  \BibitemOpen
  \bibfield  {author} {\bibinfo {author} {\bibfnamefont {Stav}\ \bibnamefont
  {Zaitsev}}, \bibinfo {author} {\bibfnamefont {Ashok~K.}\ \bibnamefont
  {Pandey}}, \bibinfo {author} {\bibfnamefont {Oleg}\ \bibnamefont
  {Shtempluck}}, \ and\ \bibinfo {author} {\bibfnamefont {Eyal}\ \bibnamefont
  {Buks}},\ }\bibfield  {title} {\enquote {\bibinfo {title} {{Forced and
  self-excited oscillations of an optomechanical cavity}},}\ }\href
  {http://link.aps.org/doi/10.1103/PhysRevE.84.046605} {\bibfield  {journal}
  {\bibinfo  {journal} {Physical Review E}\ }\textbf {\bibinfo {volume} {84}},\
  \bibinfo {pages} {046605} (\bibinfo {year} {2011})}\BibitemShut {NoStop}%
\bibitem [{\citenamefont {Marquardt}\ \emph {et~al.}(2006)\citenamefont
  {Marquardt}, \citenamefont {Harris},\ and\ \citenamefont
  {Girvin}}]{Marquardt2006}%
  \BibitemOpen
  \bibfield  {author} {\bibinfo {author} {\bibfnamefont {Florian}\ \bibnamefont
  {Marquardt}}, \bibinfo {author} {\bibfnamefont {J.G.E.}\ \bibnamefont
  {Harris}}, \ and\ \bibinfo {author} {\bibfnamefont {S.M.}\ \bibnamefont
  {Girvin}},\ }\bibfield  {title} {\enquote {\bibinfo {title} {{Dynamical
  Multistability Induced by Radiation Pressure in High-Finesse Micromechanical
  Optical Cavities}},}\ }\href
  {http://link.aps.org/doi/10.1103/PhysRevLett.96.103901} {\bibfield  {journal}
  {\bibinfo  {journal} {Physical Review Letters}\ }\textbf {\bibinfo {volume}
  {96}},\ \bibinfo {pages} {103901} (\bibinfo {year} {2006})}\BibitemShut
  {NoStop}%
\bibitem [{\citenamefont {Zaitsev}\ \emph {et~al.}(2012)\citenamefont
  {Zaitsev}, \citenamefont {Gottlieb},\ and\ \citenamefont
  {Buks}}]{Zaitsev2012}%
  \BibitemOpen
  \bibfield  {author} {\bibinfo {author} {\bibfnamefont {Stav}\ \bibnamefont
  {Zaitsev}}, \bibinfo {author} {\bibfnamefont {Oded}\ \bibnamefont
  {Gottlieb}}, \ and\ \bibinfo {author} {\bibfnamefont {Eyal}\ \bibnamefont
  {Buks}},\ }\bibfield  {title} {\enquote {\bibinfo {title} {{Nonlinear
  dynamics of a microelectromechanical mirror in an optical resonance
  cavity}},}\ }\href {http://link.springer.com/10.1007/s11071-012-0371-9}
  {\bibfield  {journal} {\bibinfo  {journal} {Nonlinear Dynamics}\ }\textbf
  {\bibinfo {volume} {69}},\ \bibinfo {pages} {1589--1610} (\bibinfo {year}
  {2012})}\BibitemShut {NoStop}%
\bibitem [{\citenamefont {Khurgin}\ \emph
  {et~al.}(2012{\natexlab{a}})\citenamefont {Khurgin}, \citenamefont
  {Pruessner}, \citenamefont {Stievater},\ and\ \citenamefont
  {Rabinovich}}]{Khurgin2012}%
  \BibitemOpen
  \bibfield  {author} {\bibinfo {author} {\bibfnamefont {J.~B.}\ \bibnamefont
  {Khurgin}}, \bibinfo {author} {\bibfnamefont {M.~W.}\ \bibnamefont
  {Pruessner}}, \bibinfo {author} {\bibfnamefont {T.~H.}\ \bibnamefont
  {Stievater}}, \ and\ \bibinfo {author} {\bibfnamefont {W.~S.}\ \bibnamefont
  {Rabinovich}},\ }\bibfield  {title} {\enquote {\bibinfo {title}
  {{Laser-Rate-Equation Description of Optomechanical Oscillators}},}\ }\href
  {http://link.aps.org/doi/10.1103/PhysRevLett.108.223904} {\bibfield
  {journal} {\bibinfo  {journal} {Physical Review Letters}\ }\textbf {\bibinfo
  {volume} {108}},\ \bibinfo {pages} {223904} (\bibinfo {year}
  {2012}{\natexlab{a}})}\BibitemShut {NoStop}%
\bibitem [{\citenamefont {Khurgin}\ \emph
  {et~al.}(2012{\natexlab{b}})\citenamefont {Khurgin}, \citenamefont
  {Pruessner}, \citenamefont {Stievater},\ and\ \citenamefont
  {Rabinovich}}]{Khurgin2012a}%
  \BibitemOpen
  \bibfield  {author} {\bibinfo {author} {\bibfnamefont {J~B}\ \bibnamefont
  {Khurgin}}, \bibinfo {author} {\bibfnamefont {M~W}\ \bibnamefont
  {Pruessner}}, \bibinfo {author} {\bibfnamefont {T~H}\ \bibnamefont
  {Stievater}}, \ and\ \bibinfo {author} {\bibfnamefont {W~S}\ \bibnamefont
  {Rabinovich}},\ }\bibfield  {title} {\enquote {\bibinfo {title} {{Optically
  pumped coherent mechanical oscillators: the laser rate equation theory and
  experimental verification}},}\ }\href
  {http://iopscience.iop.org/1367-2630/14/10/105022/article/} {\bibfield
  {journal} {\bibinfo  {journal} {New Journal of Physics}\ }\textbf {\bibinfo
  {volume} {14}},\ \bibinfo {pages} {105022} (\bibinfo {year}
  {2012}{\natexlab{b}})}\BibitemShut {NoStop}%
\bibitem [{\citenamefont {Ludwig}\ \emph {et~al.}(2008)\citenamefont {Ludwig},
  \citenamefont {Kubala},\ and\ \citenamefont {Marquardt}}]{Ludwig2008}%
  \BibitemOpen
  \bibfield  {author} {\bibinfo {author} {\bibfnamefont {Max}\ \bibnamefont
  {Ludwig}}, \bibinfo {author} {\bibfnamefont {Bj\"{o}rn}\ \bibnamefont
  {Kubala}}, \ and\ \bibinfo {author} {\bibfnamefont {Florian}\ \bibnamefont
  {Marquardt}},\ }\bibfield  {title} {\enquote {\bibinfo {title} {{The
  optomechanical instability in the quantum regime}},}\ }\href
  {http://iopscience.iop.org/1367-2630/10/9/095013/fulltext/} {\bibfield
  {journal} {\bibinfo  {journal} {New Journal of Physics}\ }\textbf {\bibinfo
  {volume} {10}},\ \bibinfo {pages} {095013} (\bibinfo {year}
  {2008})}\BibitemShut {NoStop}%
\bibitem [{\citenamefont {Vahala}(2008)}]{Vahala2008}%
  \BibitemOpen
  \bibfield  {author} {\bibinfo {author} {\bibfnamefont {K.J.}\ \bibnamefont
  {Vahala}},\ }\bibfield  {title} {\enquote {\bibinfo {title} {{Back-action
  limit of linewidth in an optomechanical oscillator}},}\ }\href
  {http://link.aps.org/doi/10.1103/PhysRevA.78.023832} {\bibfield  {journal}
  {\bibinfo  {journal} {Physical Review A}\ }\textbf {\bibinfo {volume} {78}},\
  \bibinfo {pages} {023832} (\bibinfo {year} {2008})}\BibitemShut {NoStop}%
\bibitem [{\citenamefont {Rodrigues}\ and\ \citenamefont
  {Armour}(2010)}]{Rodrigues2010}%
  \BibitemOpen
  \bibfield  {author} {\bibinfo {author} {\bibfnamefont {D.~A.}\ \bibnamefont
  {Rodrigues}}\ and\ \bibinfo {author} {\bibfnamefont {A.~D.}\ \bibnamefont
  {Armour}},\ }\bibfield  {title} {\enquote {\bibinfo {title} {{Amplitude Noise
  Suppression in Cavity-Driven Oscillations of a Mechanical Resonator}},}\
  }\href {http://link.aps.org/doi/10.1103/PhysRevLett.104.053601} {\bibfield
  {journal} {\bibinfo  {journal} {Physical Review Letters}\ }\textbf {\bibinfo
  {volume} {104}},\ \bibinfo {pages} {053601} (\bibinfo {year}
  {2010})}\BibitemShut {NoStop}%
\bibitem [{\citenamefont {Armour}\ and\ \citenamefont
  {Rodrigues}(2012)}]{Armour2012a}%
  \BibitemOpen
  \bibfield  {author} {\bibinfo {author} {\bibfnamefont {Andrew~D.}\
  \bibnamefont {Armour}}\ and\ \bibinfo {author} {\bibfnamefont {Denzil~A.}\
  \bibnamefont {Rodrigues}},\ }\bibfield  {title} {\enquote {\bibinfo {title}
  {{Quantum dynamics of a mechanical resonator driven by a cavity}},}\ }\href
  {http://dx.doi.org/10.1016/j.crhy.2012.03.006} {\bibfield  {journal}
  {\bibinfo  {journal} {Comptes Rendus Physique}\ }\textbf {\bibinfo {volume}
  {13}},\ \bibinfo {pages} {440--453} (\bibinfo {year} {2012})}\BibitemShut
  {NoStop}%
\bibitem [{\citenamefont {Qian}\ \emph {et~al.}(2012)\citenamefont {Qian},
  \citenamefont {Clerk}, \citenamefont {Hammerer},\ and\ \citenamefont
  {Marquardt}}]{Qian2012}%
  \BibitemOpen
  \bibfield  {author} {\bibinfo {author} {\bibfnamefont {Jiang}\ \bibnamefont
  {Qian}}, \bibinfo {author} {\bibfnamefont {A.~A.}\ \bibnamefont {Clerk}},
  \bibinfo {author} {\bibfnamefont {K.}~\bibnamefont {Hammerer}}, \ and\
  \bibinfo {author} {\bibfnamefont {Florian}\ \bibnamefont {Marquardt}},\
  }\bibfield  {title} {\enquote {\bibinfo {title} {{Quantum Signatures of the
  Optomechanical Instability}},}\ }\href
  {http://prl.aps.org/abstract/PRL/v109/i25/e253601} {\bibfield  {journal}
  {\bibinfo  {journal} {Physical Review Letters}\ }\textbf {\bibinfo {volume}
  {109}},\ \bibinfo {pages} {253601} (\bibinfo {year} {2012})}\BibitemShut
  {NoStop}%
\bibitem [{\citenamefont {Nation}(2013)}]{Nation2013}%
  \BibitemOpen
  \bibfield  {author} {\bibinfo {author} {\bibfnamefont {P.~D.}\ \bibnamefont
  {Nation}},\ }\bibfield  {title} {\enquote {\bibinfo {title} {{Nonclassical
  Mechanical States in a Optomechanical Micromaser Analogue}},}\ }\href
  {http://arxiv.org/abs/1308.4213} {\bibfield  {journal} {\bibinfo  {journal}
  {arXiv:1308.4213}\ } (\bibinfo {year} {2013})}\BibitemShut {NoStop}%
\bibitem [{\citenamefont {Dykman}\ and\ \citenamefont
  {Krivoglatz}(1975)}]{Dykman1975}%
  \BibitemOpen
  \bibfield  {author} {\bibinfo {author} {\bibfnamefont {M.}~\bibnamefont
  {Dykman}}\ and\ \bibinfo {author} {\bibfnamefont {M.}~\bibnamefont
  {Krivoglatz}},\ }\bibfield  {title} {\enquote {\bibinfo {title} {{Spectral
  Distribution of Nonlinear Oscillators with Nonlinear Friction Due to a
  Medium}},}\ }\href@noop {} {\bibfield  {journal} {\bibinfo  {journal}
  {Physica Status Solidi (b)}\ }\textbf {\bibinfo {volume} {68}} (\bibinfo
  {year} {1975})}\BibitemShut {NoStop}%
\bibitem [{\citenamefont {Lax}(1967)}]{Lax1967}%
  \BibitemOpen
  \bibfield  {author} {\bibinfo {author} {\bibfnamefont {Melvin}\ \bibnamefont
  {Lax}},\ }\bibfield  {title} {\enquote {\bibinfo {title} {{Classical Noise.
  V. Noise in Self-Sustained Oscillators}},}\ }\href
  {http://link.aps.org/doi/10.1103/PhysRev.160.290} {\bibfield  {journal}
  {\bibinfo  {journal} {Physical Review}\ }\textbf {\bibinfo {volume} {160}},\
  \bibinfo {pages} {290--307} (\bibinfo {year} {1967})}\BibitemShut {NoStop}%
\bibitem [{\citenamefont {Haake}\ and\ \citenamefont
  {Lewenstein}(1983)}]{Haake1983}%
  \BibitemOpen
  \bibfield  {author} {\bibinfo {author} {\bibfnamefont {Fritz}\ \bibnamefont
  {Haake}}\ and\ \bibinfo {author} {\bibfnamefont {Maciej}\ \bibnamefont
  {Lewenstein}},\ }\bibfield  {title} {\enquote {\bibinfo {title} {{Adiabatic
  expansion for the single-mode laser}},}\ }\href
  {http://link.aps.org/doi/10.1103/PhysRevA.27.1013} {\bibfield  {journal}
  {\bibinfo  {journal} {Physical Review A}\ }\textbf {\bibinfo {volume} {27}},\
  \bibinfo {pages} {1013--1021} (\bibinfo {year} {1983})}\BibitemShut {NoStop}%
\bibitem [{\citenamefont {Gardiner}\ and\ \citenamefont
  {Zoller}(2004)}]{Gardiner2004b}%
  \BibitemOpen
  \bibfield  {author} {\bibinfo {author} {\bibfnamefont {Crispin}\ \bibnamefont
  {Gardiner}}\ and\ \bibinfo {author} {\bibfnamefont {Peter}\ \bibnamefont
  {Zoller}},\ }\href
  {http://www.amazon.com/Quantum-Noise-Non-Markovian-Applications-Synergetics/dp/3540223010}
  {\emph {\bibinfo {title} {{Quantum Noise: A Handbook of Markovian and
  Non-Markovian Quantum Stochastic Methods with Applications to Quantum Optics
  (Springer Series in Synergetics)}}}}\ (\bibinfo  {publisher} {Springer},\
  \bibinfo {year} {2004})\BibitemShut {NoStop}%
\bibitem [{\citenamefont {Khurgin}(2010)}]{Khurgin2010}%
  \BibitemOpen
  \bibfield  {author} {\bibinfo {author} {\bibfnamefont {Jacob~B.}\
  \bibnamefont {Khurgin}},\ }\bibfield  {title} {\enquote {\bibinfo {title}
  {{Phonon lasers gain a sound foundation}},}\ }\href
  {http://physics.aps.org/articles/v3/16} {\bibfield  {journal} {\bibinfo
  {journal} {Physics}\ }\textbf {\bibinfo {volume} {3}},\ \bibinfo {pages} {16}
  (\bibinfo {year} {2010})}\BibitemShut {NoStop}%
\bibitem [{\citenamefont {Rabl}(2011)}]{Rabl2011}%
  \BibitemOpen
  \bibfield  {author} {\bibinfo {author} {\bibfnamefont {P.}~\bibnamefont
  {Rabl}},\ }\bibfield  {title} {\enquote {\bibinfo {title} {{Photon Blockade
  Effect in Optomechanical Systems}},}\ }\href {\doibase
  10.1103/PhysRevLett.107.063601} {\bibfield  {journal} {\bibinfo  {journal}
  {Physical Review Letters}\ }\textbf {\bibinfo {volume} {107}},\ \bibinfo
  {pages} {063601} (\bibinfo {year} {2011})}\BibitemShut {NoStop}%
\bibitem [{\citenamefont {Nunnenkamp}\ \emph {et~al.}(2011)\citenamefont
  {Nunnenkamp}, \citenamefont {B\o~rkje},\ and\ \citenamefont
  {Girvin}}]{Nunnenkamp2011}%
  \BibitemOpen
  \bibfield  {author} {\bibinfo {author} {\bibfnamefont {A.}~\bibnamefont
  {Nunnenkamp}}, \bibinfo {author} {\bibfnamefont {K.}~\bibnamefont
  {B\o~rkje}}, \ and\ \bibinfo {author} {\bibfnamefont {S.~M.}\ \bibnamefont
  {Girvin}},\ }\bibfield  {title} {\enquote {\bibinfo {title} {{Single-Photon
  Optomechanics}},}\ }\href {\doibase 10.1103/PhysRevLett.107.063602}
  {\bibfield  {journal} {\bibinfo  {journal} {Physical Review Letters}\
  }\textbf {\bibinfo {volume} {107}},\ \bibinfo {pages} {063602} (\bibinfo
  {year} {2011})}\BibitemShut {NoStop}%
\bibitem [{\citenamefont {Wilson-Rae}\ \emph {et~al.}(2007)\citenamefont
  {Wilson-Rae}, \citenamefont {Nooshi}, \citenamefont {Zwerger},\ and\
  \citenamefont {Kippenberg}}]{Wilson-Rae2007}%
  \BibitemOpen
  \bibfield  {author} {\bibinfo {author} {\bibfnamefont {I.}~\bibnamefont
  {Wilson-Rae}}, \bibinfo {author} {\bibfnamefont {N.}~\bibnamefont {Nooshi}},
  \bibinfo {author} {\bibfnamefont {W.}~\bibnamefont {Zwerger}}, \ and\
  \bibinfo {author} {\bibfnamefont {T.J.}\ \bibnamefont {Kippenberg}},\
  }\bibfield  {title} {\enquote {\bibinfo {title} {{Theory of Ground State
  Cooling of a Mechanical Oscillator Using Dynamical Backaction}},}\ }\href
  {http://prl.aps.org/abstract/PRL/v99/i9/e093901} {\bibfield  {journal}
  {\bibinfo  {journal} {Physical Review Letters}\ }\textbf {\bibinfo {volume}
  {99}},\ \bibinfo {pages} {093901} (\bibinfo {year} {2007})}\BibitemShut
  {NoStop}%
\bibitem [{\citenamefont {Marquardt}\ \emph {et~al.}(2007)\citenamefont
  {Marquardt}, \citenamefont {Chen}, \citenamefont {Clerk},\ and\ \citenamefont
  {Girvin}}]{Marquardt2007}%
  \BibitemOpen
  \bibfield  {author} {\bibinfo {author} {\bibfnamefont {Florian}\ \bibnamefont
  {Marquardt}}, \bibinfo {author} {\bibfnamefont {J.P.}\ \bibnamefont {Chen}},
  \bibinfo {author} {\bibfnamefont {A.A.}\ \bibnamefont {Clerk}}, \ and\
  \bibinfo {author} {\bibfnamefont {S.M.}\ \bibnamefont {Girvin}},\ }\bibfield
  {title} {\enquote {\bibinfo {title} {{Quantum Theory of Cavity-Assisted
  Sideband Cooling of Mechanical Motion}},}\ }\href
  {http://link.aps.org/doi/10.1103/PhysRevLett.99.093902} {\bibfield  {journal}
  {\bibinfo  {journal} {Physical Review Letters}\ }\textbf {\bibinfo {volume}
  {99}},\ \bibinfo {pages} {093902} (\bibinfo {year} {2007})}\BibitemShut
  {NoStop}%
\bibitem [{Note1()}]{Note1}%
  \BibitemOpen
  \bibinfo {note} {The zero point amplitude is $\protect \sqrt {\hbar /m\omega
  _\protect \mathrm {m}}$ for an oscillator of mass $m$}\BibitemShut {NoStop}%
\bibitem [{\citenamefont {Mahan}(2000)}]{Mahan2000}%
  \BibitemOpen
  \bibfield  {author} {\bibinfo {author} {\bibfnamefont {Gerald~D.}\
  \bibnamefont {Mahan}},\ }\href
  {http://www.amazon.com/Many-Particle-Physics-Solids-Liquids/dp/0306463385}
  {\emph {\bibinfo {title} {{Many-Particle Physics (Physics of Solids and
  Liquids)}}}}\ (\bibinfo  {publisher} {Springer},\ \bibinfo {year}
  {2000})\BibitemShut {NoStop}%
\bibitem [{\citenamefont {Drummond}\ and\ \citenamefont
  {Walls}(1980)}]{Drummond1980}%
  \BibitemOpen
  \bibfield  {author} {\bibinfo {author} {\bibfnamefont {P~D}\ \bibnamefont
  {Drummond}}\ and\ \bibinfo {author} {\bibfnamefont {D~F}\ \bibnamefont
  {Walls}},\ }\bibfield  {title} {\enquote {\bibinfo {title} {{Quantum theory
  of optical bistability. I. Nonlinear polarisability model}},}\ }\href
  {http://iopscience.iop.org/0305-4470/13/2/034} {\bibfield  {journal}
  {\bibinfo  {journal} {Journal of Physics A: Mathematical and General}\
  }\textbf {\bibinfo {volume} {13}},\ \bibinfo {pages} {725--741} (\bibinfo
  {year} {1980})}\BibitemShut {NoStop}%
\bibitem [{\citenamefont {Mari}\ and\ \citenamefont {Eisert}(2009)}]{Mari2009}%
  \BibitemOpen
  \bibfield  {author} {\bibinfo {author} {\bibfnamefont {A.}~\bibnamefont
  {Mari}}\ and\ \bibinfo {author} {\bibfnamefont {J.}~\bibnamefont {Eisert}},\
  }\bibfield  {title} {\enquote {\bibinfo {title} {{Gently Modulating
  Optomechanical Systems}},}\ }\href
  {http://apps.webofknowledge.com/full\_record.do?product=WOS\&search\_mode=GeneralSearch\&qid=16\&SID=N26hJaEMInnlLKoclKC\&page=1\&doc=1}
  {\bibfield  {journal} {\bibinfo  {journal} {Physical Review Letters}\
  }\textbf {\bibinfo {volume} {103}},\ \bibinfo {pages} {213603} (\bibinfo
  {year} {2009})}\BibitemShut {NoStop}%
\bibitem [{\citenamefont {Genes}\ \emph {et~al.}(2008)\citenamefont {Genes},
  \citenamefont {Vitali}, \citenamefont {Tombesi}, \citenamefont {Gigan},\ and\
  \citenamefont {Aspelmeyer}}]{Genes2008}%
  \BibitemOpen
  \bibfield  {author} {\bibinfo {author} {\bibfnamefont {C.}~\bibnamefont
  {Genes}}, \bibinfo {author} {\bibfnamefont {D.}~\bibnamefont {Vitali}},
  \bibinfo {author} {\bibfnamefont {P.}~\bibnamefont {Tombesi}}, \bibinfo
  {author} {\bibfnamefont {S.}~\bibnamefont {Gigan}}, \ and\ \bibinfo {author}
  {\bibfnamefont {M.}~\bibnamefont {Aspelmeyer}},\ }\bibfield  {title}
  {\enquote {\bibinfo {title} {{Ground-state cooling of a micromechanical
  oscillator: Comparing cold damping and cavity-assisted cooling schemes}},}\
  }\href {http://link.aps.org/doi/10.1103/PhysRevA.77.033804} {\bibfield
  {journal} {\bibinfo  {journal} {Physical Review A}\ }\textbf {\bibinfo
  {volume} {77}},\ \bibinfo {pages} {033804} (\bibinfo {year}
  {2008})}\BibitemShut {NoStop}%
\bibitem [{\citenamefont {Ghobadi}\ \emph {et~al.}(2011)\citenamefont
  {Ghobadi}, \citenamefont {Bahrampour},\ and\ \citenamefont
  {Simon}}]{Ghobadi2011}%
  \BibitemOpen
  \bibfield  {author} {\bibinfo {author} {\bibfnamefont {R.}~\bibnamefont
  {Ghobadi}}, \bibinfo {author} {\bibfnamefont {A.~R.}\ \bibnamefont
  {Bahrampour}}, \ and\ \bibinfo {author} {\bibfnamefont {C.}~\bibnamefont
  {Simon}},\ }\bibfield  {title} {\enquote {\bibinfo {title} {{Quantum
  optomechanics in the bistable regime}},}\ }\href
  {http://link.aps.org/doi/10.1103/PhysRevA.84.033846} {\bibfield  {journal}
  {\bibinfo  {journal} {Physical Review A}\ }\textbf {\bibinfo {volume} {84}},\
  \bibinfo {pages} {033846} (\bibinfo {year} {2011})}\BibitemShut {NoStop}%
\bibitem [{\citenamefont {Aldana}\ \emph {et~al.}(2013)\citenamefont {Aldana},
  \citenamefont {Bruder},\ and\ \citenamefont {Nunnenkamp}}]{Aldana2013}%
  \BibitemOpen
  \bibfield  {author} {\bibinfo {author} {\bibfnamefont {Samuel}\ \bibnamefont
  {Aldana}}, \bibinfo {author} {\bibfnamefont {Christoph}\ \bibnamefont
  {Bruder}}, \ and\ \bibinfo {author} {\bibfnamefont {Andreas}\ \bibnamefont
  {Nunnenkamp}},\ }\bibfield  {title} {\enquote {\bibinfo {title} {{On the
  equivalence between an optomechanical system and a Kerr medium}},}\ }\href
  {http://arxiv.org/abs/1306.0415} {\bibfield  {journal} {\bibinfo  {journal}
  {arXiv:1306.0415}\ ,\ \bibinfo {pages} {10}} (\bibinfo {year}
  {2013})}\BibitemShut {NoStop}%
\bibitem [{\citenamefont {Heinrich}\ \emph {et~al.}(2011)\citenamefont
  {Heinrich}, \citenamefont {Ludwig}, \citenamefont {Qian}, \citenamefont
  {Kubala},\ and\ \citenamefont {Marquardt}}]{Heinrich2011}%
  \BibitemOpen
  \bibfield  {author} {\bibinfo {author} {\bibfnamefont {Georg}\ \bibnamefont
  {Heinrich}}, \bibinfo {author} {\bibfnamefont {Max}\ \bibnamefont {Ludwig}},
  \bibinfo {author} {\bibfnamefont {Jiang}\ \bibnamefont {Qian}}, \bibinfo
  {author} {\bibfnamefont {Bj\"{o}rn}\ \bibnamefont {Kubala}}, \ and\ \bibinfo
  {author} {\bibfnamefont {Florian}\ \bibnamefont {Marquardt}},\ }\bibfield
  {title} {\enquote {\bibinfo {title} {{Collective Dynamics in Optomechanical
  Arrays}},}\ }\href {http://link.aps.org/doi/10.1103/PhysRevLett.107.043603}
  {\bibfield  {journal} {\bibinfo  {journal} {Physical Review Letters}\
  }\textbf {\bibinfo {volume} {107}},\ \bibinfo {pages} {043603} (\bibinfo
  {year} {2011})}\BibitemShut {NoStop}%
\bibitem [{\citenamefont {Golubev}\ and\ \citenamefont
  {Sokolov}(1984)}]{Sokolov1984}%
  \BibitemOpen
  \bibfield  {author} {\bibinfo {author} {\bibfnamefont {Yu.~M.}\ \bibnamefont
  {Golubev}}\ and\ \bibinfo {author} {\bibfnamefont {I.~V.}\ \bibnamefont
  {Sokolov}},\ }\bibfield  {title} {\enquote {\bibinfo {title} {{Photon
  antibunching in a coherent light source and suppression of the photorecording
  noise}},}\ }\href@noop {} {\bibfield  {journal} {\bibinfo  {journal} {Journal
  of Experimental and Theoretical Physics}\ }\textbf {\bibinfo {volume} {60}},\
  \bibinfo {pages} {234--238} (\bibinfo {year} {1984})}\BibitemShut {NoStop}%
\bibitem [{\citenamefont {Golubev}(1995)}]{Golubev1995}%
  \BibitemOpen
  \bibfield  {author} {\bibinfo {author} {\bibfnamefont {Yu.~M.}\ \bibnamefont
  {Golubev}},\ }\bibfield  {title} {\enquote {\bibinfo {title} {{Excitation of
  the active medium of a micromaser by light from a sub-Poissonian optical
  laser}},}\ }\href@noop {} {\bibfield  {journal} {\bibinfo  {journal} {Journal
  of Experimental and Theoretical Physics}\ }\textbf {\bibinfo {volume} {80}},\
  \bibinfo {pages} {212--218} (\bibinfo {year} {1995})}\BibitemShut {NoStop}%
\bibitem [{\citenamefont {McKeever}\ \emph {et~al.}(2003)\citenamefont
  {McKeever}, \citenamefont {Boca}, \citenamefont {Boozer}, \citenamefont
  {Buck},\ and\ \citenamefont {Kimble}}]{McKeever2003}%
  \BibitemOpen
  \bibfield  {author} {\bibinfo {author} {\bibfnamefont {J}~\bibnamefont
  {McKeever}}, \bibinfo {author} {\bibfnamefont {A}~\bibnamefont {Boca}},
  \bibinfo {author} {\bibfnamefont {A~D}\ \bibnamefont {Boozer}}, \bibinfo
  {author} {\bibfnamefont {J~R}\ \bibnamefont {Buck}}, \ and\ \bibinfo {author}
  {\bibfnamefont {H~J}\ \bibnamefont {Kimble}},\ }\bibfield  {title} {\enquote
  {\bibinfo {title} {{Experimental realization of a one-atom laser in the
  regime of strong coupling.}}}\ }\href@noop {} {\bibfield  {journal} {\bibinfo
   {journal} {Nature}\ }\textbf {\bibinfo {volume} {425}},\ \bibinfo {pages}
  {268--271} (\bibinfo {year} {2003})}\BibitemShut {NoStop}%
\bibitem [{\citenamefont {Kilin}\ and\ \citenamefont
  {Mikhalychev}(2012)}]{Kilin2012}%
  \BibitemOpen
  \bibfield  {author} {\bibinfo {author} {\bibfnamefont {S.}~\bibnamefont
  {Kilin}}\ and\ \bibinfo {author} {\bibfnamefont {A.}~\bibnamefont
  {Mikhalychev}},\ }\href {\doibase 10.1103/PhysRevA.85.063817} {\enquote
  {\bibinfo {title} {{Single-atom laser generates nonlinear coherent
  states}},}\ } (\bibinfo {year} {2012}),\ \Eprint
  {http://arxiv.org/abs/1206.3459} {arXiv:1206.3459} \BibitemShut {NoStop}%
\bibitem [{\citenamefont {Dalibard}\ \emph {et~al.}(1992)\citenamefont
  {Dalibard}, \citenamefont {Castin},\ and\ \citenamefont
  {M\o~lmer}}]{Dalibard1992}%
  \BibitemOpen
  \bibfield  {author} {\bibinfo {author} {\bibfnamefont {Jean}\ \bibnamefont
  {Dalibard}}, \bibinfo {author} {\bibfnamefont {Yvan}\ \bibnamefont {Castin}},
  \ and\ \bibinfo {author} {\bibfnamefont {Klaus}\ \bibnamefont {M\o~lmer}},\
  }\bibfield  {title} {\enquote {\bibinfo {title} {{Wave-function approach to
  dissipative processes in quantum optics}},}\ }\href {\doibase
  10.1103/PhysRevLett.68.580} {\bibfield  {journal} {\bibinfo  {journal}
  {Physical Review Letters}\ }\textbf {\bibinfo {volume} {68}},\ \bibinfo
  {pages} {580--583} (\bibinfo {year} {1992})}\BibitemShut {NoStop}%
\bibitem [{\citenamefont {Dum}\ \emph {et~al.}(1992)\citenamefont {Dum},
  \citenamefont {Zoller},\ and\ \citenamefont {Ritsch}}]{Dum1992}%
  \BibitemOpen
  \bibfield  {author} {\bibinfo {author} {\bibfnamefont {R}~\bibnamefont
  {Dum}}, \bibinfo {author} {\bibfnamefont {P}~\bibnamefont {Zoller}}, \ and\
  \bibinfo {author} {\bibfnamefont {H}~\bibnamefont {Ritsch}},\ }\bibfield
  {title} {\enquote {\bibinfo {title} {{Monte Carlo simulation of the atomic
  master equation for spontaneous emission}},}\ }\href {\doibase
  10.1103/PhysRevA.45.4879} {\bibfield  {journal} {\bibinfo  {journal}
  {Physical Review A}\ }\textbf {\bibinfo {volume} {45}},\ \bibinfo {pages}
  {4879--4887} (\bibinfo {year} {1992})}\BibitemShut {NoStop}%
\bibitem [{\citenamefont {M\o~lmer}\ \emph {et~al.}(1993)\citenamefont
  {M\o~lmer}, \citenamefont {Castin},\ and\ \citenamefont
  {Dalibard}}]{Molmer1993}%
  \BibitemOpen
  \bibfield  {author} {\bibinfo {author} {\bibfnamefont {Klaus}\ \bibnamefont
  {M\o~lmer}}, \bibinfo {author} {\bibfnamefont {Yvan}\ \bibnamefont {Castin}},
  \ and\ \bibinfo {author} {\bibfnamefont {Jean}\ \bibnamefont {Dalibard}},\
  }\bibfield  {title} {\enquote {\bibinfo {title} {{Monte Carlo wave-function
  method in quantum optics}},}\ }\href {\doibase 10.1364/JOSAB.10.000524}
  {\bibfield  {journal} {\bibinfo  {journal} {Journal of the Optical Society of
  America B}\ }\textbf {\bibinfo {volume} {10}},\ \bibinfo {pages} {524}
  (\bibinfo {year} {1993})}\BibitemShut {NoStop}%
\bibitem [{\citenamefont {Johansson}\ \emph {et~al.}(2011)\citenamefont
  {Johansson}, \citenamefont {Nation},\ and\ \citenamefont
  {Nori}}]{Johansson2011a}%
  \BibitemOpen
  \bibfield  {author} {\bibinfo {author} {\bibfnamefont {J.~R.}\ \bibnamefont
  {Johansson}}, \bibinfo {author} {\bibfnamefont {P.~D.}\ \bibnamefont
  {Nation}}, \ and\ \bibinfo {author} {\bibfnamefont {Franco}\ \bibnamefont
  {Nori}},\ }\bibfield  {title} {\enquote {\bibinfo {title} {{QuTiP: An
  open-source Python framework for the dynamics of open quantum systems}},}\
  }\href {\doibase 10.1016/j.cpc.2012.02.021} {\bibfield  {journal} {\bibinfo
  {journal} {Computer Physics Communications}\ }\textbf {\bibinfo {volume}
  {183}},\ \bibinfo {pages} {1760} (\bibinfo {year} {2011})},\ \Eprint
  {http://arxiv.org/abs/1110.0573} {arXiv:1110.0573} \BibitemShut {NoStop}%
\bibitem [{\citenamefont {Johansson}\ \emph {et~al.}(2013)\citenamefont
  {Johansson}, \citenamefont {Nation},\ and\ \citenamefont
  {Nori}}]{Johansson2013}%
  \BibitemOpen
  \bibfield  {author} {\bibinfo {author} {\bibfnamefont {J.R.}\ \bibnamefont
  {Johansson}}, \bibinfo {author} {\bibfnamefont {P.D.}\ \bibnamefont
  {Nation}}, \ and\ \bibinfo {author} {\bibfnamefont {Franco}\ \bibnamefont
  {Nori}},\ }\bibfield  {title} {\enquote {\bibinfo {title} {{QuTiP 2: A Python
  framework for the dynamics of open quantum systems}},}\ }\href {\doibase
  10.1016/j.cpc.2012.11.019} {\bibfield  {journal} {\bibinfo  {journal}
  {Computer Physics Communications}\ }\textbf {\bibinfo {volume} {184}},\
  \bibinfo {pages} {1234--1240} (\bibinfo {year} {2013})}\BibitemShut {NoStop}%
\bibitem [{\citenamefont {Oliphant}(2007)}]{Oliphant2007}%
  \BibitemOpen
  \bibfield  {author} {\bibinfo {author} {\bibfnamefont {Travis~E.}\
  \bibnamefont {Oliphant}},\ }\bibfield  {title} {\enquote {\bibinfo {title}
  {{Python for Scientific Computing}},}\ }\href {\doibase 10.1109/MCSE.2007.58}
  {\bibfield  {journal} {\bibinfo  {journal} {Computing in Science \&
  Engineering}\ }\textbf {\bibinfo {volume} {9}},\ \bibinfo {pages} {10--20}
  (\bibinfo {year} {2007})}\BibitemShut {NoStop}%
\bibitem [{\citenamefont {Barnett}\ and\ \citenamefont
  {Knight}(1986)}]{Barnett1986}%
  \BibitemOpen
  \bibfield  {author} {\bibinfo {author} {\bibfnamefont {SM}~\bibnamefont
  {Barnett}}\ and\ \bibinfo {author} {\bibfnamefont {PL}~\bibnamefont
  {Knight}},\ }\bibfield  {title} {\enquote {\bibinfo {title} {{Dissipation in
  a fundamental model of quantum optical resonance}},}\ }\href
  {http://pra.aps.org/abstract/PRA/v33/i4/p2444\_1} {\bibfield  {journal}
  {\bibinfo  {journal} {Physical Review A}\ }\textbf {\bibinfo {volume} {33}}
  (\bibinfo {year} {1986})}\BibitemShut {NoStop}%
\bibitem [{\citenamefont {Cirac}(1992)}]{Cirac1992}%
  \BibitemOpen
  \bibfield  {author} {\bibinfo {author} {\bibfnamefont {J.~Ignacio}\
  \bibnamefont {Cirac}},\ }\bibfield  {title} {\enquote {\bibinfo {title}
  {{Interaction of a two-level atom with a cavity mode in the bad-cavity
  limit}},}\ }\href@noop {} {\bibfield  {journal} {\bibinfo  {journal} {Phys.
  Rev. A}\ }\textbf {\bibinfo {volume} {46}},\ \bibinfo {pages} {4354}
  (\bibinfo {year} {1992})}\BibitemShut {NoStop}%
\end{thebibliography}%

\end{document}